\documentclass{aa}  

\usepackage{graphicx}
\usepackage{txfonts}

\usepackage{natbib,twoopt}
\usepackage[colorlinks,citecolor=blue,urlcolor=blue]{hyperref} 
\bibpunct{(}{)}{;}{a}{}{,}             
\makeatletter
  \newcommandtwoopt{\citeads}[3][][]{\href{http://adsabs.harvard.edu/abs/#3}%
    {\def\hyper@linkstart##1##2{}%
     \let\hyper@linkend\@empty\citealp[#1][#2]{#3}}}
  \newcommandtwoopt{\citepads}[3][][]{\href{http://adsabs.harvard.edu/abs/#3}%
    {\def\hyper@linkstart##1##2{}%
     \let\hyper@linkend\@empty\citep[#1][#2]{#3}}}
  \newcommandtwoopt{\citetads}[3][][]{\href{http://adsabs.harvard.edu/abs/#3}%
    {\def\hyper@linkstart##1##2{}%
     \let\hyper@linkend\@empty\citet[#1][#2]{#3}}}
  \newcommandtwoopt{\citeyearads}[3][][]%
    {\href{http://adsabs.harvard.edu/abs/#3}
    {\def\hyper@linkstart##1##2{}%
     \let\hyper@linkend\@empty\citeyear[#1][#2]{#3}}}
\makeatother
\graphicspath{{./}{figures/}}
 \def\mso{\,\mathrm{M}_\odot}
 
 \def\lso{\,{\rm L}_\odot}
 \def\zso{\,{\rm Z}_\odot}
 \def\kms{\, {\rm km}\, {\rm \end{figure}

s}^{-1}}

 \def\simle{\mathrel{\hbox{\rlap{\hbox{\lower4pt\hbox{$\sim$}}}\hbox{$<$}}}}
 \def\simgr{\mathrel{\hbox{\rlap{\hbox{\lower4pt\hbox{$\sim$}}}\hbox{$>$}}}}

\begin{document}

   \title{Stripped-envelope stars in different metallicity environments}
   \subtitle{II. Type I supernovae and compact remnants}
   \titlerunning{Stripped-envelope stars in different metallicity environments: Type I SNe and compact remnants}
   \authorrunning{Aguilera-Dena et al.}
   \author{David R. Aguilera-Dena\inst{1,2,3}
          \and
          Bernhard M\"uller\inst{4}
          \and
          John Antoniadis\inst{1,3,2}
          \and
          Norbert Langer\inst{2,3}
          \and
          Luc Dessart\inst{5}
          \and
          Alejandro Vigna-G\'omez\inst{6,7}
          \and
          Sung-Chul Yoon\inst{8}
          }

   \institute{Institute of Astrophysics, FORTH, Dept. of Physics, University of Crete, Voutes, University Campus, GR-71003 Heraklion, Greece\\
              \email{davidrad@ia.forth.gr}
         \and
            Argelander-Institut f\"ur Astronomie, Universit\"at Bonn, Auf dem H\"ugel 71, 53121 Bonn, Germany
         \and
             Max-Planck-Institut f\"ur Radioastronomie, Auf dem H\"ugel 69, 53121 Bonn, Germany
        \and
             School of Physics and Astronomy, Monash University, VIC 3800, Australia
         \and
             Institut d'Astrophysique de Paris, CNRS-Sorbonne Universit\'e, 98 bis boulevard Arago, F-75014 Paris, France
         \and
            Niels Bohr International Academy, Niels Bohr Institute, Blegdamsvej 17, DK-2100,  Copenhagen, Denmark
         \and
             DARK, Niels Bohr Institute, University of Copenhagen, Jagtvej 128, DK-2200, Copenhagen, Denmark
         \and
             Department of Physics and Astronomy, Seoul National University, Gwanak-gu, Seoul, 151-742, Republic of Korea
             }

  \abstract{
Stripped-envelope stars can be observed as  Wolf-Rayet (WR) stars, or as less luminous hydrogen-poor stars with low mass loss rates and transparent winds. Both types are potential progenitors of Type I core-collapse supernovae (SNe). We use grids of core-collapse models obtained from helium stars at different metallicities to study the effects of metallicity on the transients and remnants these stars produce. We characterise the surface and core properties of our core collapse models, and investigate their explodability employing three criteria. In cases where explosions are predicted, we estimate the ejecta mass, explosion energy, nickel mass and neutron star (NS) mass. Otherwise, we predict the mass of the resulting black hole (BH). We construct a simplified population model, and find that the properties of SNe and compact objects depend strongly on metallicity. Ejecta masses and explosion energies for Type Ic SNe are best reproduced by models with $Z=0.04$ which exhibit strong winds during core helium burning. This implies that either their mass loss rates are underestimated, or that Type Ic SN progenitors experience mass loss through other mechanisms before exploding. The distributions of ejecta masses, explosion energies and nickel mass for Type Ib  SNe are not well reproduced by progenitor models with WR mass loss, but are better reproduced if we assume no mass loss in progenitors with luminosities below the minimum WR star luminosity. We find that Type Ic SNe become more common as metallicity increases, and that the vast majority of progenitors of Type Ib SNe must be transparent-wind stripped-envelope stars. We find several models with pre-collapse CO-masses of up to $\sim 30 \mso$ may form  $\sim 3 \mso$ BHs in fallback SNe. This may carry important consequences for our understanding of SNe, binary BH and NS systems, X-ray binary systems and gravitational-wave transients.}

   \keywords{Stars: massive -- Stars: Wolf-Rayet -- Supernovae: general
               }

   \maketitle
   \defcitealias{2021arXiv211206948A}{Paper I}

%

\section{Introduction} \label{sec:intro}

Most massive stars are thought to be members of binary or multiple systems that  eventually interact 
via mass transfer \citep[e.g.][]{2011IAUS..272..474S,2012Sci...337..444S,2014ApJS..215...15S,2017ApJS..230...15M}. Such binary interaction is expected to at least partially  remove the hydrogen envelope  \citep[e.g.][]{1992ApJ...391..246P,2010ApJ...725..940Y,
2019MNRAS.486.4451G,2020A&A...638A..39L,2020A&A...637A...6L,2020A&A...638A..55K,2021arXiv211110271K,2021arXiv211103329S}, leaving behind a stripped-envelope star. 
Stripped-envelope stars differ substantially from their hydrogen-rich 
counterparts in their observational properties, their evolution, and the possible outcomes that they lead to. The most massive and 
luminous among them are observed as classical Wolf-Rayet (WR) type stars. These objects are distinguished by very 
strong mass-loss rates and optically-thick winds, which lead to their characteristic emission 
line-dominated spectra,  with either strong nitrogen (WN subclass), carbon (WC subclass) or oxygen features (WO subclass, see e.g. \citealt{2007ARA&A..45..177C} for a review). It has been empirically found that there is a metallicity-dependent threshold luminosity above which stripped-envelope 
stars may become  observable as WR stars. 
\citep{2020A&A...634A..79S}. Stars below this luminosity limit are thought to have weaker, optically thin 
winds \citep{2017A&A...607L...8V,2020MNRAS.499..873S}. Such stars are also  UV-bright, but 
optically-faint, which renders them difficult to observe \citep{2001A&A...369..939W,2017A&A...608A..11G}.

In \cite{2021arXiv211206948A}, hereafter referred to as \citetalias{2021arXiv211206948A}, we presented a grid of detailed one-dimensional evolutionary models of stripped-envelope stars, and proposed a method to characterise the minimum WR luminosity as a function of metallicity. We then used stellar evolution models to explore the influence of metallicity during late evolutionary phases of stripped-envelope stars. We found that the properties of our models depend sensitively on the metallicity-dependent mass loss. Combining our theoretical and numerical findings, we constructed a model population of stripped-envelope stars, finding that the populations of WR and transparent-wind stripped-envelope stars are shaped by a combined effect of the minimum WR luminosity, and the strength of their mass loss, both of which change as a function of metallicity at different rates.

Stripped-envelope stars are also progenitors of hydrogen-poor supernovae (SNe, of Type IIb, Ib and Ic),
which are characterised by very weak or absent hydrogen features in their spectra near maximum light  \citep[e.g.][]{1997ARA&A..35..309F}.
Envelope stripping has been found to have a strong influence in the evolution of massive stars, and in the properties that determine what kind of SN explosion they  lead to \citep[e.g.][]{2019ApJ...878...49W,2021A&A...645A...5S,2021A&A...656A..58L}. With this in mind, in this paper we aim to study the effect of metallicity on stripped-envelope stars through the study of hydrogen-poor SNe, which can be observed in detail at much larger distances than their progenitors. These transients are relatively frequent in the local Universe. Most common among them are the ``normal'' Type IIb, Type Ib and Type Ic; accounting for around 30\% of all nearby core-collapse SNe \citep{1999A&A...351..459C,2005PASP..117..773V,2008ApJ...673..999P,2009MNRAS.395.1409S,2011MNRAS.412.1473L,2017PASP..129e4201S}. Less frequent sub-classes of stripped-envelope SNe include Type Ic-BL SNe, often found in association with long gamma-ray bursts \citep{2012grb..book..169H}, and Type I superluminous SNe \citep[SLSNe;][]{2016ApJ...830...13P,2016A&A...593A.115J,2018MNRAS.473.1258S}, typically found in environments of lower metallicity \citep{2014ApJ...787..138L}. A common origin for these transients has been proposed, that relies in an evolutionary channel that is distinct from the ordinary stripped-envelope SNe \citep{2018ApJ...858..115A,2020ApJ...901..114A}. 

The absence of hydrogen lines in the spectra of ordinary Type Ib and Type Ic SNe indicates that their progenitors have completely lost their hydrogen-rich envelope by the time they collapse \citep{2011MNRAS.414.2985D,2012MNRAS.422...70H}.
Both SN types have light curves that are often indistinguishable from each other \citep{2011ApJ...741...97D}. Type Ib SNe however are characterised by helium lines, which are absent in Type Ic SNe. The spectra of Type Ib SNe are consistent with stripped-envelope progenitors that end their evolution with a helium envelope that is rich in nitrogen.  Conversely, Type Ic SNe originate in stripped-envelope stars that have lost this layer through winds, and have exposed the products of helium burning  \citep[e.g.][]{2012MNRAS.424.2139D,2020A&A...642A.106D}. 
\cite{2011ApJ...731L...4M} found that Type Ib and Type IIb SNe typically originate in environments that are less metal-rich than those of Type Ic SNe. These properties indicate that metallicity-dependent wind mass loss is a key factor in the late evolution of their progenitors. 

While massive WR stars are often regarded as  progenitors of hydrogen-poor SNe \citep{1986ApJ...306L..77G},  obervationally-inferred ejecta masses  \citep{2018A&A...609A.136T,2021A&A...651A..81B} are lower than typical WR masses \citep[e.g.][]{2019A&A...621A..92S}. Rather, their observed light-curves and spectra are better reproduced by explosion models of relatively low-mass helium stars \citep[e.g.][]{2011MNRAS.414.2985D,2020A&A...642A.106D}. Some SNe have been found to have properties that suggest WR stars as more likely progenitor candidates \citep[e.g.][]{2022Natur.601..201G}, but their properties are unusual compared to the bulk of the population of stripped-envelope SNe.

Several studies of evolutionary channels for interacting binaries as likely progenitors of Type I core-collapse SNe have also been carried out in recent years. Several approaches have been taken, including detailed binary evolution \citep[e.g.][]{1999A&A...350..148W,2010ApJ...725..940Y,2017ApJ...840...10Y}, rapid binary population synthesis \cite[e.g.][]{2017ApJ...842..125Z,2018MNRAS.481.1908K,2018MNRAS.481.4009V}, and detailed evolutionary calculations of single \citep[e.g.][]{1974ApJ...194..373A,1995ApJ...448..315W,2016MNRAS.459.1505M,2017MNRAS.470.3970Y,2019ApJ...878...49W,2020ApJ...890...51E} and binary helium stars  \citep[e.g.][]{2002MNRAS.331.1027D,2003MNRAS.344..629D,2015MNRAS.451.2123T}. Evolutionary channels for Type I core-collapse SNe have been widely found in these studies, but explaining the rate, distribution of ejecta mass and SNe types remains challenging.

In a recent study, \cite{2017MNRAS.470.3970Y} found that an increase in the metallicity dependent mass-loss rate that is observed to in carbon-rich WC stars, could account for the formation of the faintest WC and WO type stars in our Galaxy. The detailed SN light curve and spectral models of \cite{2020A&A...642A.106D} confirm that an increase in the metallicity-dependent mass loss rate of WR stars can lead to the production of Type Ic SNe, and results in a sharp dichotomy in the chemical compositions and a change of trend in the initial-to-final mass relation of progenitors of Type Ic and Type Ib SNe.

Unlike the case of Type II SNe, where several progenitors have been directly identified in observations, most attempts to recognise the progenitor systems to type Ibc SNe in archival data of their fields have been unsuccessful, setting only upper limits \citep{2009ARA&A..47...63S}. So far, only two progenitor identifications for type Ib SNe \citep{2013ApJ...775L...7C,2021MNRAS.504.2073K} and a candidate for a type Ic SN \citep{2018MNRAS.480.2072K} exist. Non-detections place constrains on the progenitor stars to these explosions (and the  binary companions, if present), that suggest them to likely be hot and compact  \citep{2012A&A...544L..11Y}. The lack of hydrogen in their spectra also sets a stringent constraint on the pre-collapse surface hydrogen abundance of less than $0.001 \mso$ \citepads{2011MNRAS.414.2985D}. With this in mind, two main evolutionary channels have been proposed to explain the inferred abundances of these SNe: (i) Single massive stars becoming stripped via strong stellar winds (e.g. \citealt{2009A&A...502..611G}), and (ii) binary stars that have been stripped due to interaction \citepads{1992ApJ...391..246P,2012ARA&A..50..107L}.

Evidently, variations in surface abundances and metallicity in WR mass loss have a very strong impact on the observed properties of WR stars, as well as in SNe that might result from them. Motivated by this, we study the evolution of helium stars as a proxy for stripped-envelope stars including wind mass loss, in a similar fashion to \cite{2017MNRAS.470.3970Y} and \cite{2019ApJ...878...49W}. We extend the initial conditions to cover different metallicity environments, focusing on high metallicities, which are not often addressed in the literature, but that can account for changes in properties of populations of WR stars and Type I core-collapse SNe. We also pay particular attention to the lower luminosity limit of WR stars through the method found in \citetalias{2021arXiv211206948A}, and explore its effect in the properties of stripped-envelope SNe.

We have divided paper as follows: In Sect.~\ref{sec:methods} we briefly describe the models of stripped-envelope stars at the pre-collapse stage that we employ, and the methods we employ to analyse them. In Sect.~\ref{sec:final} we present our main results and in Sect.~\ref{sec:sn_pop} we discuss our predictions for SN and compact-object populations. We finalise our paper with a discussion in Sects.~\ref{sec:discussion} and conclusions in Sect. \ref{sec:conclusions}.

\section{Method}\label{sec:methods}

\subsection{Pre-supernova evolution of stripped-envelope stars}

To quantify the effect of metallicity on populations of stripped-envelope SNe, we study the structure and composition of grids of stellar evolutionary calculations of helium stars at core collapse, defined as the time where the iron core reaches an infall velocity of more than 1000 km s$^{-1}$. These core-collapse stellar structure models correspond to the end-points of the evolutionary calculations presented in \citetalias{2021arXiv211206948A}. Here, we give a short description of the models, but refer the reader to \citetalias{2021arXiv211206948A} for more details on the numerical and physical parameters employed in the stellar evolution calculations. 
The grids of core collapse models of non-rotating helium stars models were computed using the Modules for Experiments in Stellar Astrophysics (MESA) code \citep{MESAI,MESAII,MESAIII,2018ApJS..234...34P}, version 10398.

The  grids of evolutionary calculations analysed here correspond to seven  metallicities, from 0.01 to 0.04 in steps of 0.005, scaled from solar abundances found by \cite{1996ASPC...99..117G}. Each grid has models with initial masses between 1.5 and 70\,$\mso$ in steps of 0.5\,$\mso$. However, only the evolutionary sequences with an initial mass above $4.0-4.5$\,$\mso$ (depending on metallicity) were calculated up to  core collapse. Evolutionary sequences with lower initial masses either failed to converge to core collapse due to the presence of off-centre carbon and neon burning layers  that are numerically difficult to resolve, or do not experience to core collapse at all.

The initial models of these calculations were generated through pre-main sequence models of standard composition that were evolved with artificial mixing and no mass loss until core-hydrogen depletion. After that, they were relaxed until reaching thermal equilibrium, without mass loss, but with standard mixing. This guarantees that the surface abundances of our models correspond to CNO equilibrium abundances, with enhanced nitrogen and reduced amounts of carbon and oxygen.

Convection was modeled through mixing length theory \citep{1958ZA.....46..108B}, with $\alpha_\mathrm{MLT} = 2.0$. We adopted the Ledoux criterion for convection, and we modeled semiconvection using $\alpha_\mathrm{SC} = 1.0$. To calculate the energy generation rate and changes in chemical composition, we use the \texttt{approx21} nuclear network in MESA, which includes 21 species from $^{1}$H to $^{56}$Ni, and neutrons.

To avoid convergence issues, we employ MESA's \texttt{mlt++} \citep{MESAII}, and exclude radiative acceleration in the envelope by setting the velocity to 0 in layers with $T>10^8$ K during the late evolution.
We use the mass loss rates for WN and WC stars of \cite{2017MNRAS.470.3970Y}, who adapts the observed relations of each class, as measured by \cite{2014A&A...565A..27H} and \cite{2016ApJ...833..133T}, respectively. The mass loss rates are given by
\begin{equation}\label{eq:windWN}
\dot{\text{M}}_{\text{WN}} = f_{\text{WR}} \left(\frac{\text{L}}{\lso}\right)^{1.18}  \left(\frac{\text{Z}_\mathrm{init}}{\zso}\right)^{0.6} 10^{-11.32} \frac{{\text{M}_{\odot}}}{\text{yr}},
\end{equation}
for Y=1-Z$_{\text{init}}$, and by 
\begin{equation}\label{eq:windWC}
\dot{\text{M}}_{\text{WC}} = f_{\text{WR}} \left(\frac{\text{L}}{\lso}\right)^{0.85}  \left(\frac{\text{Z}_\mathrm{init}}{\zso}\right)^{0.25} \text{Y}^{0.85} 10^{-9.2} \frac{{\text{M}_{\odot}}}{\text{yr}}
\end{equation}
for Y$<$0.9. The intermediate range is linearly interpolated between the two regimes, using
\begin{equation}
\dot{\text{M}} = (1-x)\dot{\text{M}}_{\text{WN}} + x\dot{\text{M}}_{\text{WC}},
\end{equation}
with $x=(1-\text{Z}_{\text{init}}-\text{Y})/(1-\text{Z}_{\text{init}}-0.9)$. We set $f_{\text{WR}}$=1.58, and scale these equations assuming $\zso=$0.02.

\subsection{Explodability criteria and observable supernova properties}\label{sec:methods_exp}

To assess the explodability of the core-collapse models, we employ four different tests. First, we measure the so-called compactness parameter  at core collapse, as proposed by \cite{2011ApJ...730...70O}. It is defined as
\begin{equation}
\xi_{M}   = \frac{M/\mso}{R/1000 \ \mathrm{km}},
\end{equation}
and evaluated at mass coordinate 2.5\,$\mso$ (and therefore labelled as $\xi_{2.5}$). A high value of this quantity is indicative of a lower probability for a successful SN  within the neutrino-driven explosion paradigm. However, a clear boundary between successful and failed SN explosion cannot be drawn from this one-parameter check alone.

To complement this, we employ the two-parameter criterion proposed by \cite{2016ApJ...818..124E}. They propose an explodability test based on a relation between proxies of the core mass and the core density gradient of stars at collapse via the following parameters:
\begin{equation}
M_4 = M(s=4)/\mso,
\end{equation}
the mass coordinate at the location where the specific entropy $s$ is equal to 4;  and 
\begin{equation}
\mu_4 = \left.\frac{\mathrm{d}m/\mso}{\mathrm{d}r/1000 \ \mathrm{km}}\right|_{s=4}.
\end{equation}
The former is thought to be a good proxy of the core mass, while the latter is a measure of the steepness of the density gradient at the same location. \cite{2016ApJ...818..124E} provide a test using a combination of these two values to determine the final fate of a stellar model.

As a final approach we use the semi-analytical model proposed by \cite{2016MNRAS.460..742M}, and the updated version of the same model presented by \cite{2020MNRAS.499.3214M}, which includes the effect of fallback onto the newly formed compact object. These semi-analytical models were constructed within the context of the neutrino-driven SN explosion mechanism, and were calibrated against results of 3D  numerical simulations. These models also yield predictions for several SN and compact-object observables, such as neutron star (NS) and black hole (BH) gravitational masses, explosion energy and nickel mass produced in the resulting explosions. Using the predictions for explosion energy and remnant mass, we also calculate the kick velocity imparted on the newly-formed compact object following \cite{2018MNRAS.481.4009V}. If the initial  energy of the explosion is small, then the asymmetric inner ejecta is expected to fall back completely. This results in a weak explosion launched when the initial ejecta become subsonic. In this case, the transport of energy by the sound pulse becomes decoupled from the transport of matter. The sound pulse quickly becomes spherical and any explosion asymmetry becomes attenuated. Therefore, in cases where explosion energies are low, the kick velocity is set to 0\,km\,s$^{-1}$, as it is expected to be very small.

The model of \cite{2016MNRAS.460..742M} is a set of equations that treats the pre-explosion and explosion phases of a core collapse event in a simplified form. 
The model takes the density, chemical composition, binding energy, sound speed and entropy profiles of a stellar model at core collapse, and computes the amount of mass that is accreted by the proto-NS, and the location of the formed shock as a function of time. It then computes the location of the heating region behind the shock, and estimates if and when the neutrino heating conditions become sufficient to trigger an explosion. If the gain region acquires enough energy to produce an explosion, then it calculates the evolution of the NS mass and the explosion energy until they settle to their final state. During the explosion phase, the model accounts for energy input by neutrinos, and energy loss and gain by nuclear disassociation and explosive nuclear burning.

In the version of the model presented by \cite{2020MNRAS.499.3214M}, the shock is allowed to propagate further out even if the NS reaches the maximum mass threshold due to ongoing accretion after an explosion has been triggered. The shock is left to evolve until it either exceeds escape velocity (resulting in the termination of accretion onto the remnant) or is attenuated to a weak sound pulse that ejects a small part of the envelope.

These two explosion models depend on several variable parameters that affect the explodability of each stellar model, and the properties of the successful SN explosions. The main variable parameters in the explosion model are $\alpha_\mathrm{out}$, the volume fraction occupied by neutrino-driven outflows far away from the gain radius; $\alpha_\mathrm{turb}$, a correction to the expansion of the shock radius due to turbulent stresses; $\beta_\mathrm{expl}$, the shock compression ratio during the explosion phase; $\zeta$, the efficiency factor for the conversion of accretion energy into neutrino luminosity; $\tau_{1.5}$, the cooling timescale for a 1.5 $\mso$ NS, and the maximum gravitational mass of a NS.

We calculate the explosion properties of our models with a large range of values for all parameters. In the main text, we present the outcome of these calculations using a fiducial set of parameters. They correspond to $\alpha_\mathrm{out}=0.4$, $\alpha_\mathrm{turb}=1.18$, $\beta_\mathrm{expl}=4$, $\zeta=0.75$ and $\tau_{1.5}=1.2$. We set the maximum NS gravitational mass to {$2.05 \mso$}. This choice corresponds to a lower value of $\alpha_\mathrm{out}$ and $\zeta$, compared with the standard values used by \cite{2016MNRAS.460..742M}. We use this particular combination of parameters, as it more closely resembles the outcome expected from employing the \cite{2016ApJ...818..124E} explodability criterion. We discuss the impact of parameter variation, as well as the impact of the binding energy that is used to calculate the outcome of the explosions in Appendix \ref{sec:app_parameters}. The models from Set B, presented in Appendix\,\ref{sec:app_params}, correspond to those used by \cite{2022A&A...657L...6A}.

\subsection{Population synthesis models}\label{sec:pop_method}

To gain a better understanding of how the measurable parameters of Type I SN explosions evolve as a function of metallicity, we have generated populations of helium stars in a similar way as described in \citetalias{2021arXiv211206948A}. We randomly sample main sequence stars from a Salpeter initial mass function \citep[IMF][]{1955ApJ...121..161S}, and use relation between zero-age main-sequence (ZAMS) and helium-core masses from \cite{2019ApJ...878...49W}, given by
\begin{eqnarray}\label{eq:zams_he}
\mathrm{M}_{\mathrm{He,ini}}  & \approx&
 \left\{
\begin{array}{cc}
0.0385 \ \mathrm{M}_{\mathrm{ZAMS}}^{1.603}, & \mathrm{if}  \ \mathrm{M}_{\mathrm{ZAMS}} < 30 \mso \\
0.5 \textrm{M}_{\mathrm{ZAMS}} - 5.87 \mso, & \textrm{if} \ \mathrm{M}_{\mathrm{ZAMS}} \geq 30 \mso \\
\end{array}     \right. .
\end{eqnarray}

We count the number of transients generated, independently of the lifetime of helium stars, assuming a constant star-formation rate, and convolve this with our results from Sect.~\ref{sec:supernovae}. Counting every transient generated from this sampling, assuming that they are all stripped, and then looking at the relative distribution of their properties, is equivalent to assuming that the stripping probability is the same at each mass, and ignores the contribution of Type Ib and Type Ic SNe that are generated through other channels. While the assumptions in this population model are an oversimplification, they allow us to crudely estimate how the fate of stripped-envelope stars change across cosmic time.

A caveat in our calculations is that our models with initial masses below $4-4.5$\,$\mso$ do not reach core collapse, since they ignite nuclear burning above a degenerate core and experience convergence problems. Therefore, we cannot assess the minimum mass at which stars produce SNe, and stars with masses smaller than our least massive core collapse model that produce SNe cannot be analyzed using the \cite{2016MNRAS.460..742M} model; which does not allow us to infer their explosion properties. To account for the number of successful supernovae regardless of the results from our evolutionary calculations, we follow \cite{savvas2021} and set the lower limit in initial helium-core mass for core collapse at 3\,$\mso$, which they find to be metallicity independent if no core overshooting is included during helium burning. This is further justified since stripped-envelope stars of this mass are always below $\mathrm{L}^\mathrm{tau}_\mathrm{min,WN}$, implying that their mass loss rates are very small during core helium burning (although they may experience intense mass loss after core helium depletion, see \cite{savvas2021}). Consequently the evolution of stripped-envelope stars near the boundary between core collapse progenitors and WDs or thermonuclear explosions is  similar, independent of their initial chemical composition. We estimate the distribution of observables in this regime by extrapolating our finding for higher-mass stars. More specifically, NS gravitational masses are assigned a random value between 1.22 and 1.3\,$\mso$, explosion energies are in the range of 0.5-0.8$\times 10^{51}$\,erg, and nickel masses are in the range 0.04-0.15\,$\mso$. Kick velocities are assigned through these values, according to Eq.~2 of \cite{2020MNRAS.499.3214M}, and ejecta mass is inferred by extrapolating the initial mass--final mass relation, and removing the corresponding baryonic mass that was lost to the formation of the NS (see below for further justification).

\section{Impact of metallicity on Type I supernovae and compact remnants}\label{sec:final}

In this Section, we study the effect of metallicity on the structure and chemical composition of stripped-envelope stars at the end of their evolution, and how it affects the transients that they produce. To do this, we characterize the final outputs of the models described \citetalias{2021arXiv211206948A}. To remain consistent in our notation, in Table\,\ref{tab:notation} we recall the definitions of key quantities from \citetalias{2021arXiv211206948A}, which are important for the analysis of core collapse models.

\begin{table*}
\caption{Notation definitions.}\label{tab:notation}
\centering
\begin{tabular}{ll}
\hline \hline 
Symbol & Definition \\
\hline 
$\mathrm{L}^\mathrm{tau}_\mathrm{min,WN}$ & Minimum luminosity above which a stellar model is classified as a WN star (see \citetalias{2021arXiv211206948A})\\
$\mathrm{L}^\mathrm{tau}_\mathrm{min,WC}$ &  Minimum luminosity above which a stellar model is classified as a WC star (see \citetalias{2021arXiv211206948A})\\
$\mathrm{L}^\mathrm{evo}_\mathrm{min,WC}$ & Minimum luminosity at which a WC is produced by winds in our evolutionary calculations \\
$\mathrm{L}^\mathrm{evo}_\mathrm{min,WO}$ & Minimum luminosity at which a WO is produced by winds in our evolutionary calculations \\
$\mathrm{t}_\mathrm{total}$ & Total lifetime of a helium star from core helium ignition to core collapse \\
$\mathrm{t}_\mathrm{He-N}$ & Lifetime of a helium star spent with an outer helium-nitrogen shell \\
$\mathrm{t}_\mathrm{He-C}$ & Lifetime of a helium star spent with an outer helium-carbon shell\\
$\mathrm{t}_\mathrm{He-O}$ & Lifetime of a helium star spent with oxygen surface abundance higher than 0.05 after core helium depletion\\
\hline
\end{tabular}
\end{table*}

To determine the stellar models for which our wind prescriptions are accurate, we employ the analytical method derived in \citetalias{2021arXiv211206948A} to find the metallicity-dependent transition luminosity between WR stars, and their low mass counterparts, which have optically thin winds and lower mass loss rates. Namely, we set these luminosities as
\begin{equation}\label{eq:fit_wn}
    \mathrm{L}^\mathrm{tau}_\mathrm{min,WN} = 6.85 \times 10^{4} \left(\frac{Z}{0.02}\right)^{-0.71} \lso.
\end{equation}
For WC-type stars, the fit to the observations takes the form
\begin{equation}\label{eq:fit_wc}
    \mathrm{L}^\mathrm{tau}_\mathrm{min,WC} = 7.51 \times 10^{3} \left(\frac{Z}{0.02}\right)^{-1.55} \lso.
\end{equation}
These luminosities are converted to stellar mass estimates using the mass-luminosity relation of \cite{1989A&A...210...93L}. The explosion properties of stripped-envelope stars with luminosities below the minimum WR luminosities are also shown, but the implications of lower mass loss rates on transparent-wind stripped-envelope stars are considered a posteriori.

We have subdivided this Section into the following parts: in Sect.~\ref{sec:comps} we present the distribution of final masses and surface chemical compositions in our models. In Sect.~\ref{sec:supernovae} we characterize the internal structure of our models at core collapse and apply several tests to assess which models are more likely to produce SNe, and which ones are more likely to form BHs as a result of core collapse.

\subsection{Pre-collapse masses and chemical compositions of Type I supernova progenitors}\label{sec:comps}
\begin{figure}
\resizebox{\hsize}{!}{\includegraphics{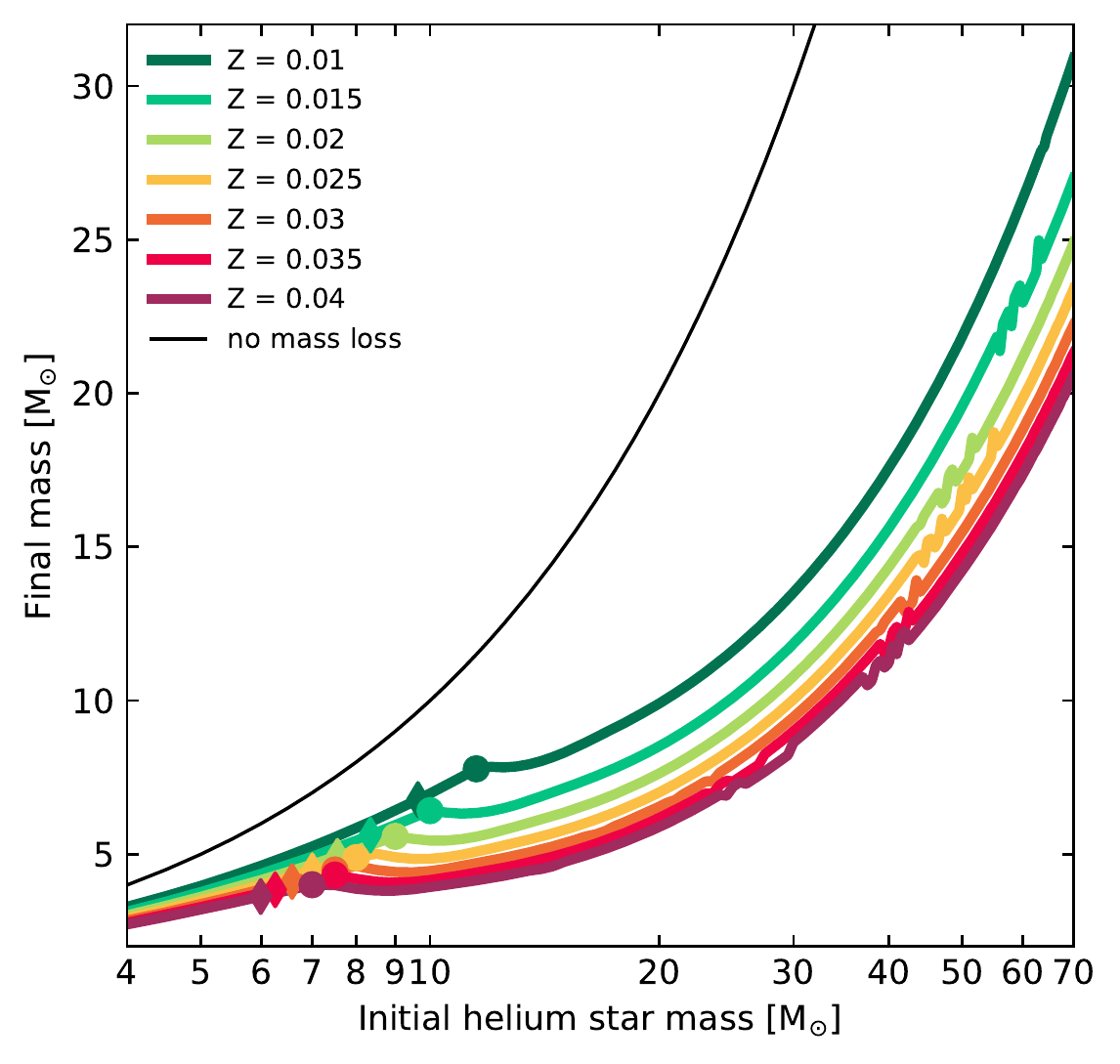}}
\caption{Final mass (at core-collapse) as a function of initial mass for helium star models of different metallicities. The dots on each line indicate, for each metallicity, the initial helium star mass at which the transition between WN- and WC-type mass loss is observed to occur in the models. The rhombi indicate the value of the minimum luminosity WN-type models, obtained Eq. \ref{eq:fit_wn}, and using the mass-luminosity relation of \cite{1989A&A...210...93L}. Helium stars below this limit do not have optically thick WR winds, and have overestimated mass loss rates.} \label{fig:finalmass}
\end{figure}

The pre-collapse masses of our helium-star models, presented in Fig.~\ref{fig:finalmass}, are determined by an interplay between their lifetime, and the intensity of their winds. A change in the trend in their final mass occurs between around 7 and 11\,$\mso$, indicated by dots in Fig.~\ref{fig:finalmass}, and it is also observed in the models of \cite{2017MNRAS.470.3970Y} and \cite{2019ApJ...878...49W}. As discussed in \citetalias{2021arXiv211206948A}, it corresponds to the metallicity-dependent transition mass between stars that lose mass as WNs for most of their lives, and those that spend a significant time as WC and WO stars with stronger mass-loss rates. The ejecta mass of SNe can be estimated, to first order, from Fig.~\ref{fig:finalmass}, but a thorough analysis of this is presented in Sects. \ref{sec:supernovae} and \ref{sec:sn_pop}.

 The type of SN that helium stars may produce not only depends on the final or ejecta mass, but also on the mass and composition of their envelope \citep{2020A&A...642A.106D}. According to \cite{2012MNRAS.422...70H}, approximately only 0.01\,$\mso$ of helium in the envelope of a hydrogen-poor star at core collapse is enough to produce detectable helium lines in the SN spectrum. This would render all of our models Type Ib SNe progenitor candidates. However, \cite{2011MNRAS.414.2985D,2012MNRAS.424.2139D} determined that the spectral type is not only sensitive to helium mass, but also to the envelope composition at core-collapse and the mixing that takes place during the SN explosion. Furthermore, the model spectra of \cite{2020A&A...642A.106D}, produced by progenitors similar to ours, show that Type Ic SNe spectra can be produced by helium stars if their mass loss is strong enough to remove the helium-nitrogen envelope. Thus, there is a sharp transition between the evolution of progenitors of Type Ib and Ic SNe, which roughly corresponds to the transition between WN-  and WC-type stars. A peculiar large-scale mixing of nickel and helium might make some WN-type progenitors appear as Type Ic SNe. While helium-deficient WR stars can only produce Type Ic SNe, the complexity of 3D mixing and of the excitation of HeI lines by non-thermal processes does not exclude the possibility that some helium-rich WR progenitors appear as helium-deficient, Type Ic SNe.

\begin{figure}
\resizebox{\hsize}{!}{\includegraphics{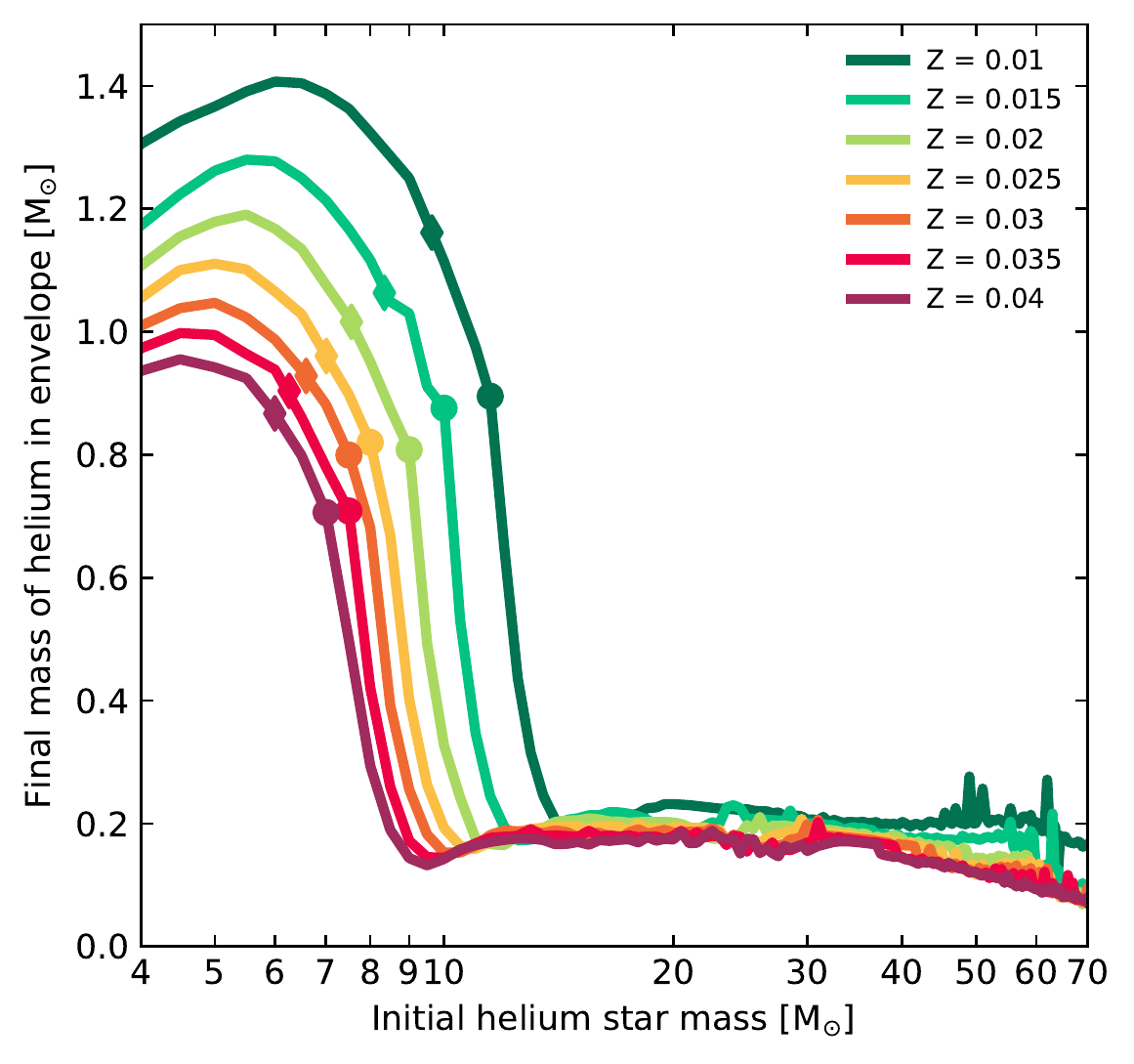}}
\caption{Final mass of helium in the envelope, as calculated by Eq. \ref{eq:he-env}, as a function of initial mass for helium stars of different metallicities. The dots on each line indicate, for each metallicity, the initial helium star mass at which the transition between WN- and WC-type mass loss, is observed to occur in the models. The rhombi indicate the value of the minimum initial mass above which helium stars are observable as WN-type stars, obtained Eq. \ref{eq:fit_wn}. Helium stars below this limit do not have optically thick WR winds, and have overestimated mass loss rates.} \label{fig:helium}
\end{figure}

\begin{figure*}
\centering
\includegraphics[width=8cm]{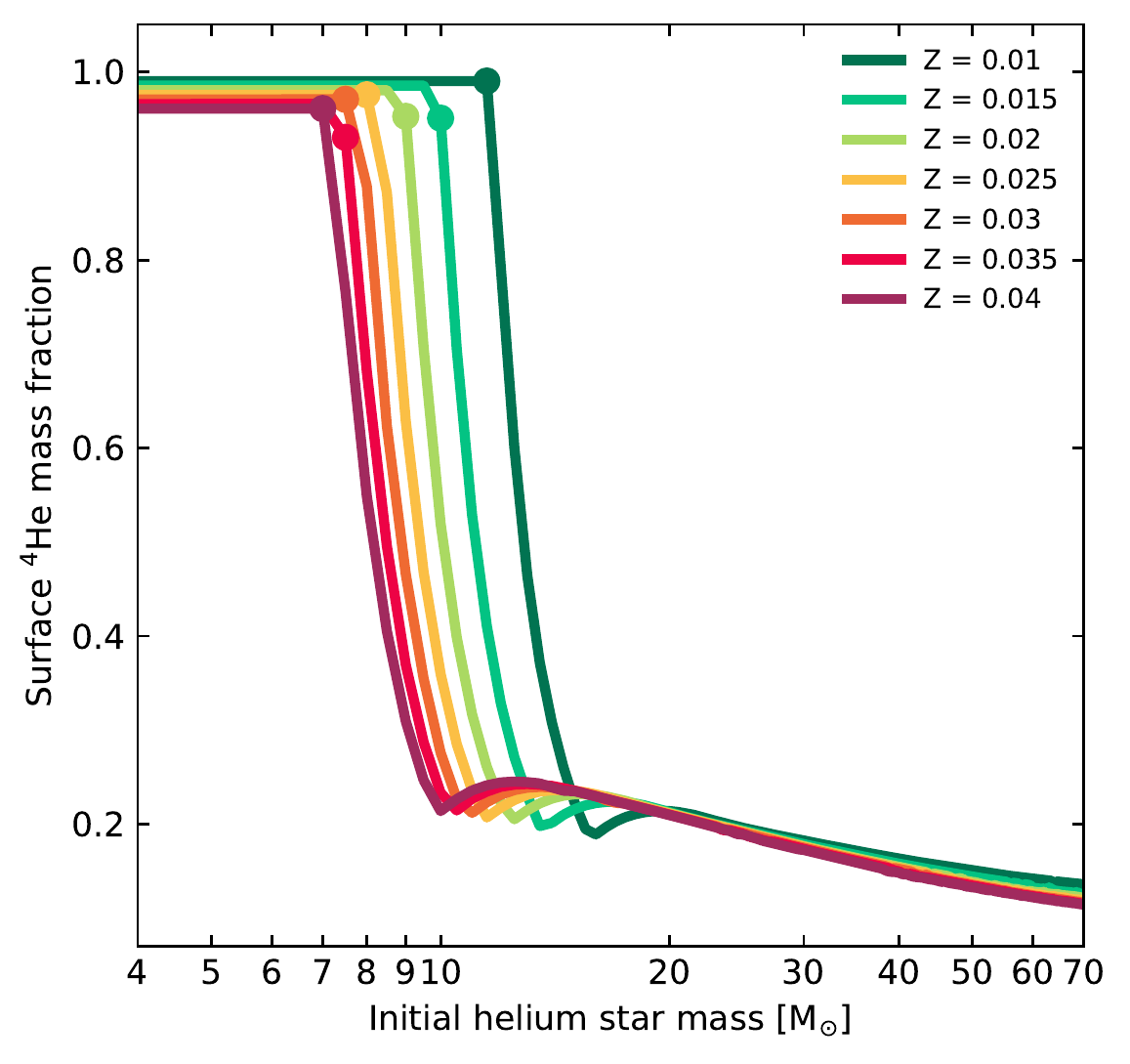}
\includegraphics[width=8cm]{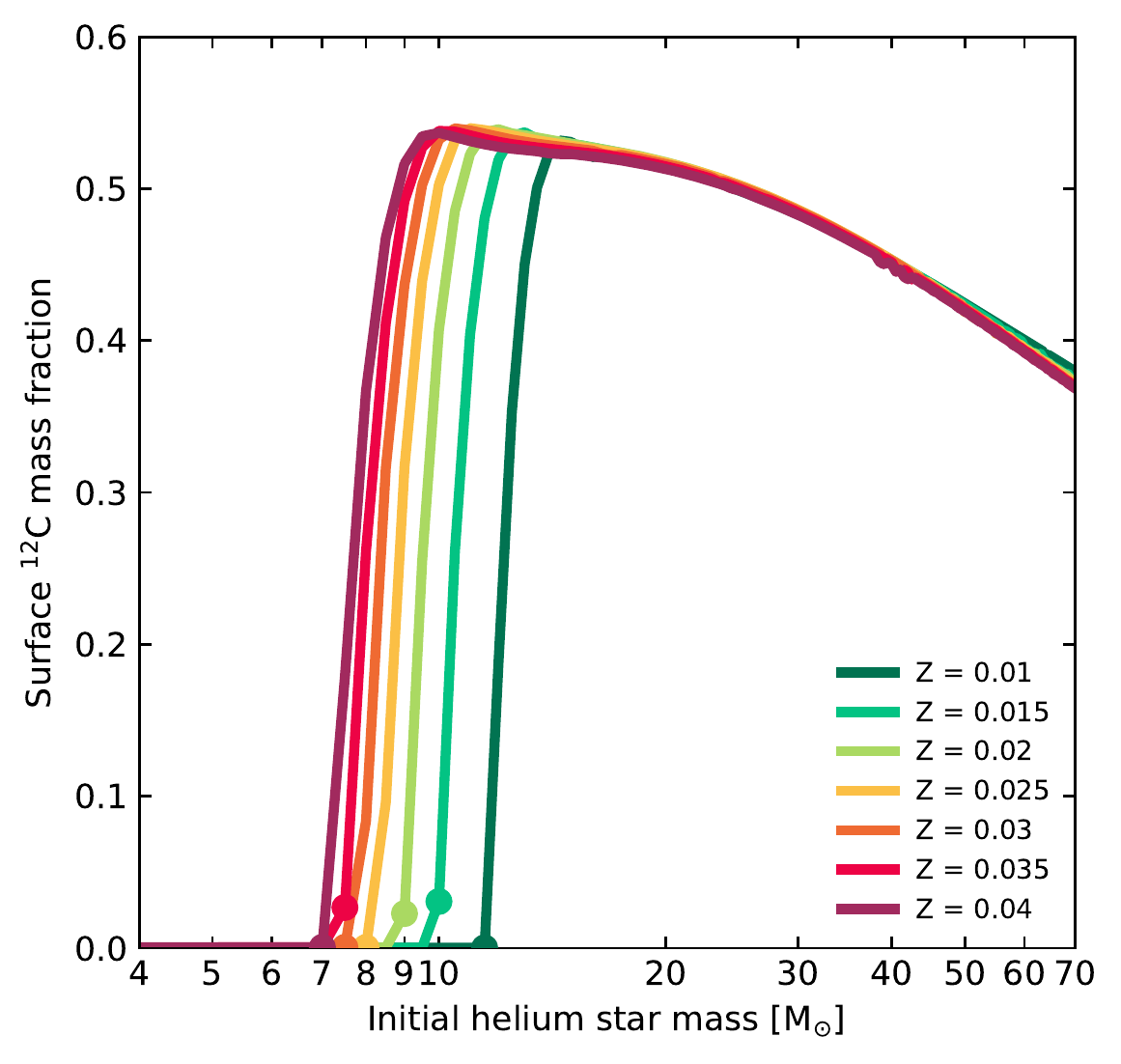}
\includegraphics[width=8cm]{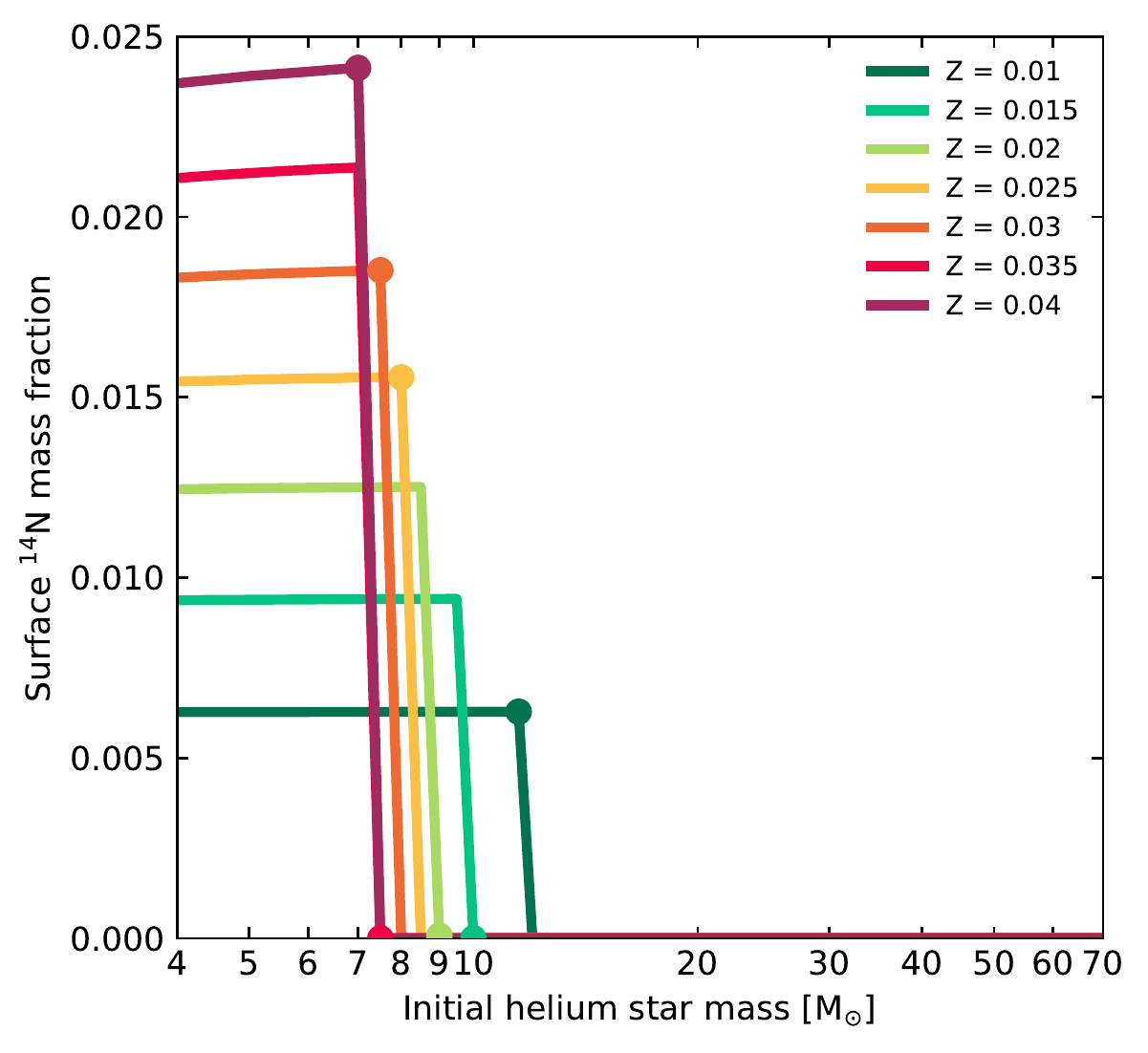}
\includegraphics[width=8cm]{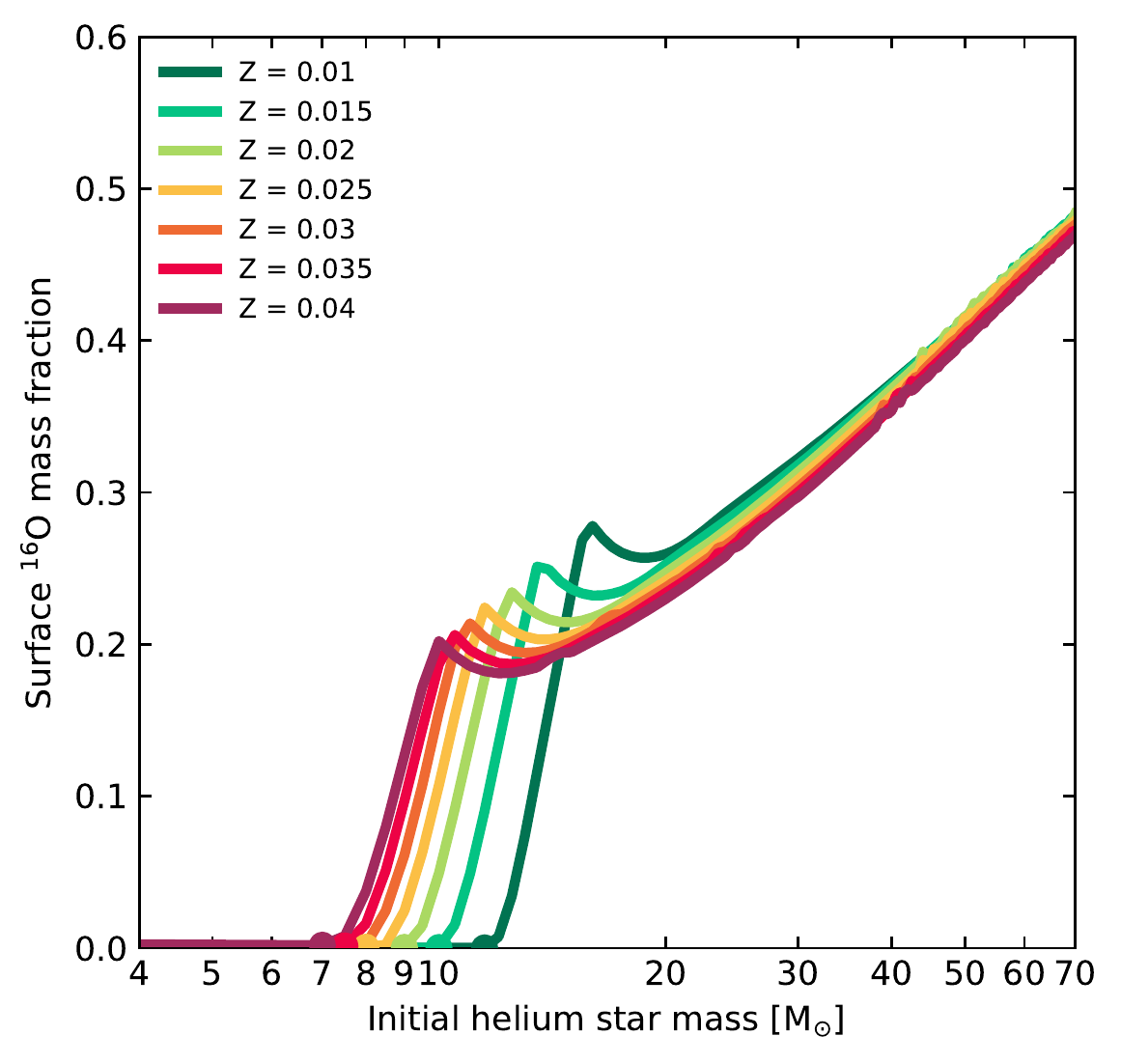}
\caption{Surface mass fractions of $^{4}$He, $^{12}$C, $^{14}$N and $^{16}$O of our helium star models at core-collapse, as a function of initial mass, for different metallicities. The dots on each line indicate, for each metallicity, the initial helium star mass at which the transition between WN- and WC-type mass loss, is observed to occur in the models.
\label{fig:surf_abundances}}
\end{figure*}

Fig.~\ref{fig:helium} shows the remaining mass of helium in the envelope as a function of initial mass, at the time of core collapse. We calculate this quantity as
\begin{equation}\label{eq:he-env}
\mathrm{M}_\mathrm{He} = \int_{\mathrm{T}<10^8 \ \text{K}} \mathrm{Y}(m) \, \mathrm{d}m,
\end{equation}
where, by only considering the regions with temperature smaller than $10^8$ K, we exclude the helium formed in the core during collapse, thereby only accounting for the envelope content. Note that this is different from the envelope mass (i.e. the total mass above the carbon-oxygen core), since only the mass in the form of helium is accounted for.

As shown in Fig.~\ref{fig:helium}, our helium star models sharply transition from having abundant helium in their envelope, to having only very little. Most models that lose mass as WN stars throughout their evolution have between 0.8 and 1.4 $\mso$ of helium in their envelopes. On the other hand, the amount of helium in the envelope converges at a value of around 0.2 $\mso$ for models with initial masses large enough to produce WC-type stars (denoted by dots in the Figure), with very few models reaching core collapse between these two regimes. The position where this transition occurs depends on metallicity, but models with intermediate helium masses occur at all metallicities, and may lead to SNe with ``intermediate'' spectra between a Type Ib and a Type Ic \citep[see model he9 from][]{2020A&A...642A.106D}. We highlight, however, that the exact spectral properties of a Type I SN depend on nickel mixing, and some WN-type stars may also produce Type Ic SNe.

Contrary to the convective helium-burning cores of hydrogen-rich stars, the convective cores of helium stars shrink in size as they evolve. As their convective cores decrease in mass, a smooth composition gradient develops in the formerly convective region, followed by a sharp transition between the helium and nitrogen envelope and the layers enriched with carbon and oxygen. The layers directly above the helium-free core later become part of a convective helium burning shell, which can grow in mass due to the smooth composition gradient left behind by the gradual retreat of the convective helium core. This means that most of the helium is stored in the nitrogen-rich envelope, and stars that lose these layers have a similarly low amount of helium at core collapse.

Total helium mass is, however, not the only factor that determines whether a star  explodes as a Type Ib or a Type Ic SN. Fig.~\ref{fig:surf_abundances} shows the behavior of the surface abundances of He, C, N and O in our core-collapse models as a function of initial mass. As can be seen, there is a clear dichotomy in surface chemical composition. In particular, we find a steep change in N abundance, but also a gradual reduction of He and an increase of C and O as a function of initial mass, transitioning from CNO equilibrium abundances to the abundances that result from helium burning. Our models have a sharp divide between these two populations of helium stars, with no intermediate cases. Surface nitrogen abundance has the sharpest drop since it burns at a lower temperature than helium, leading to a sharp transition in its abundance between the initially convective region and the envelope. The absence of nitrogen-rich layers corresponds to the transition between Type Ib and the intermediate Type Ibc and Type Ic spectra in \cite{2020A&A...642A.106D}. We therefore use that as the dividing line between the two cases in the following sections.

As initial mass increases, there is a trend for the carbon abundance to initially increase, and then decrease again. A similar trend is observed for surface oxygen, except the trend has a local maximum and minimum before continuing to increase for the highest final masses. Above $\sim$20$\mso$, final surface abundances seem to depend mostly on initial mass. This implies that this quantity is determined early in the evolution, and leads to small variations in the surface abundances for models of similar final mass, independent of metallicity.

\subsection{Properties of supernova explosions from stripped-envelope stars}\label{sec:supernovae}

Since the final masses and surface compositions of models of stripped-envelope stars with similar initial masses change with metallicity,  the relative rate of different SN types coming from them is also metallicity-dependent. This is a consequence of how the SN type depends on surface chemical composition  \citep{2020A&A...642A.106D}. However, SN rates also depend on the number of stars that actually become SNe, instead of collapsing into a black hole directly \citep{2020MNRAS.499.2803P}. The change in surface chemical composition with mass and metallicity has been addressed in Sect.\,\ref{sec:comps}. The latter issue is discussed here. First, we discuss the carbon-oxygen core masses and central carbon content at the end of helium burning in our models. Then, to gauge the explodability of our core collapse models, we employ the four methods described in Sect.\,\ref{sec:methods}. The results from these analyses are discussed here, and a different visualisation of the outcome of these tests is presented in the Appendix.

\begin{figure}
\resizebox{\hsize}{!}{\includegraphics{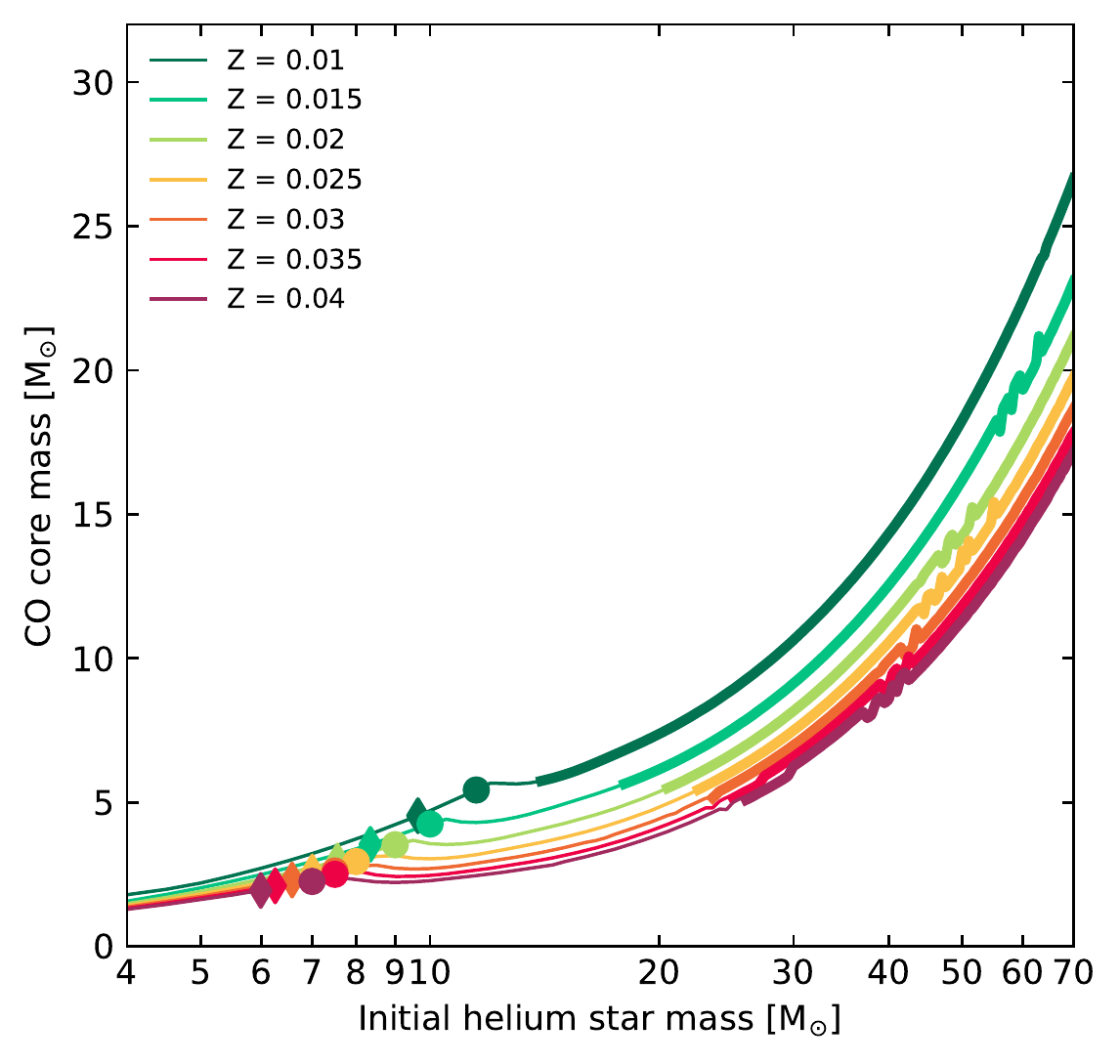}}
\resizebox{\hsize}{!}{\includegraphics{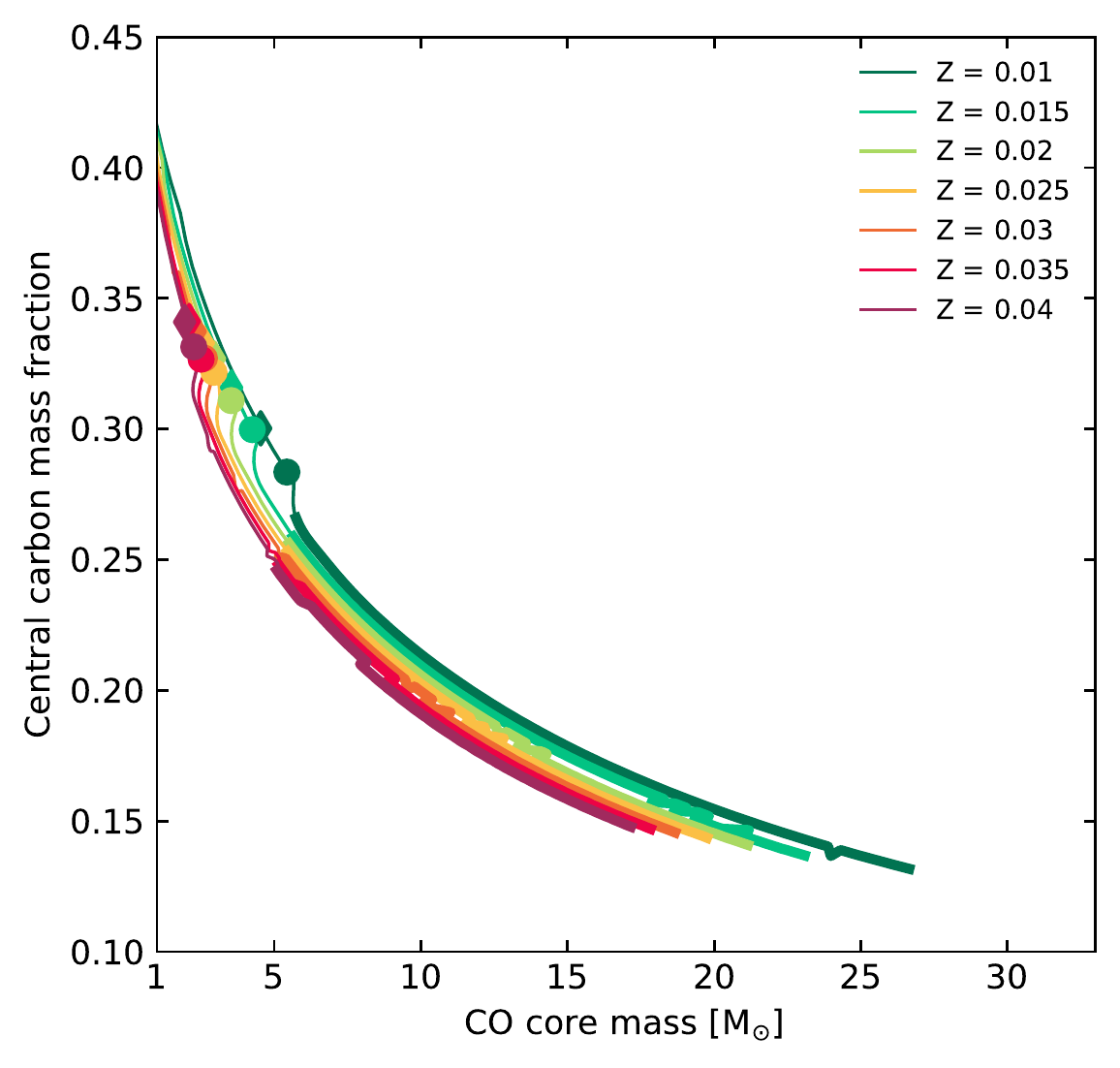}}
\caption{Top: Carbon-oxygen core mass at the end of helium burning, as a function of initial helium star mass. Bottom: Central carbon abundance at the end of helium burning, as a function of carbon-oxygen core mass. Thin lines represent models where core carbon burning occurs convectively, whereas thick lines represent models where carbon burning occurs radiatively. The dots on each line indicate, for each metallicity, the initial helium star mass at which the transition between WN- and WC-type mass loss, is observed to occur in the models. The rhombi indicate the value of the minimum initial mass above which helium stars are observable as WN-type stars, obtained Eq. \ref{eq:fit_wn}. Helium stars below this limit do not have optically thick WR winds, and have overestimated mass loss rates.} \label{fig:mco}
\end{figure}

Carbon core mass and the mass fraction of carbon at the end of helium burning are quantities which will influence the subsequent carbon and neon burning stages, and final fates of helium stars of different helium masses and metallicities \citep{2020ApJ...890...43C, 2020MNRAS.499.2803P,2020MNRAS.492.2578S}. Lower core carbon abundances at the end of helium burning result in weaker carbon burning, which does not require convection to transport the energy generated away from the burning region. This will impact the structure of the core, making it more compact, resulting in a higher gravitational binding energy, and therefore making SN explosions less likely to take place as metallicity increases \citep{2014ApJ...783...10S}. We find that our helium star models of the same initial mass will progressively have less massive carbon-oxygen cores as metallicity increases, as shown in the top panel of Fig. \ref{fig:mco}. More massive carbon-oxygen core masses will in turn result in a lower central carbon mass fraction at the end of helium burning, as shown in the bottom panel. 

The threshold mass above which models spend some time as WC-type stars before core collapse corresponds to the change in the trend of final mass as a function of initial helium mass (see Fig. \ref{fig:finalmass}), and this division is accompanied by different trends in carbon-oxygen core masses and central carbon mass fractions at the end of helium burning as a function of initial mass. This is broadly consistent with the results of \cite{2021A&A...645A...5S}, but demonstrates that the dependence of WR winds on metallicity and surface composition (or WR type) have a strong impact on the core evolution of helium stars.

\begin{figure}[ht!]
\resizebox{\hsize}{!}{\includegraphics{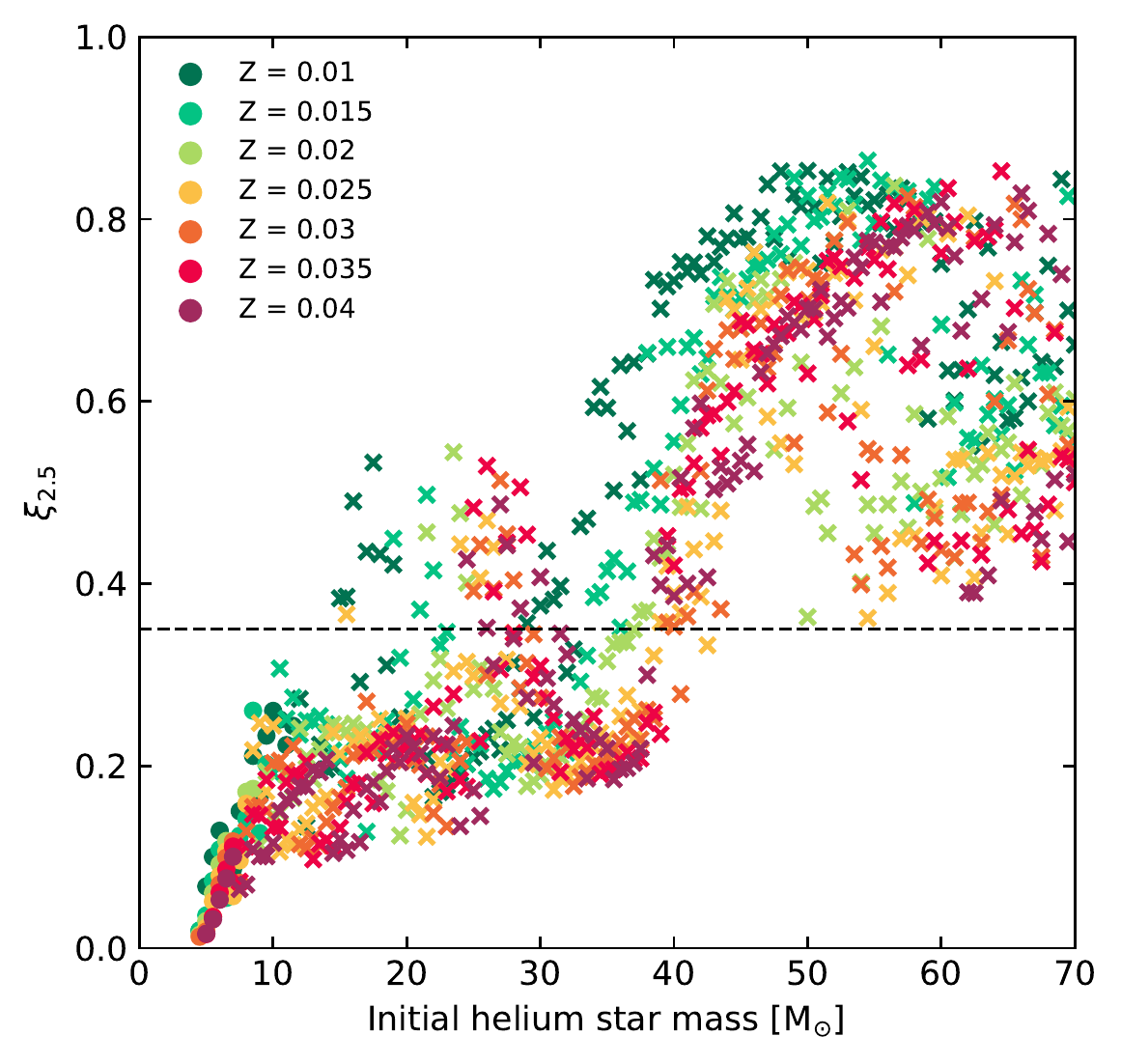}}
\resizebox{\hsize}{!}{\includegraphics{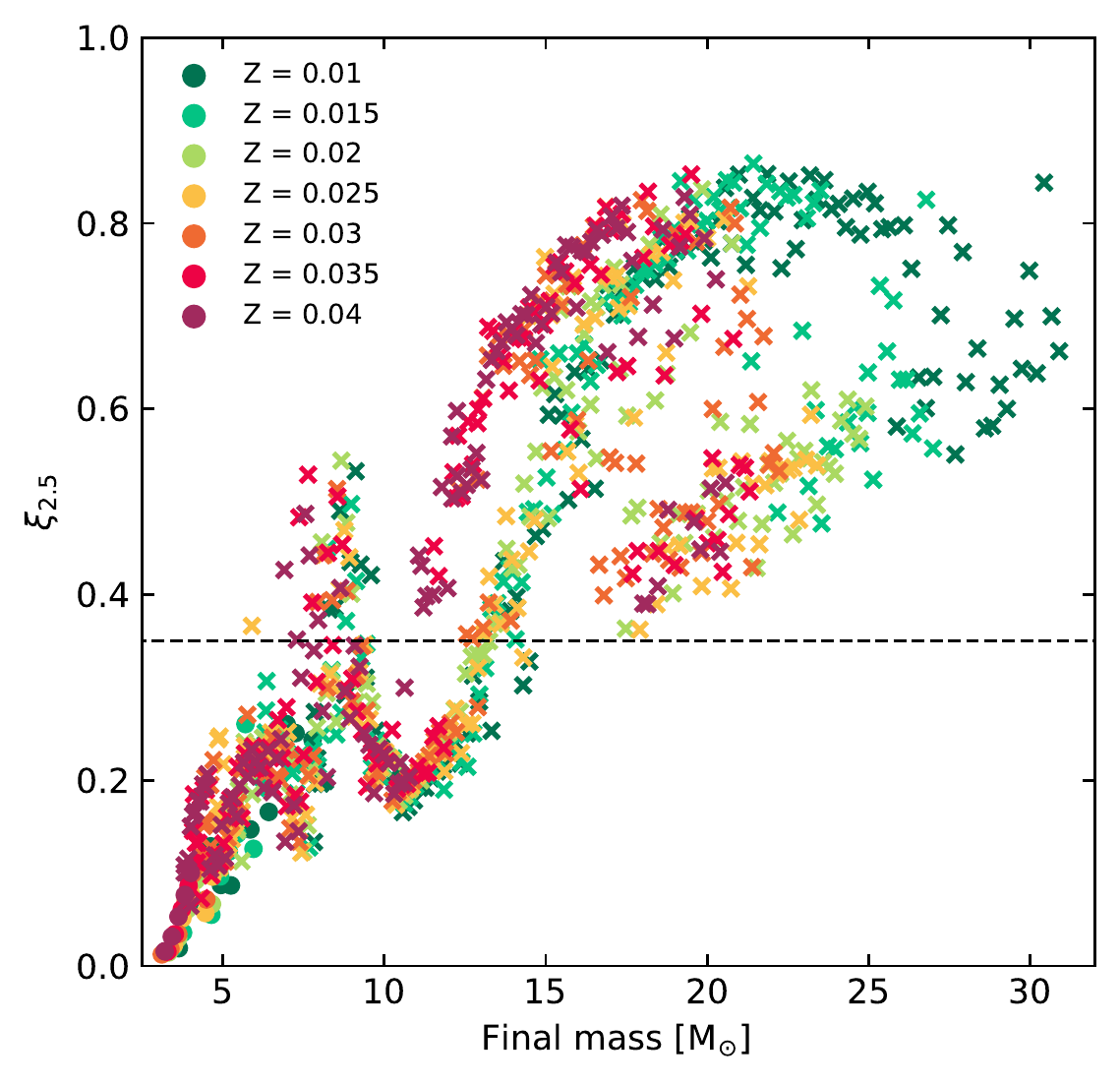}}
\caption{Compactness parameter as a function of initial mass (top) and final mass (bottom) for helium star models of different metallicities. Circles represent models with final surface abundances that correspond to Type Ib SN progenitors, whereas crosses represent models with surface abundances that correspond to Type Ic SN progenitors. The dashed horizontal line at $\xi_{2.5}$ = 0.35 divides the majority of exploding models, from those that produce BHs.
\label{fig:compactness}}
\end{figure}

Figure \ref{fig:compactness} shows the compactness parameter at core collapse of our models, as a function of both initial helium star mass and final mass. Because helium stars experience shrinking of their convective core during helium burning, a helium star of a given initial mass will often have a lower compactness than a hydrogen rich, non-stripped star whose ZAMS mass correspond to the same helium core mass at the beginning of core helium burning, and the first sharp increase in the value of $\xi_{2.5}$ will occur at a corresponding lower ZAMS mass than for stars with a hydrogen envelope \citep{2021A&A...645A...5S,2021A&A...656A..58L}. 

We observe a general trend for features in the behaviour of compactness as a function of initial helium star mass, such as peaks and valleys, to be displaced to higher initial helium star masses with increasing metallicity, correlated with the drop in final mass for similar initial masses, as shown in the bottom panel of Fig.\,\ref{fig:compactness}.

Although the compactness parameter alone is not enough to determine whether a stellar model at core collapse will lead to a successful neutrino-driven SN explosion or not, we find that most cases with $\xi_{2.5}<0.35$ are predicted to explode according to the \cite{2016ApJ...818..124E} test, as well as with the \cite{2016MNRAS.460..742M} test (see Fig. \ref{fig:all_tests}). At every metallicity, we find that most helium star models with initial mass below $\sim$ 35 - 40 $\mso$ will produce successful explosions, with a few exceptions located mostly at the peak in $\xi_{2.5}$, found at a final mass of around 8 $\mso$. The similarity in $\xi_{2.5}$ for models with similar final masses is due to the fact that the final mass is roughly determined before the beginning of carbon burning. The remaining lifetime is too short for stars to lose a significant amount of mass, and since core structure is mostly determined by the mass of the carbon oxygen core, features such as the transition between models with convective and radiative core carbon burning occur at similar final masses, correlated to the first peak in $\xi_{2.5}$ \citep{2014ApJ...783...10S}. However, variations occur between models with similar masses at different metallicities, likely related to differences in their core composition and mass, and although subtle, they have an effect in the outcome of SN and compact object populations that result from our models (see Sect. \ref{sec:sn_pop}).

An interesting feature in the behaviour of $\xi_{2.5}$ appears at final masses above 15 $\mso$, where we find a region in which two solutions to the compactness parameter appear. The two branches of values of the compactness parameter models in this region correspond to values of $\xi_{2.5}$ between 0.4 - 0.6 and 0.6 - 0.8. All values are above the limit of $\xi_{2.5} \sim 0.35$ where we expect to find exploding models, and all models are predicted to form BHs without SNe, according to both the \cite{2016ApJ...818..124E} and the \cite{2016MNRAS.460..742M} tests with our choice of parameters. However, variations in the values of parameters in the \cite{2016MNRAS.460..742M} model can lead to SN explosions in this regime, and some of these progenitor models are expected to produce fallback SNe and BHs in the lower mass gap, when analysed with the \cite{2020MNRAS.499.3214M} model (see Fig. \ref{fig:all_tests_fallback}). Because all of the explosions predicted in this regime are powered by fallback, we refer to this region as the ``island of fallback explodability'' \citep[see][for a discussion]{2022A&A...657L...6A}. The exact number of fallback SNe, as well as their properties, depend sensitively on the model parameters. However, every set of parameters we explored produces at least some fallback SNe in this regime (see Appendix\,\ref{sec:app_parameters}). Explosions could also be facilitated in this regime by the presence of strong rotation \citep[e.g.][]{2020ApJ...901..114A}, which can produce successful SN explosions even when neutrino emission is not enough to revive the SN shock \citep[e.g.][]{2018ApJ...852...28S}. However, rotation is not expected to be high in the cores of stripped-envelope stars formed by case A and case B mass transfer due to the loss of angular momentum during this phase \citep{2010ApJ...725..940Y}.

\begin{figure*}
\centering
\includegraphics[width=14.4cm]{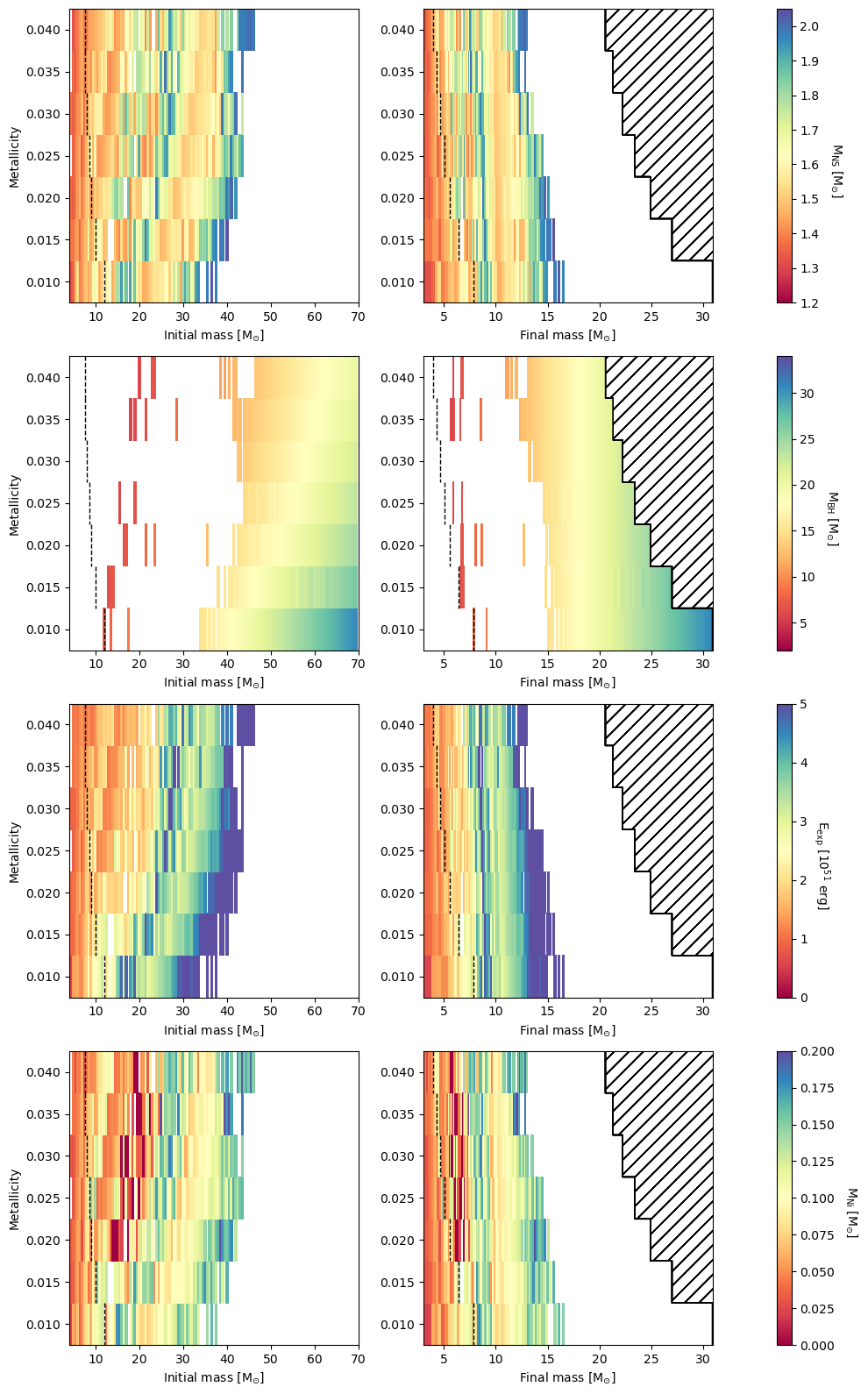}
\caption{Summary of the key parameters obtained from the core collapse models of helium stars through the explosion model of \cite{2016MNRAS.460..742M}, as a function of initial mass (left) and final (right) mass. We include the NS gravitational mass obtained for core collapse models that successfully explode, the BH mass for models that directly collapse (which corresponds to their final mass), the explosion energy and the nickel mass. Black dashed lines indicate the division between Type Ib and Type Ic progenitors, and the hashed region corresponds to areas that our model grid does not cover. The horizontal, dashed line corresponds to the minimum mass above which models reach core collapse as a WC- or WO-type star, and models to the right of this line are therefore expected to be Type Ic SN progenitors.} \label{fig:panel_mueller}
\end{figure*}

\begin{figure*}
\centering
\includegraphics[width=14.4cm]{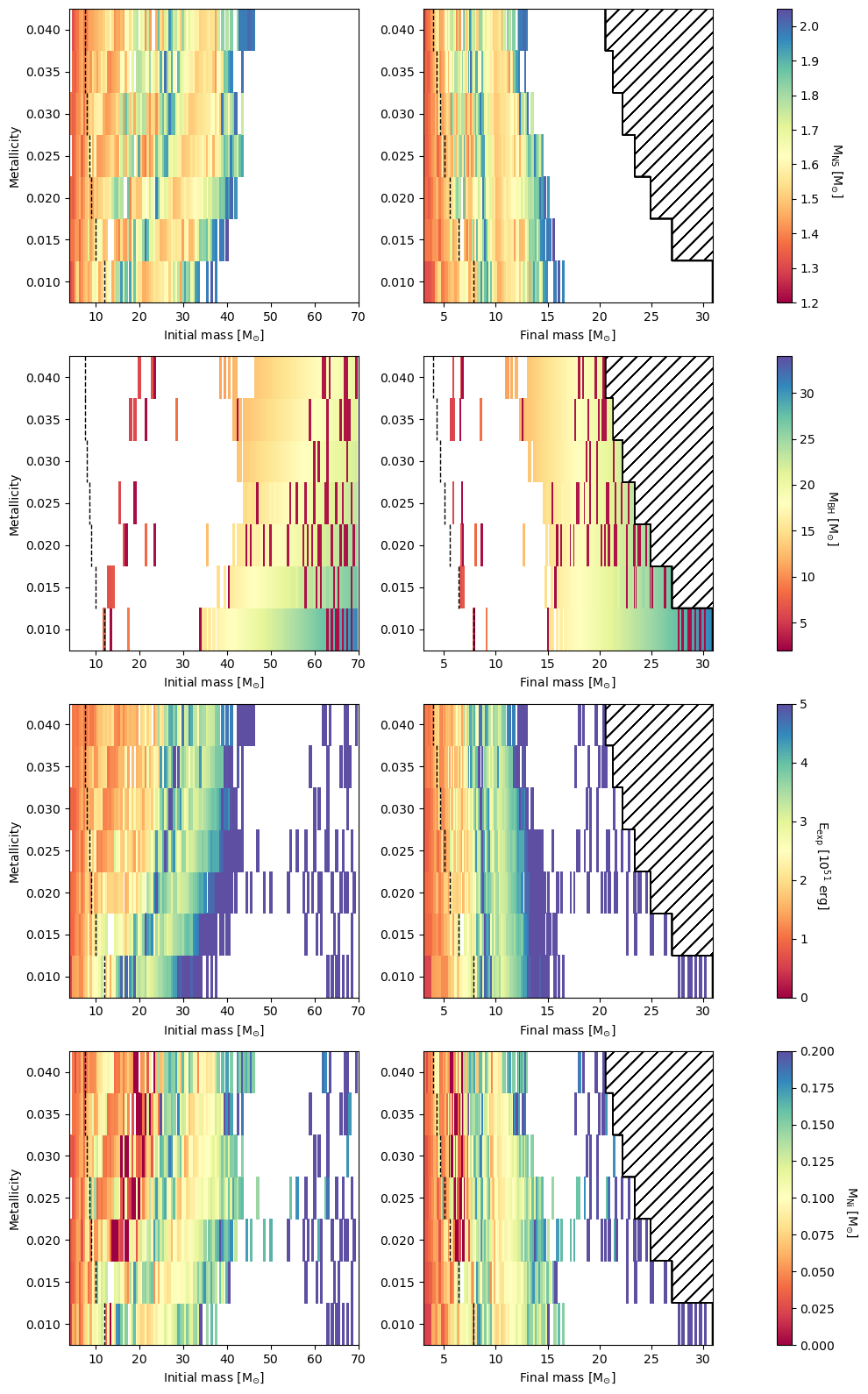}
\caption{Same as Fig. \ref{fig:panel_mueller}, but using the explosion model of \cite{2020MNRAS.499.3214M}, which includes the effect of fallback.} \label{fig:panel_fallback}
\end{figure*}

For the fiducial set of parameters we employed (see Sect\,\ref{sec:methods_exp}), the explodability, and the estimates for explosion energies, NS gravitational masses, BH masses and nickel masses obtained by analysing our models with the \cite{2016MNRAS.460..742M} and \cite{2020MNRAS.499.3214M} explosion models are summarized in Figs.\,\ref{fig:panel_mueller} and \ref{fig:panel_fallback}, respectively. The black line in the Figures indicates the value of the minimum mass above which models evolve as WC-type stars, which we take as a threshold below which successfully exploding stars are expected to be observable as Type Ib SNe, and above as Type Ic. We find that most of these quantities are comparable to those inferred from stripped-envelope SN observations \citep[e.g.][]{2018A&A...609A.136T,2021A&A...651A..81B}, but discuss this in more detail in Sect. \ref{sec:sn_pop}.

As \cite{2020ApJ...890...51E} had already found with their models with enhanced mass loss, the initial (helium) masses of stars located in the so called ``islands of explodability'' become displaced if wind mass loss rates are different. However, as Figs. \ref{fig:panel_mueller} and \ref{fig:panel_fallback} suggest, final mass is not the only defining factor that determines the explodability of a model and the properties it will have upon explosion. A region at around final masses of $\sim$8 $\mso$ where low mass BHs can form exists for most metallicities, but the location and extent of this region varies significantly as a function of metallicity. There is a tendency for the highest final mass where explosions take place to increase with decreasing metallicity (see Fig. \ref{fig:panel_mueller} and Sect. \ref{sec:sn_pop}). This implies that explosions with high ejecta masses are more likely to occur in low metallicity environments. Such explosions also tend to be more energetic. However, the exact morphology and location of these regions depends on the choice of parameters in the \cite{2016MNRAS.460..742M} model. Since the maximum ejecta mass of Type Ic SNe decreases in high metallicity environments, and the minimum mass of BHs that come from these very massive stars decreases twofold, both because of the increase in mass loss but also due to this effect.

As shown in Fig. \ref{fig:panel_fallback}, the number and the mass of models that experience fallback also varies as a function of metallicity, but it is a relatively small number. Many fallback SNe are predicted to be more energetic explosions (with energies ranging from $6\times 10^{51}$) erg, and up to a few times $10^{52}$ erg), with higher nickel masses than the rest of the sample (often in excess of 0.2 $\mso$). Including the effect of fallback in our model does not change the predictions of the properties of the explosions we calculated without including fallback, but produces models that successfully explode in places where explosions were not originally expected. However, this depends on the input parameters of the model, and we may underestimate transients that originate from progenitors of lower mass, which are likely more numerous in the Universe (see Appendix \ref{sec:app_parameters}).

NS gravitational masses predicted for our models range between 1.32 and 2.03 $\mso$. It is noteworthy that we observe a tight correlation between compactness and NS mass (see Fig. \ref{fig:everything}), and that the smallest NS mass is larger than the least massive known NS \citep{2016arXiv160501665A}. We attribute this to the lack of core-collapse models below 4 $\mso$ in our grid. The maximum NS mass we find is close to the maximum known NS mass \citep{2020NatAs...4...72C}, and it is also close to the limit of 2.05 $\mso$ that is imposed in the \cite{2016MNRAS.460..742M} model. Exploring the effect of variations on this maximum mass in the resulting SN parameters might help understand the origin of the distribution of the NS mass distribution, but is beyond the scope of this work.

The real distribution of compact object masses cannot be directly drawn from our single stellar models. One reason for this, in particular for BH masses, is that the range of final masses in our models, which will roughly correspond to the range in BH masses, is metallicity dependent. Another reason is that we do not have models in the range of (pulsational) pair instability, which will in turn determine the maximum mass of BHs. Furthermore, binarity might have a role in determining the distribution of BH masses in the Universe beyond that of creating stripped-envelope stars. However, this distribution has been computed by other authors in different works \citep{2020ApJ...902L..36F,2021ApJ...912L..31W}, and it is particularly relevant at metallicities lower than that of our models. It is believed that low-metallicity environments are more efficient at producing the binary BHs we observe as mergers today \citep[e.g.][]{2019MNRAS.490.3740N}.

Regardless, we can draw some conclusions from the distributions in Figs. \ref{fig:panel_mueller} and \ref{fig:panel_fallback}. For instance, a lower mass BH population comes for models with final masses of around 6--9 $\mso$. This distribution will dominate the BH mass distribution, as it is favored by the IMF, and the number and mass of said BHs depends on metallicity. The BHs in this regime produced at low metallicity will peak at the final masses of low metallicity models since they correspond to stars of lower initial mass. Similarly, the most massive BHs formed by helium stars will be formed at low metallicity, regardless of whether the maximum BH mass depends or not in metallicity. In the case where the effect of fallback is considered, we also expect a population of mass gap BHs to appear.

The explosion energies predicted for our models range from about 0.5 $\times 10^{51}$ erg, up to about 5.6 $\times 10^{51}$ erg; and the kinetic energy is correlated with the final mass (see Fig. \ref{fig:everything}), likely due to its dependence on the binding energy of the progenitor at core collapse. We find that the most energetic explosions correspond to those models that have higher values of $\xi_{2.5}$, which also correlates with high values of $\mu_4\mathrm{M}_4$ (in the regime where models explode), implying that high energy explosions come from models that are on the verge between a successful explosion, and collapsing into a BH. Progenitors that are predicted to form a low mass BH, but explode nonetheless due to fallback accretion according to the \cite{2020MNRAS.499.3214M} model have extremely high energies that reach up to $\sim 1.2 \times 10^{52}$ erg. Such explosions, however, come from very high mass stars, and thus might form a rare sub-class of Type Ic SN. 

The nickel masses that we predict are predominantly in the range of 0.025--0.15 $\mso$; somewhat lower than what is inferred for most stripped envelope SNe \citep[e.g.][but see Sect. \ref{sec:energies} for a discussion about potential biases]{2018A&A...609A.136T,2021A&A...651A..81B}. These nickel yields have a large spread as a function of initial and final helium star mass, but they are correlated very tightly with $\mu_4$, which may reflect the size of the region where nickel is synthesized during the explosion. Fallback SNe found in our models are predicted to have higher nickel masses, in the range of 0.2--0.3 $\mso$, with the exception of a model that produces $\sim$0.45 $\mso$.

\begin{figure}
\centering
\resizebox{\hsize}{!}{\includegraphics{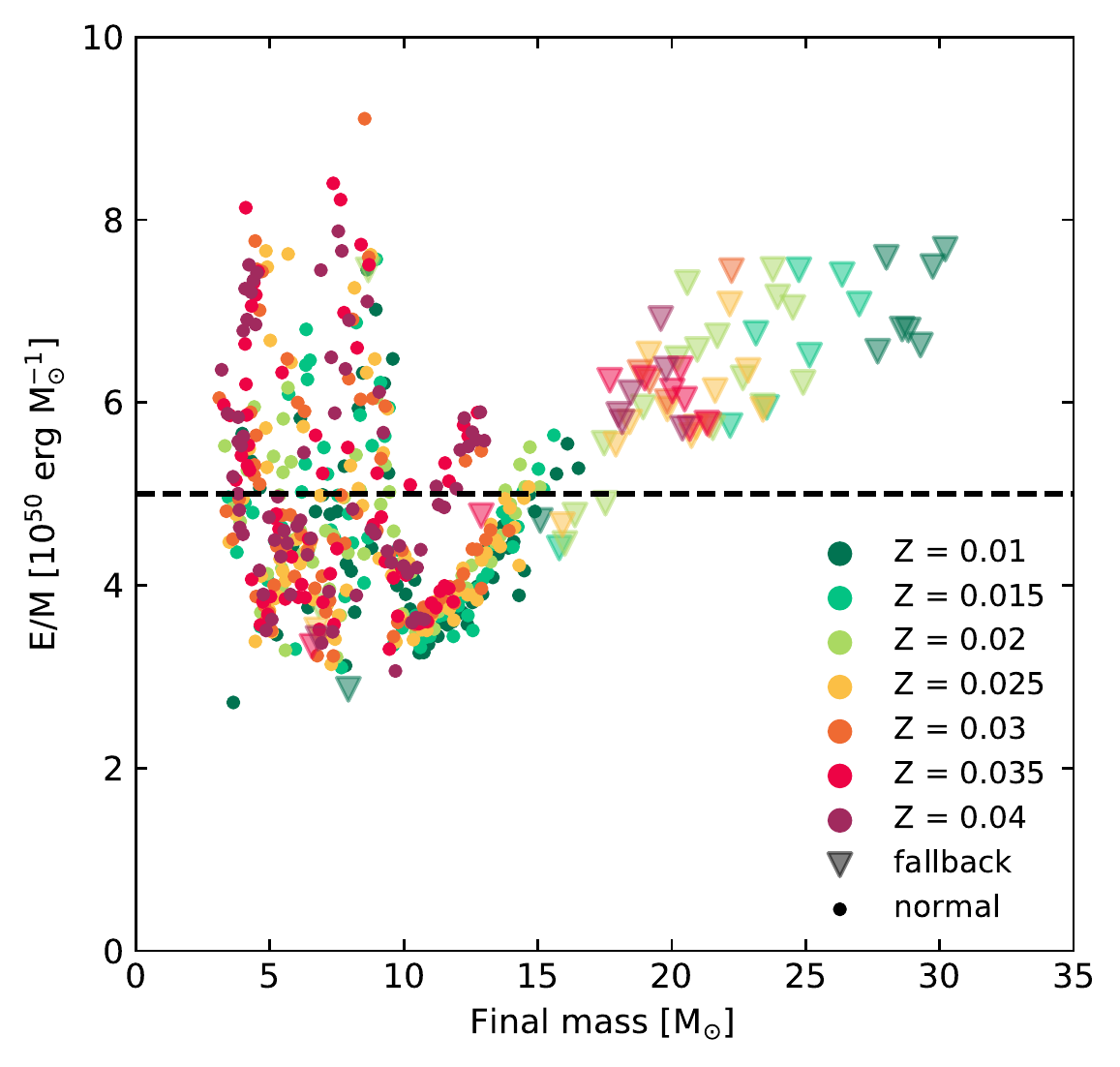}}
\caption{Ratio of explosion energy to ejecta mass  inferred using the \cite{2020MNRAS.499.3214M} model,as a function of final mass. The horizontal line at $5 \times 10^{50} \ \mathrm{erg}/ \mso$ shows the value of $\mathrm{E}/\mathrm{M}$ used in \cite{2020A&A...642A.106D} to produce light curves of Type I SNe.
\label{fig:e_over_m}}
\end{figure}

In Fig.\,\ref{fig:e_over_m} we present the ratio of explosion energy to ejecta mass, $\mathrm{E}/\mathrm{M}$, that we find using the \cite{2020MNRAS.499.3214M} model. We find that most models have a value of this quantity that ranges between 3 and $8\times 10^{50} \ \mathrm{erg}/\mso$. Models that are predicted to produce fallback SNe are found to have only large values of this quantity, while ordinary SNe cover the whole range. This is similar to the value of $5 \times 10^{50} \ \mathrm{erg}/ \mso$ used in \cite{2020A&A...642A.106D}, that was found to reproduce the light curves of Type I SNe, but we predict a significant variation of this parameter, allowing for SNe to have varying light curve properties depending on their progenitor.

The narrow range of values $\mathrm{E}/\mathrm{M}$, along with the narrow range in nickel masses predicted for the majority of our SN models, regardless of whether they are Type Ib or Ic, is consistent with the fact that the light curves of Type Ib and Ic SNe are often indistinguishable from each other \citep{2011ApJ...741...97D}.

We highlight, however, that the uncertainties in the predictions of explodability, explosion energy, nickel mass and remnant mass are more uncertain in BH forming SNe than in NS forming explosions. However, we find that the trends and correlations we present are  robust, but vary depending on the choice of parameters (see Appendix\,\ref{sec:app_parameters}). Ejecta masses predicted from our models are found to be larger at low metallicity for a fixed initial mass. This fact, combined with the behaviour of the transition mass between nitrogen- and helium-rich progenitors, and nitrogen-poor ones, implies that the maximum ejecta mass at which Type Ib SNe are observed will be larger at low metallicities. The maximum ejecta mass at which Type Ic SNe can be observed in different metallicity environments is also expected to decrease at high metallicity, due to the fact that the maximum final mass at which explosions take place decreases with increasing metallicity.

All of these results correspond purely to the outcome of our models, but do not correspond to number of observable events, or the real distribution of SN observables. An estimate of how these distributions will change is estimated in the next Section through a simplified population model.

\section{Type I supernova and compact object populations}\label{sec:sn_pop}

In this Section, we discuss the properties of Type I SN and compact remnants that result from the simple population synthesis model described in Sect. \ref{sec:pop_method}. In Sect. \ref{sec:nsbh_pop} we present the resulting compact object populations, and in Sect. \ref{sec:sne_pop} we discuss the resulting distribution of SN observables.

\subsection{Compact object populations}\label{sec:nsbh_pop}

\begin{figure}
\centering
\resizebox{0.9\hsize}{!}{\includegraphics{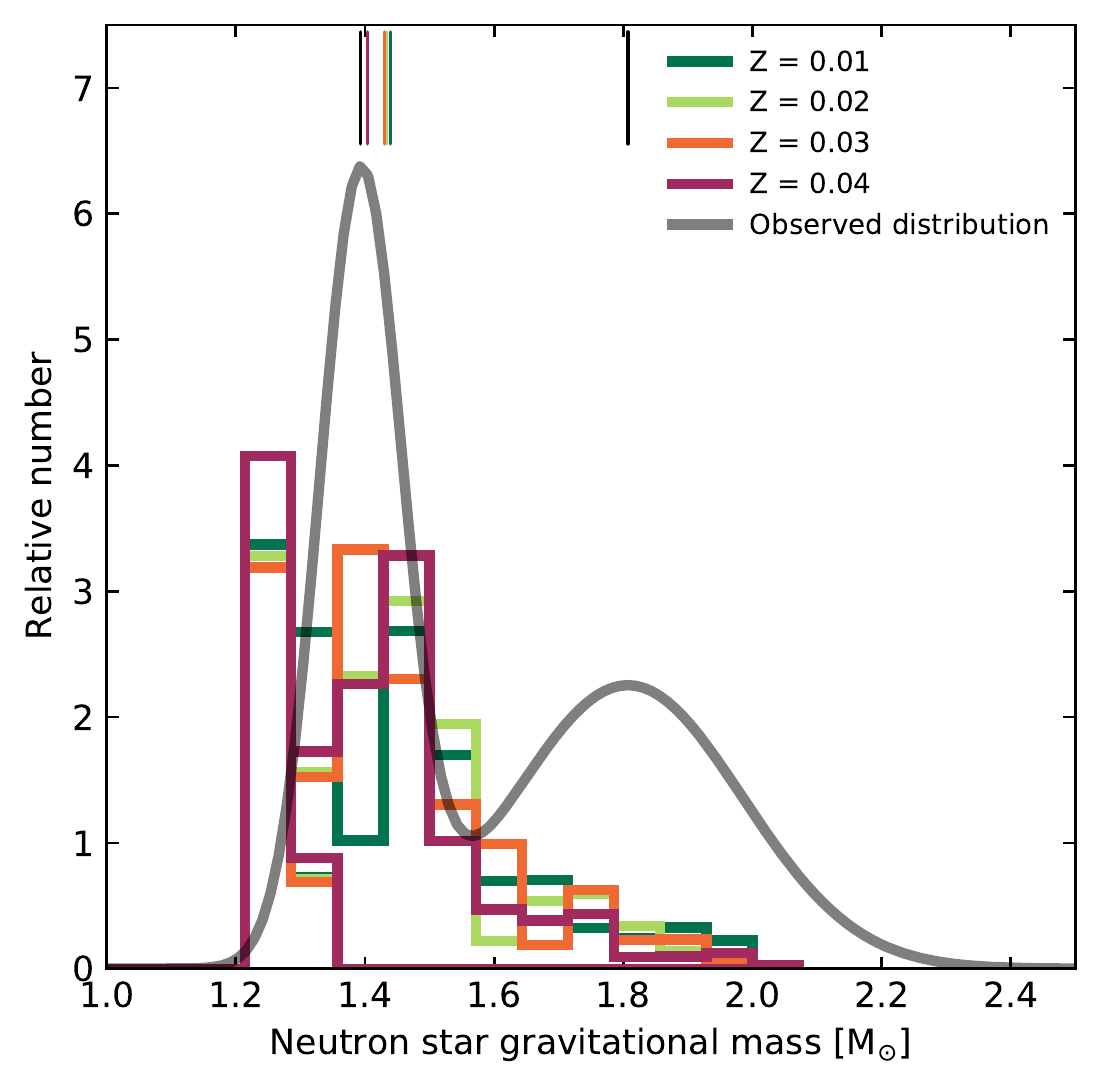}}
\resizebox{0.9\hsize}{!}{\includegraphics{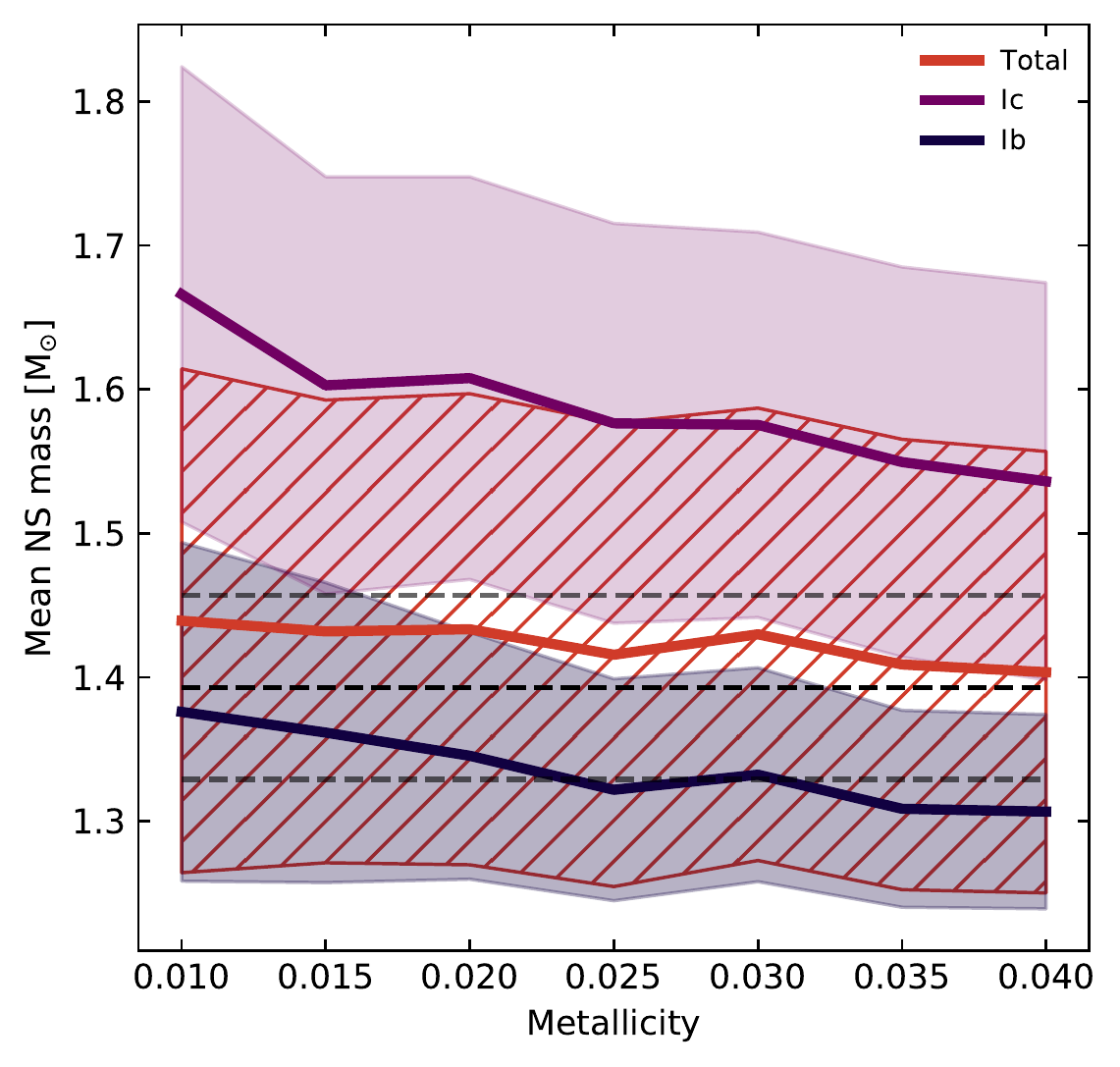}}
\caption{Top: Distribution of NS gravitational masses that result from analyzing our core-collapse models with the \cite{2020MNRAS.499.3214M} model, weighted by the IMF from \cite{1955ApJ...121..161S}, compared to the observed distribution inferred by \citep{2016arXiv160501665A}. The average value of the distributions are represented by vertical lines at the top of the Figure. The lowest mass neutron stars correspond to progenitors outside of our model grid, and their resulting mass has been assumed to be between 1.22 and 1.3 $\mso$. Bottom: Mean value (solid lines) and standard deviation (hatched, shaded regions) of the total distributions, and the distributions separated by SN type. The mean value (black dashed line) and standard deviation (grey dashed lines) of the first NS mass peak of the observed distribution of \citep{2016arXiv160501665A} is shown for comparison.
\label{fig:NS_hist}}
\end{figure}

The resulting distributions of NS masses, presented in Fig. \ref{fig:NS_hist}, depend only weakly on metallicity. We find that the average NS mass is around 1.42 $\mso$, but the distributions are skewed towards slightly lower NS masses as metallicity increases. The distributions resemble the low mass peak in the distribution of millisecond pulsars inferred by \cite{2016arXiv160501665A}, but their contribution to the high mass peak is negligible. These distributions are not affected by our implementation of fallback, but we do not rule out that some massive NS may be formed by late fallback in models where we do not predict it with our formulation. We also find that the lowest mass NSs in our populations are formed in Type Ib SNe, whereas the most massive components originate in Type Ic SNe, as shown in the lower panel of Fig. \ref{fig:NS_hist} (see Fig. \ref{fig:everything}). This is due to the correlation between $\xi_{2.5}$ and resulting NS mass. Type Ib SNe are produced in models whose final masses are predicted to be below the first compactness peak, and most of them have low values of $\xi_{2.5}$ due to their smaller final mass. Progenitors of Type Ic SNe, on the other hand, have a larger scatter of values of $\xi_{2.5}$ since many of them come from models in the region where core carbon burning transitions from radiative to convective, and have a larger scatter in $\xi_{2.5}$ \citep{2014ApJ...783...10S}, and consequently in NS mass. Alternatively, the calculations presented in Sect.\,\ref{sec:app_ener} have a few cases where high mass NSs are produced in low energy explosions, that likely do not receive sizable  kicks. In this scenario, although rare, these systems may be favourably created in binary systems that statistically are much more likely to stay bound after explosion, therefore biasing the sample of NSs in binary systems.

\begin{figure}
\centering
\resizebox{\hsize}{!}{\includegraphics{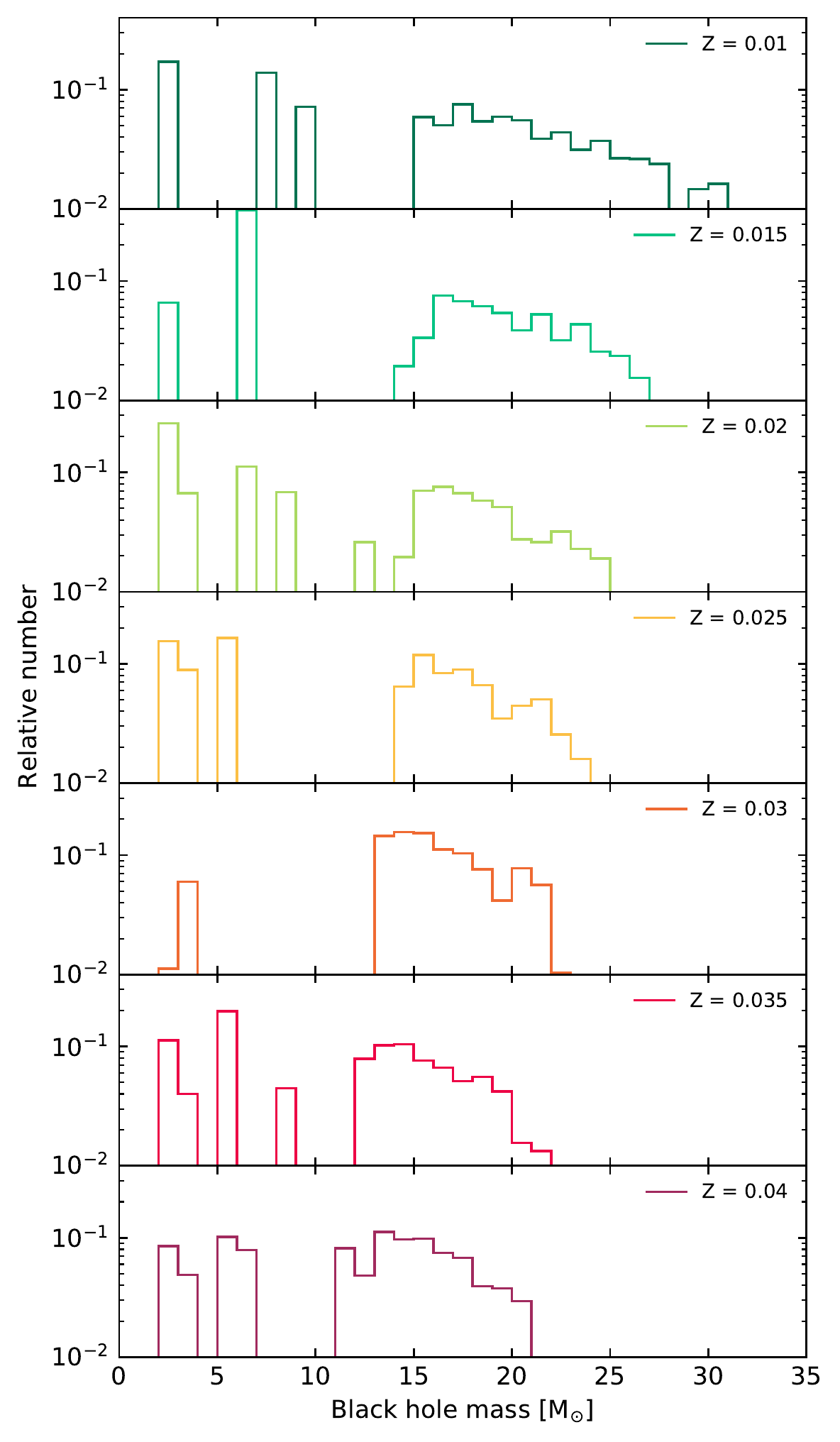}}
\caption{Distribution of BH masses at different metallicities, that result from analyzing our model with the \cite{2020MNRAS.499.3214M} model, weighted by the IMF from from \cite{1955ApJ...121..161S}.
\label{fig:BH_hist}}
\end{figure}

The distribution of BH masses that results from our calculations, shown in Fig. \ref{fig:BH_hist}, has a stronger metallicity dependence. The peak that corresponds to the high mass BHs, with masses larger than $\sim 10 \mso$, comes from the models in our grid with the highest initial masses. This peak tends to be at a higher mass for lower metallicities, owing both to the decrease in mass loss, and to the decrease in final mass of the location of the threshold between models that form BHs and models that form SNe.

The more abundant intermediate mass peak, located at 5--8$\mso$, corresponds to models that form BHs in the first high compactness peak. Given our assumptions on the population, we expect that BHs of this mass should be much more frequent than the high mass models. 

There is a gap between these two regions, also present in the results of other groups \citep[e.g.,][]{2020ApJ...890...51E,2020ApJ...896...56W}, but the precise width and location of this gap is found to be metallicity dependant. Extrapolating from our results might suggest predicting the location and shape of this gap is a difficult (but necessary) effort. The location of this gap at some metallicity may be masked by BHs coming from different environments, so establishing a metallicity behaviour for this gap, and convolving it with our knowledge of the metallicity evolution of galaxies of different redshifts, is key to properly interpret the observations of BH masses in the Universe, which will become more numerous in the future, with the advent of ongoing and future gravitational-wave observatories.

Our models show also an apparent dearth of BHs above a maximum mass, but this limit is artificial and corresponds only to the maximum mass of the models in our grid.

Fallback supernovae produce a low mass peak in the distribution, predicted to be at around 3 $\mso$, and generally less populated than the other two. However, this is sensitive to the number of fallback SNe produced, which may vary in reality. If fallback SNe tend to appear in a certain range of masses, they can create patterns in the distribution of high mass BHs, which manifests itself as valleys where fewer BHs of a certain mass are found, since they instead inhabit the lowest mass peak. If this is indeed the case, then the need arises to study this effect in more detail, to be able to better interpret the distributions of BH masses that will become available from future observations of binary BH mergers. The formation of such systems may help explain the asymmetric compact object mergers that have been recently detected by LIGO \citep{2022A&A...657L...6A}.

A final trend that is noteworthy in the BH mass distribution is a change in the slope of the distribution of BHs above a certain mass, also found by \cite{2020ApJ...896...56W}, caused by the change in the implemented mass loss rate from WN-type mass loss to WC-type mass loss; which corresponds to the change in trend in final mass shown in Fig. \ref{fig:finalmass}. If this is resolved with the increasing number of gravitational wave sources observed, it could potentially aid in constraining the mass loss rates of very massive stars.

\subsection{Distribution of supernova properties}\label{sec:sne_pop}

\begin{figure*}
\centering
\includegraphics[width=6cm]{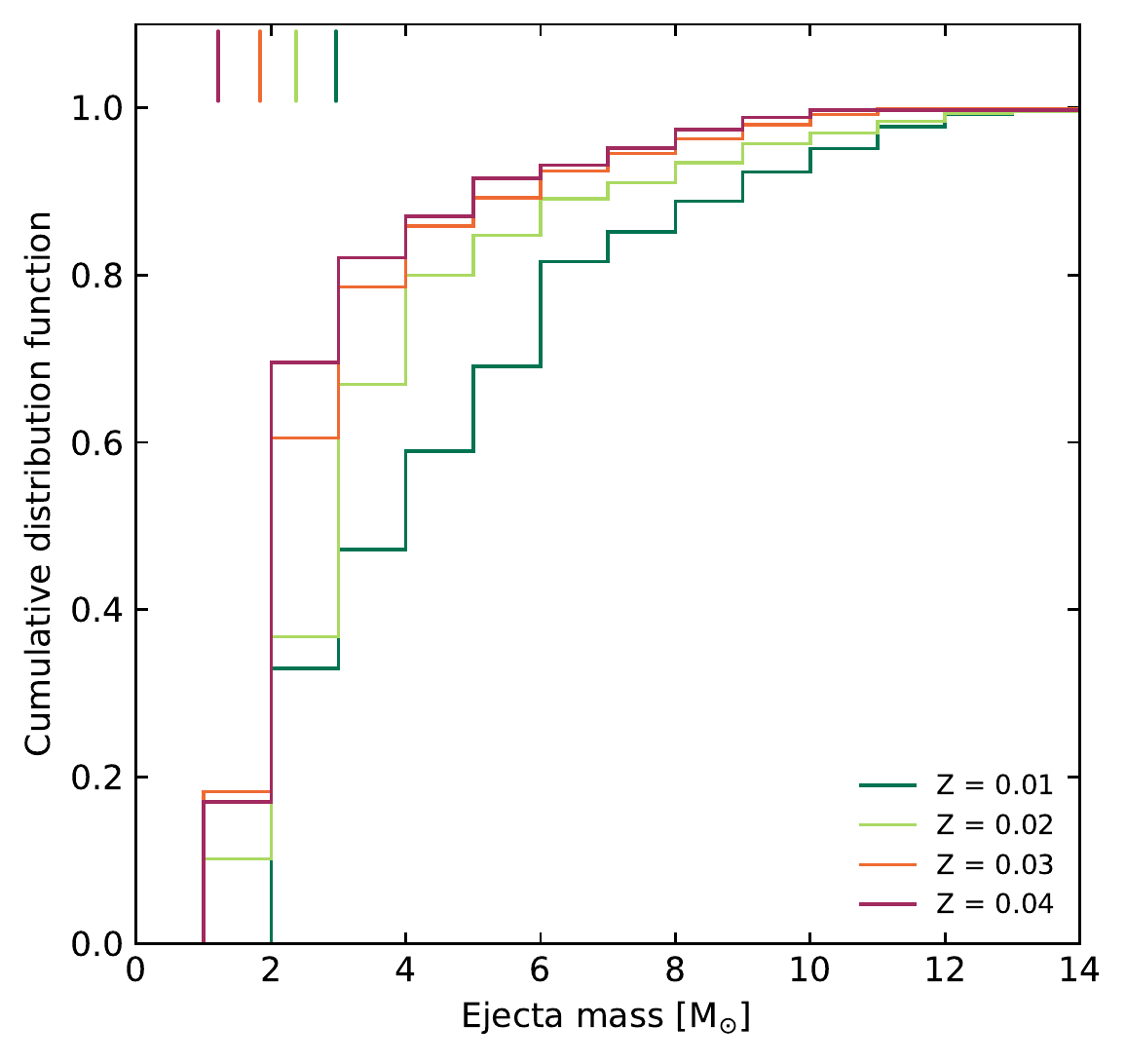}
\includegraphics[width=6cm]{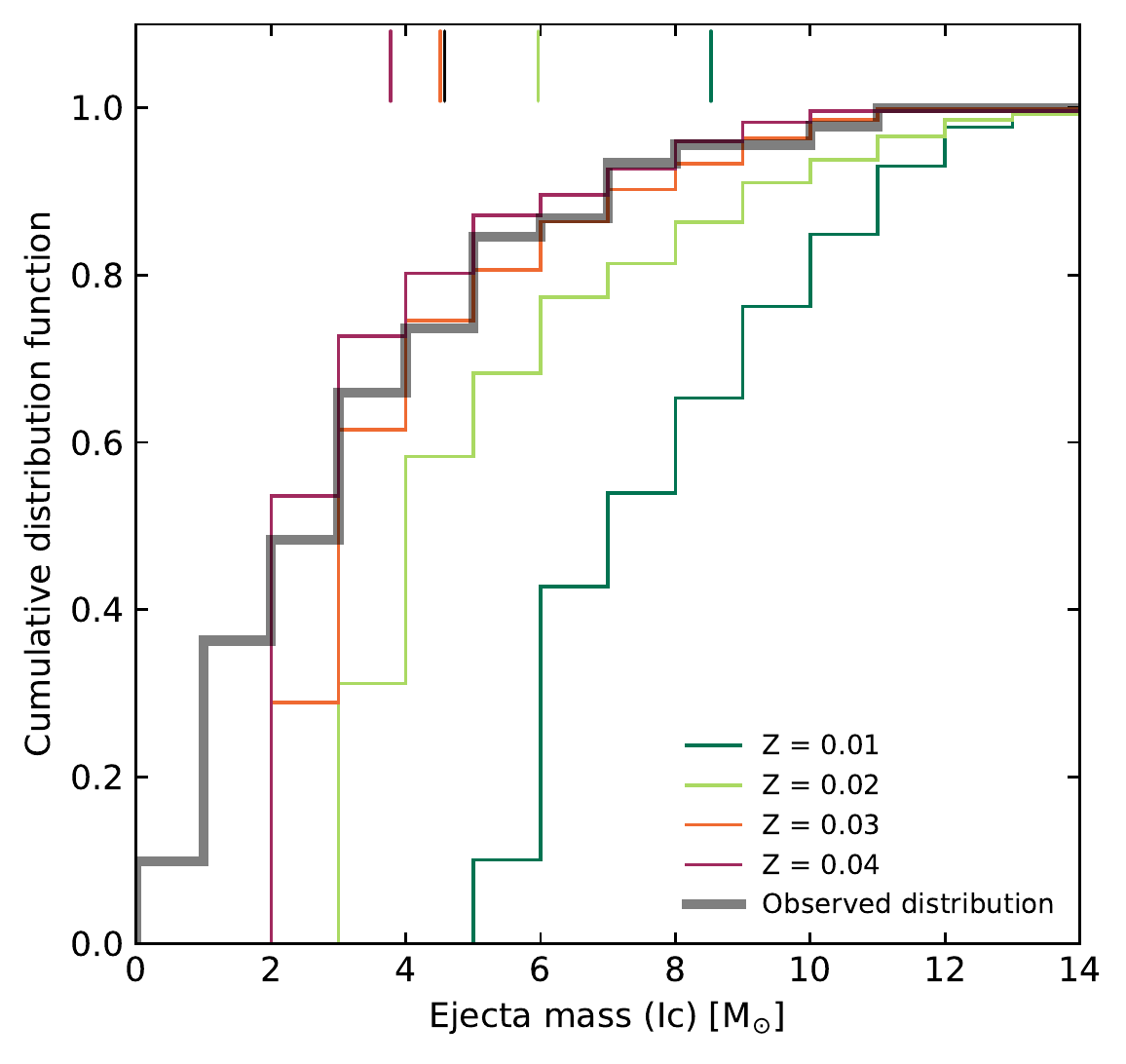}
\includegraphics[width=6cm]{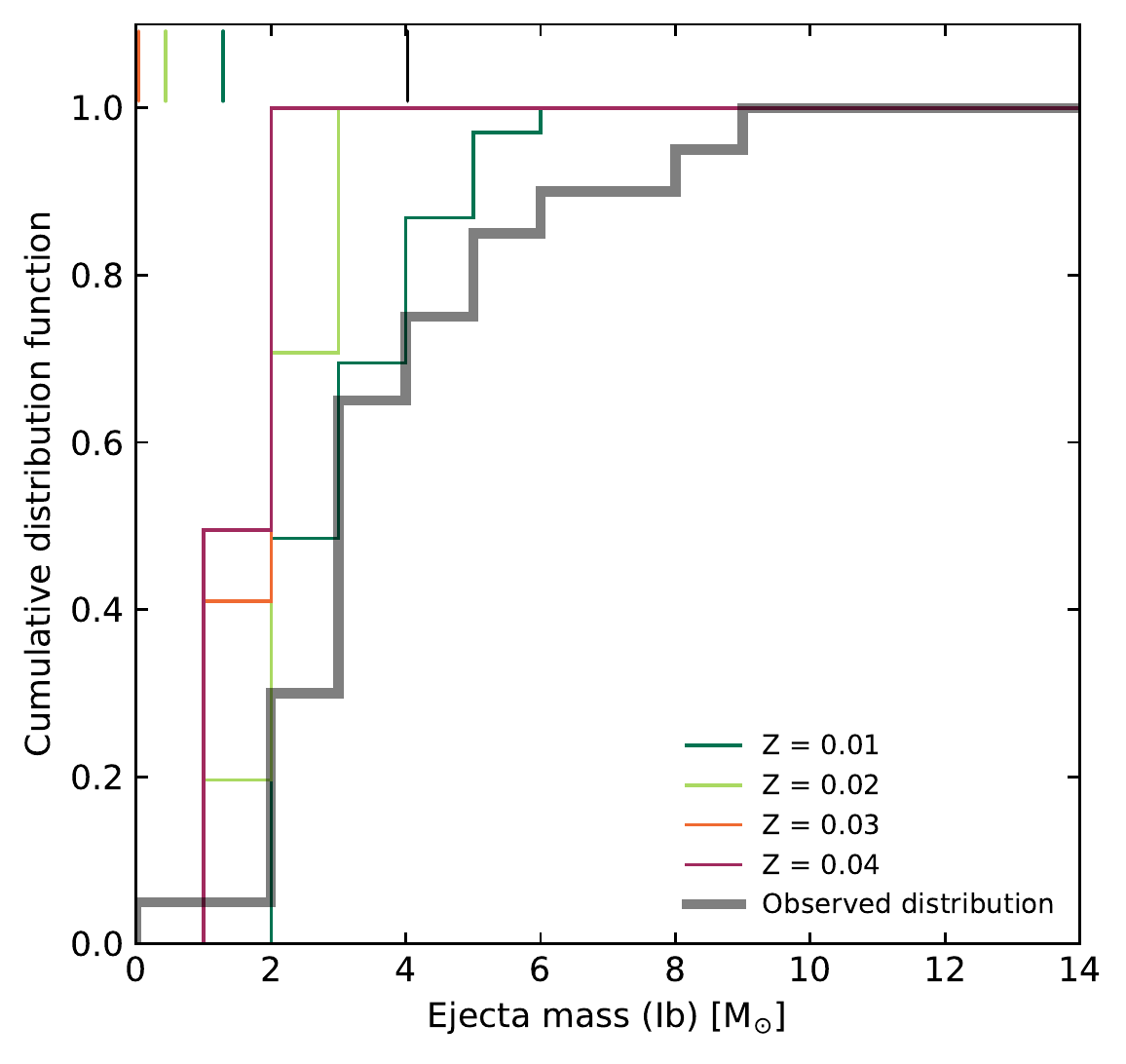}
\includegraphics[width=6cm]{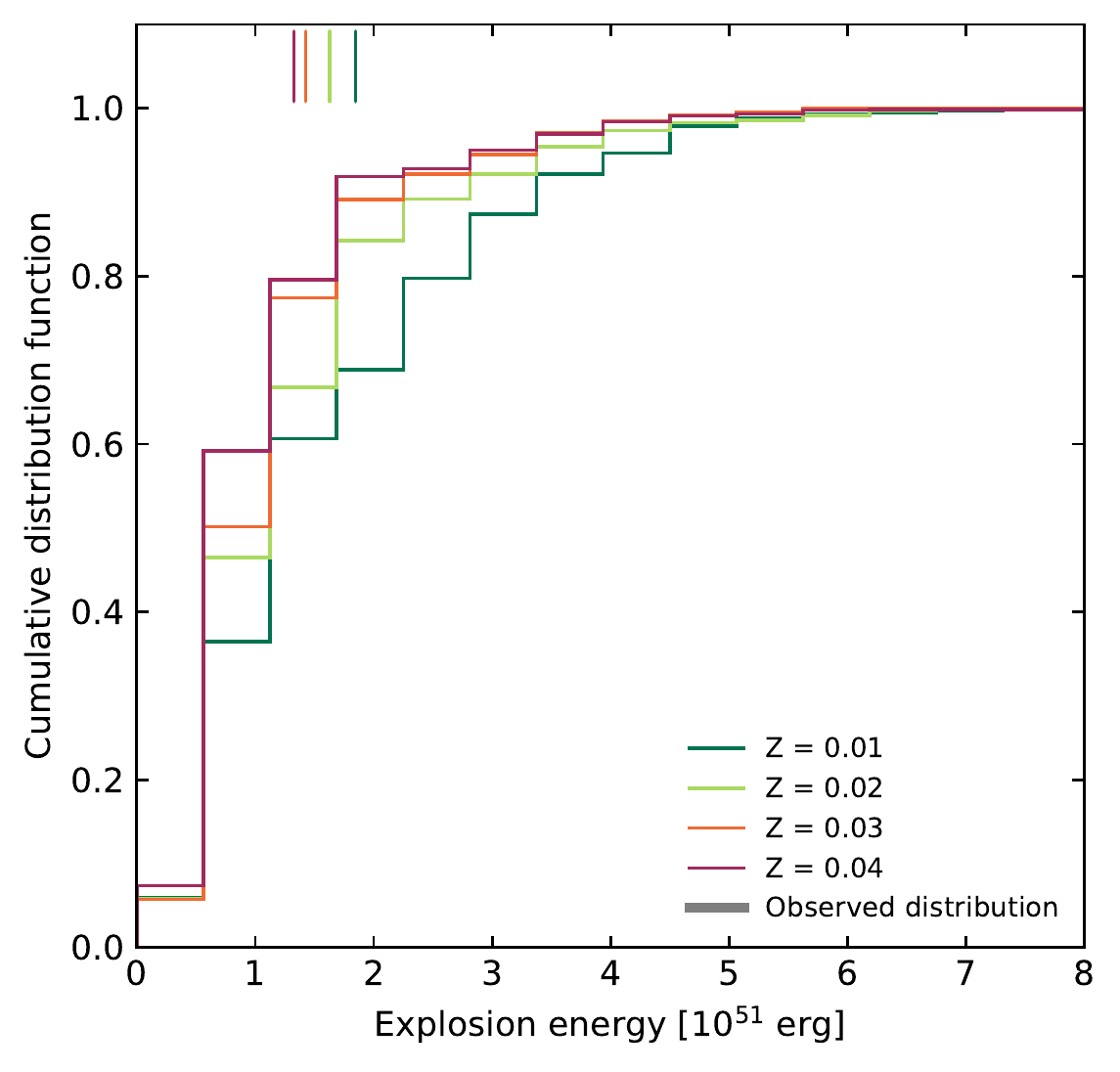}
\includegraphics[width=6cm]{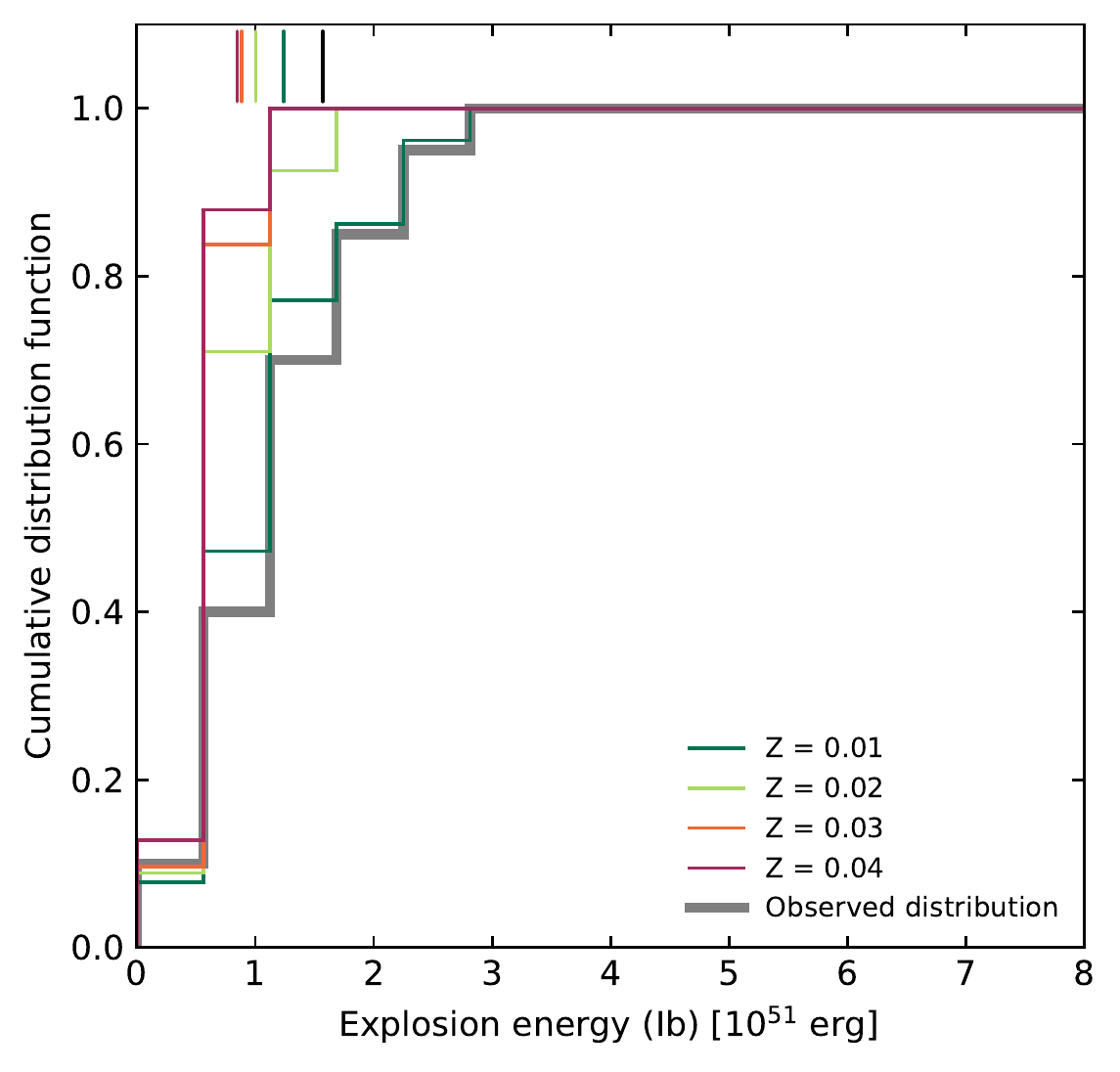}
\includegraphics[width=6cm]{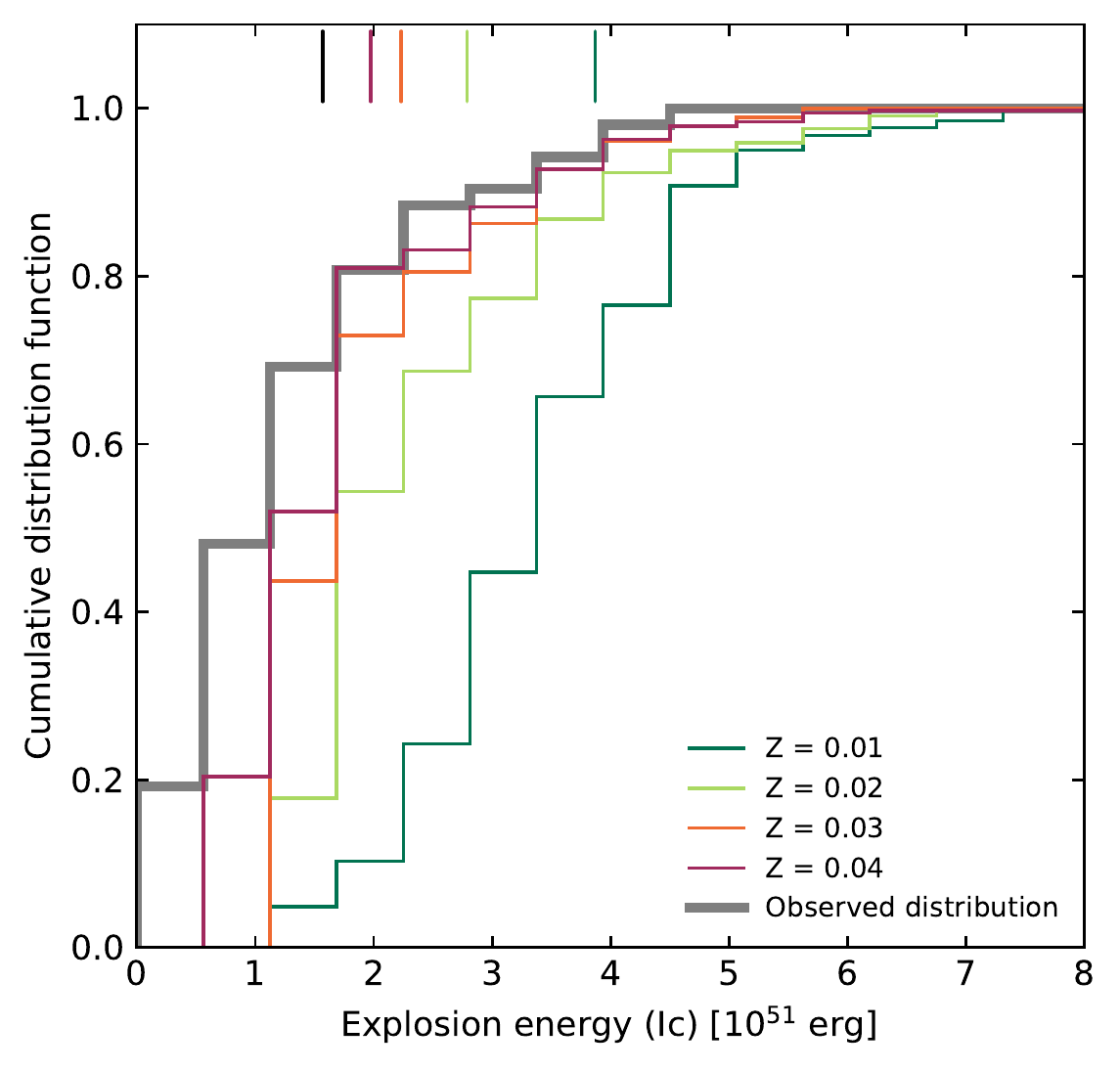}
\includegraphics[width=6cm]{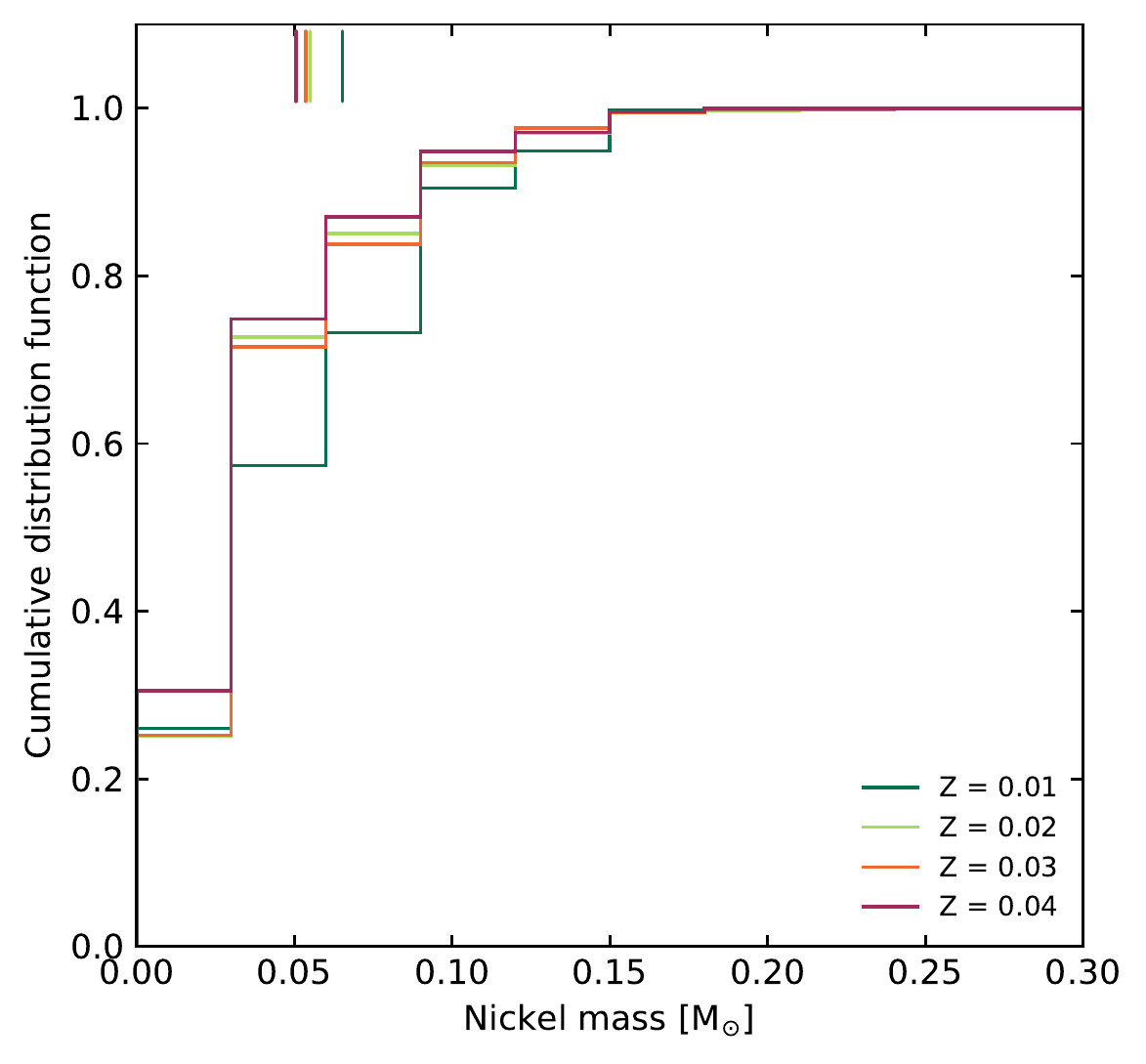}
\includegraphics[width=6cm]{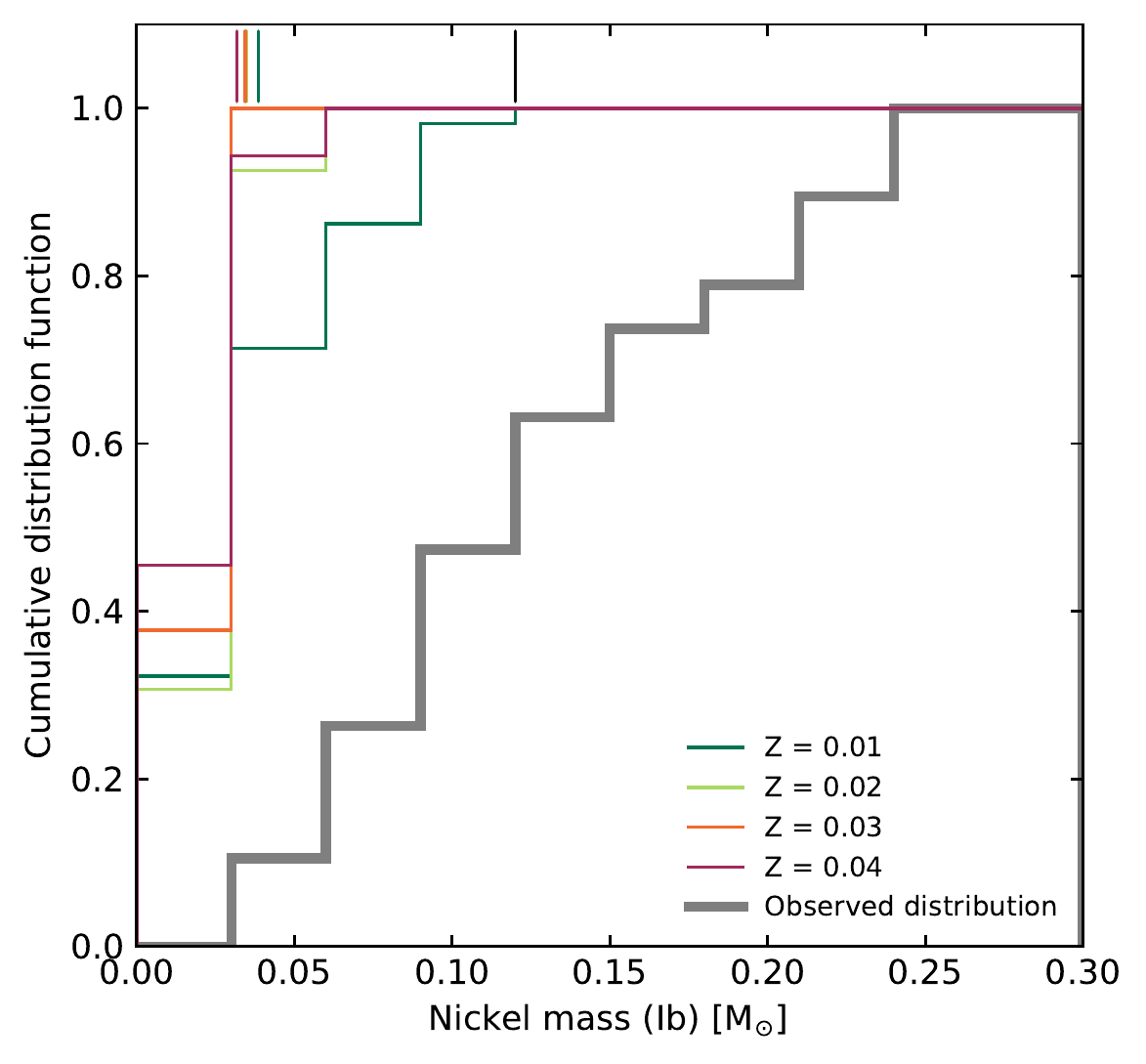}
\includegraphics[width=6cm]{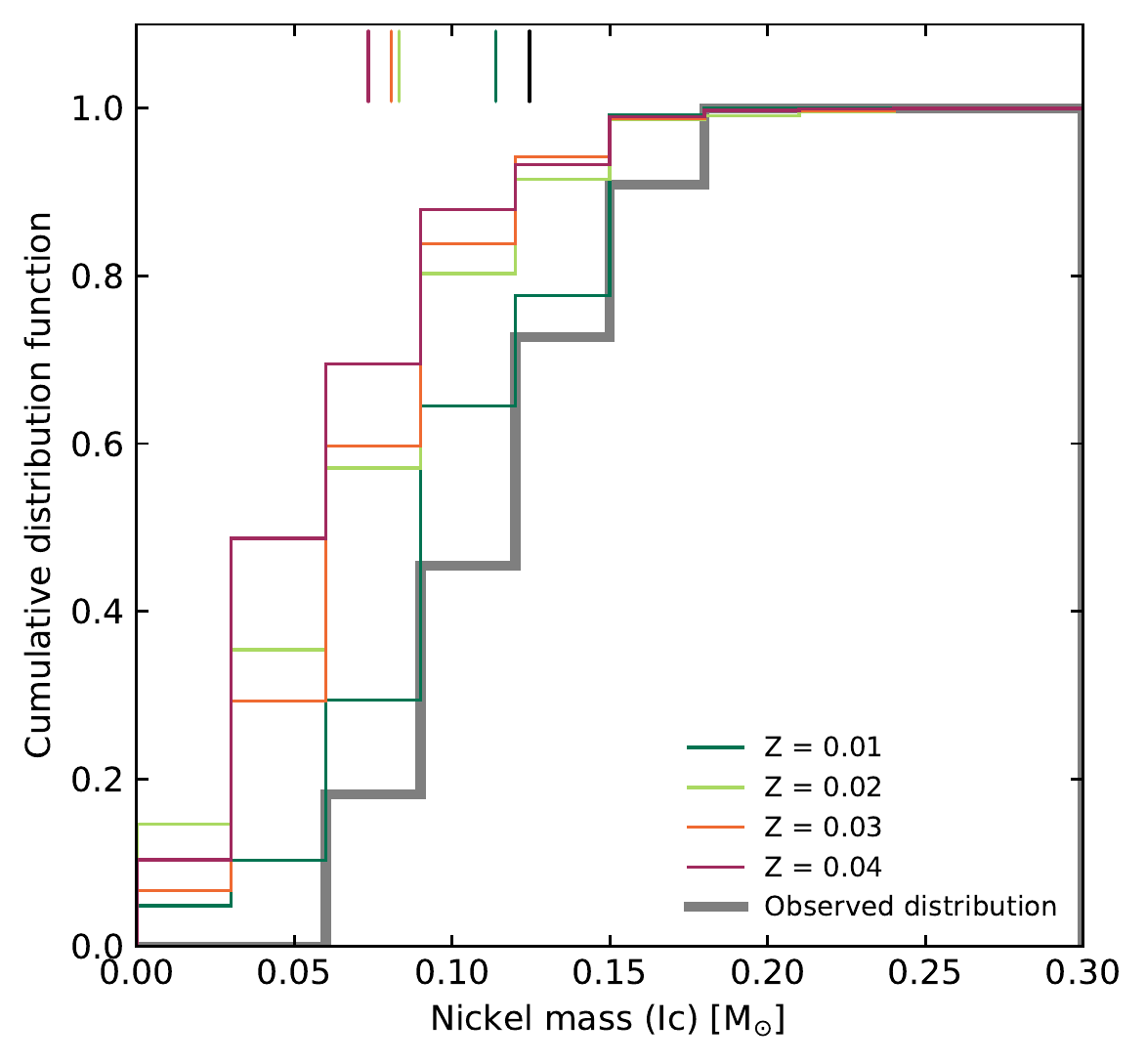}
\includegraphics[width=6cm]{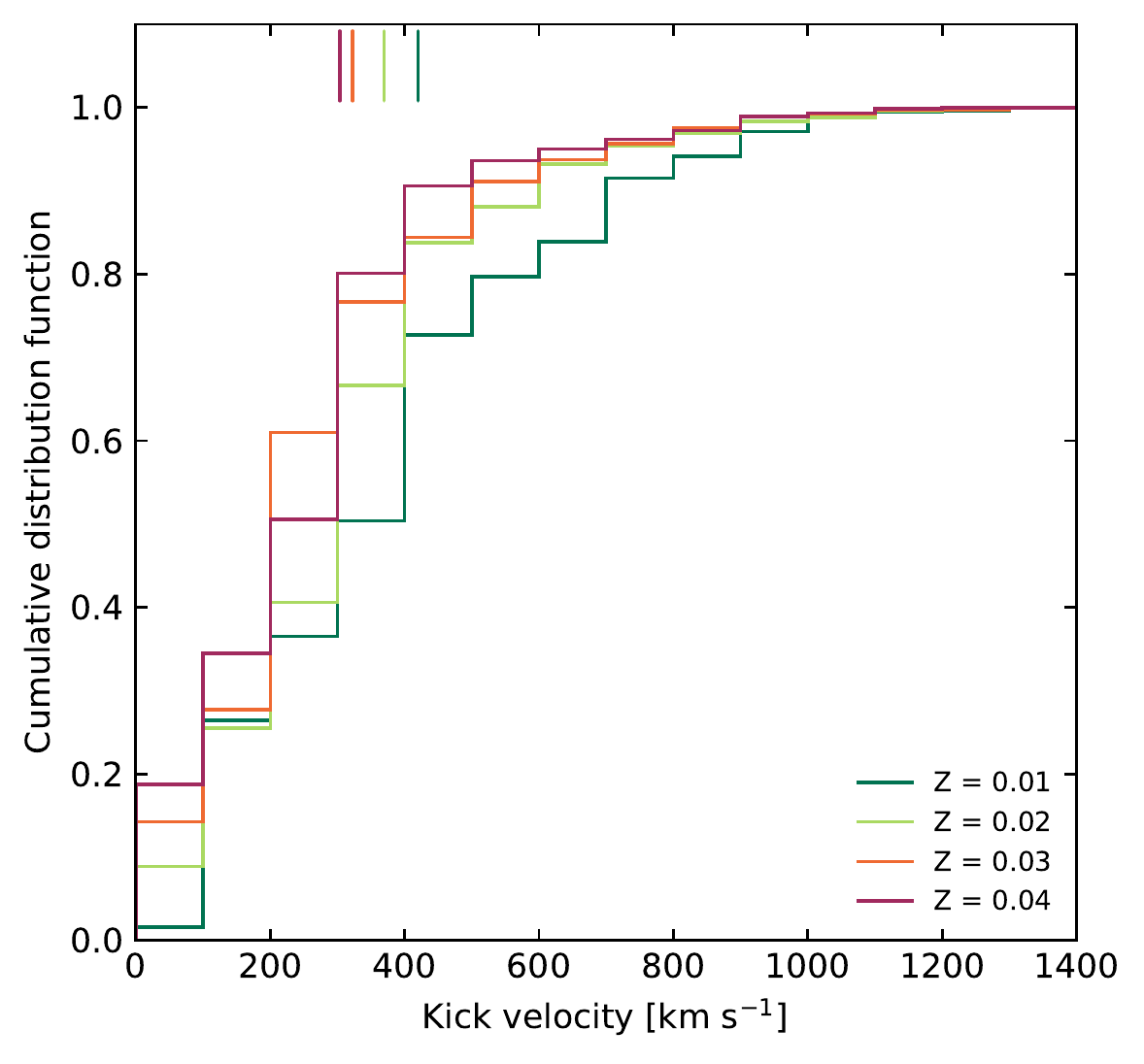}
\includegraphics[width=6cm]{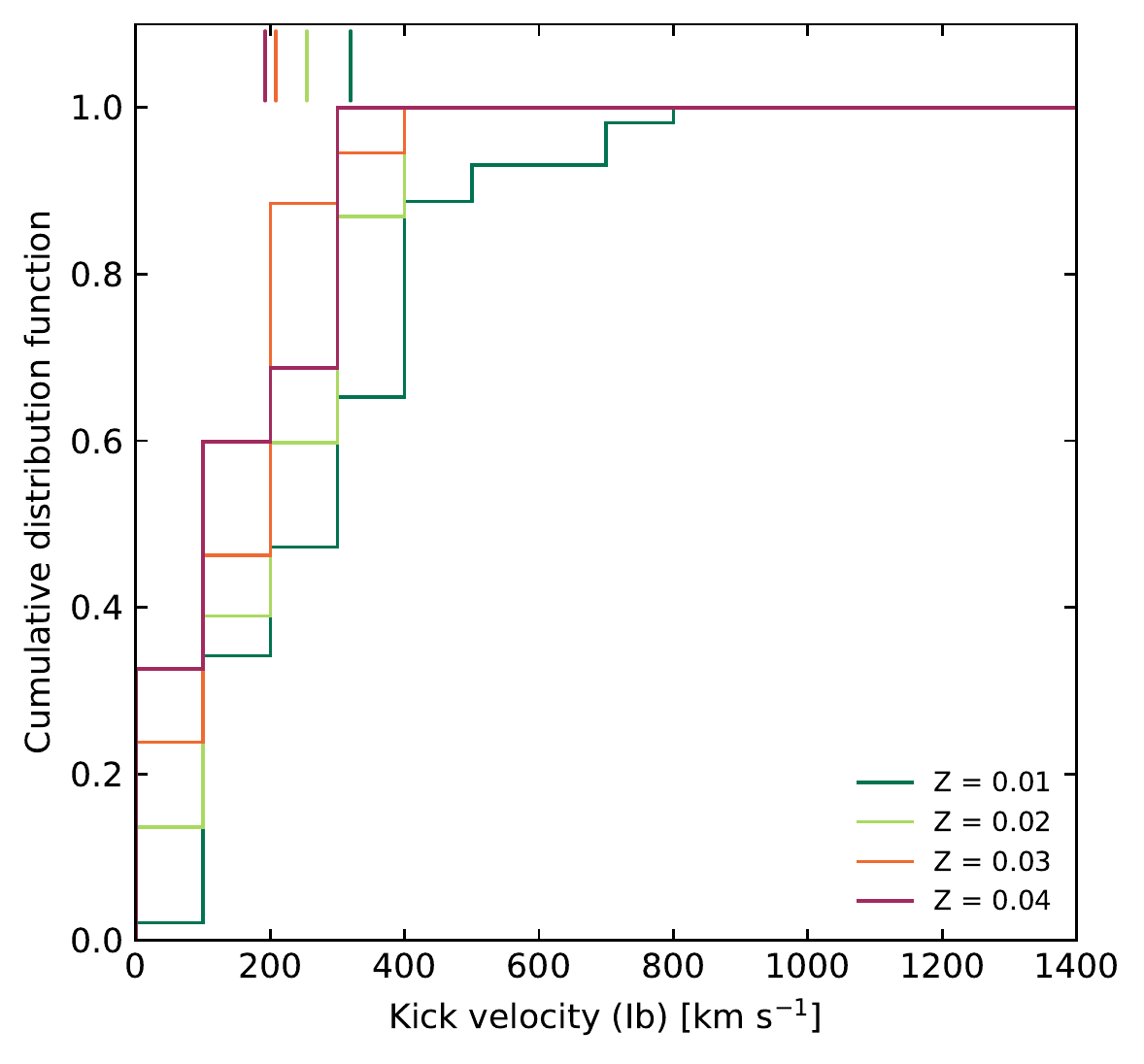}
\includegraphics[width=6cm]{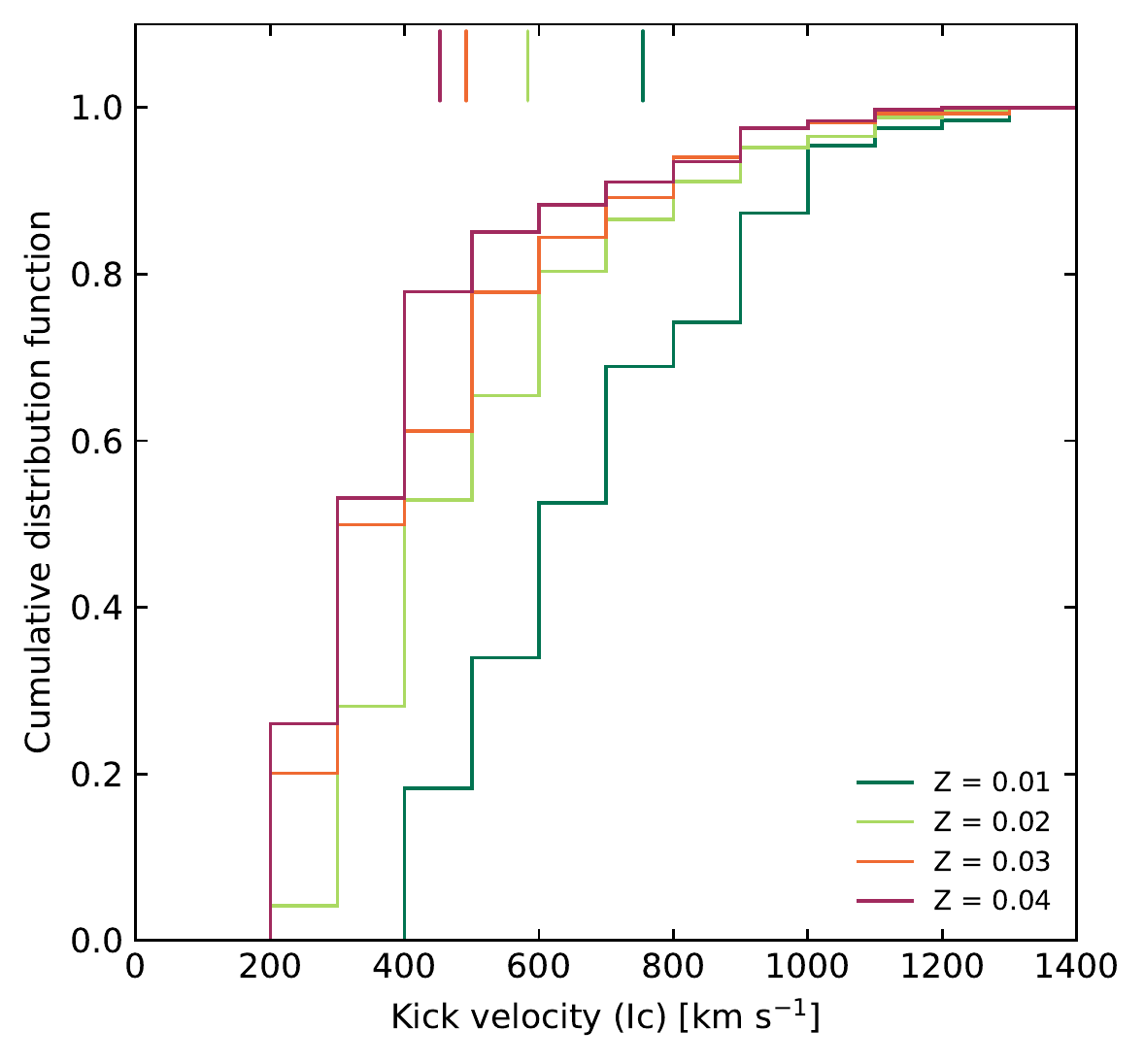}
\caption{Cumulative distribution functions of ejecta masses, explosion energies, nickel masses and kick velocities, separated for the full sample of models and separated between Type Ib and Ic SNe that result from analyzing our core-collapse models with the \cite{2020MNRAS.499.3214M} model, weighted by the IMF from \cite{1955ApJ...121..161S}. Different color lines represent different metallicities. Observed distributions of ejecta masses, explosion energies and nickel masses are drawn from \citep{2018A&A...609A.136T} for Type Ib, IIb (grouped with Type Ib) and Ic SNe, and additionaly from \cite{2021A&A...651A..81B} for a larger sample of Type Ic SNe. The average values of each distribution are indicated with vertical lines at the top of each panel. 
\label{fig:cdf}}
\end{figure*}

We present the obtained distributions for ejecta masses, explosion energies, kick velocities and nickel masses for Type Ib and Ic SNe, as cumulative distribution functions in Fig. \ref{fig:cdf}. Observed distributions obtained for Type Ib SNe are shown in comparison with the observed sample of Type Ib and Type IIb SNe from \cite{2018A&A...609A.136T}, and Type Ic SNe are compared to the previous sample, combined with that from \cite{2021A&A...651A..81B}, who performed a similar study, focused on Type Ic SNe. We compare our results to observations only after separating them by SN Type in hopes that the results of \cite{2018A&A...609A.136T} and \cite{2021A&A...651A..81B} are representative for the distributions of explosion properties of each SN Type, but refrain from comparing them to the combined sample since the observed SN properties do not reflect the relative occurrence of different SN Types.

Many trends become apparent in our estimation of the distribution of SN observables. The predicted distribution of ejecta masses of Type Ic SNe resembles the distributions we obtain at high metallicity, but likely the metallicity range that corresponds well with the observed distribution is much too high to explain most Type Ic SNe. We also find that, regardless of the metallicity, no Type Ic SNe are produced within our models with ejecta masses lower than about 2 $\mso$.

\begin{figure}
\centering
\resizebox{\hsize}{!}{\includegraphics{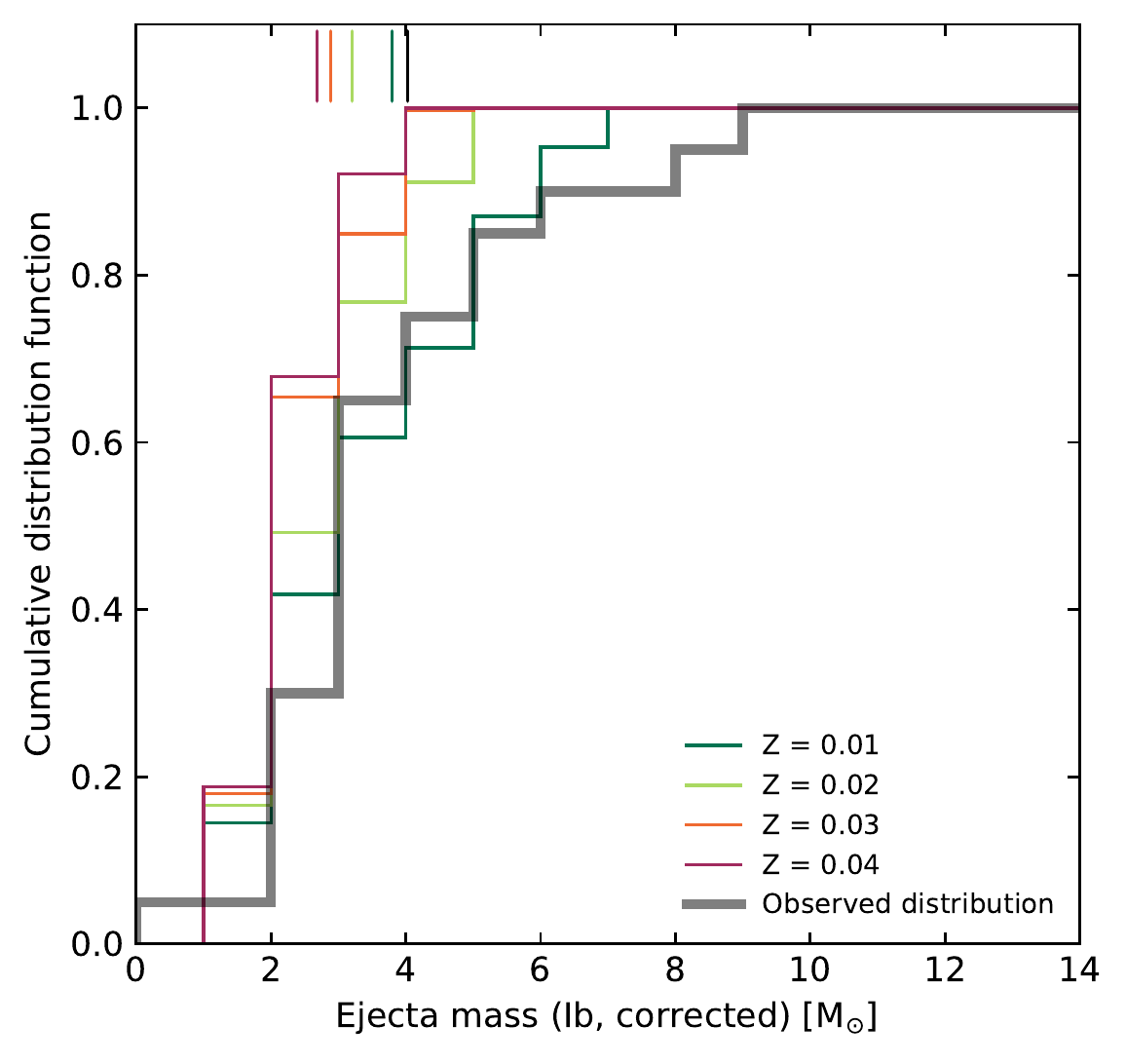}}
\resizebox{\hsize}{!}{\includegraphics{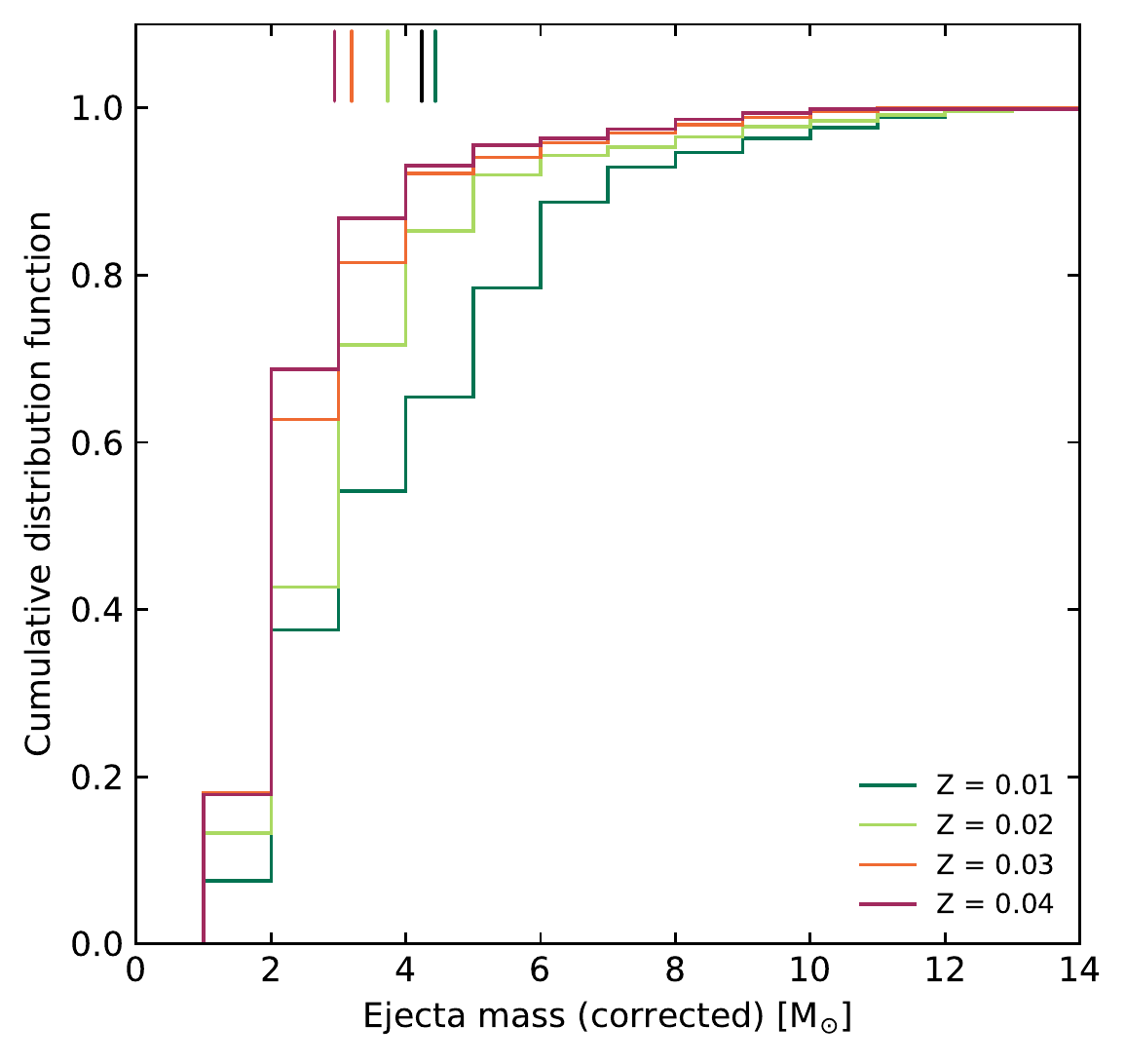}}
\caption{Cumulative distribution of ejecta masses for Type Ib SNe (top) and for all stripped-envelope SNe (bottom) that result from our models, but corrected for the strong variation in wind between WR stars and low-mass helium stars. Here, ejecta masses for stars with initial mass above the minimum mass of WN-type stars are drawn from our model results, and for the rest we assume no mass loss, but the same NS mass as predicted from its original model.
}\label{fig:corrected}
\end{figure}

The predicted distribution of ejecta masses of Type Ib SNe is skewed to considerably lower values than the observations. These results, however, have been taken at face value from the outcome of our stellar evolution tracks, and many of them overestimate the mass loss rates of the progenitors of Type Ib SNe, since they have luminosities below $\mathrm{L}_\mathrm{min,WN}^\mathrm{tau}$. To remedy this, we repeat our calculation using the initial mass as a better proxy of the final mass, but retaining the same distribution of NS masses. This results in the distribution shown in Fig. \ref{fig:corrected}, yielding a better fit to the data.

\begin{figure*}
\centering
\includegraphics[width=6cm]{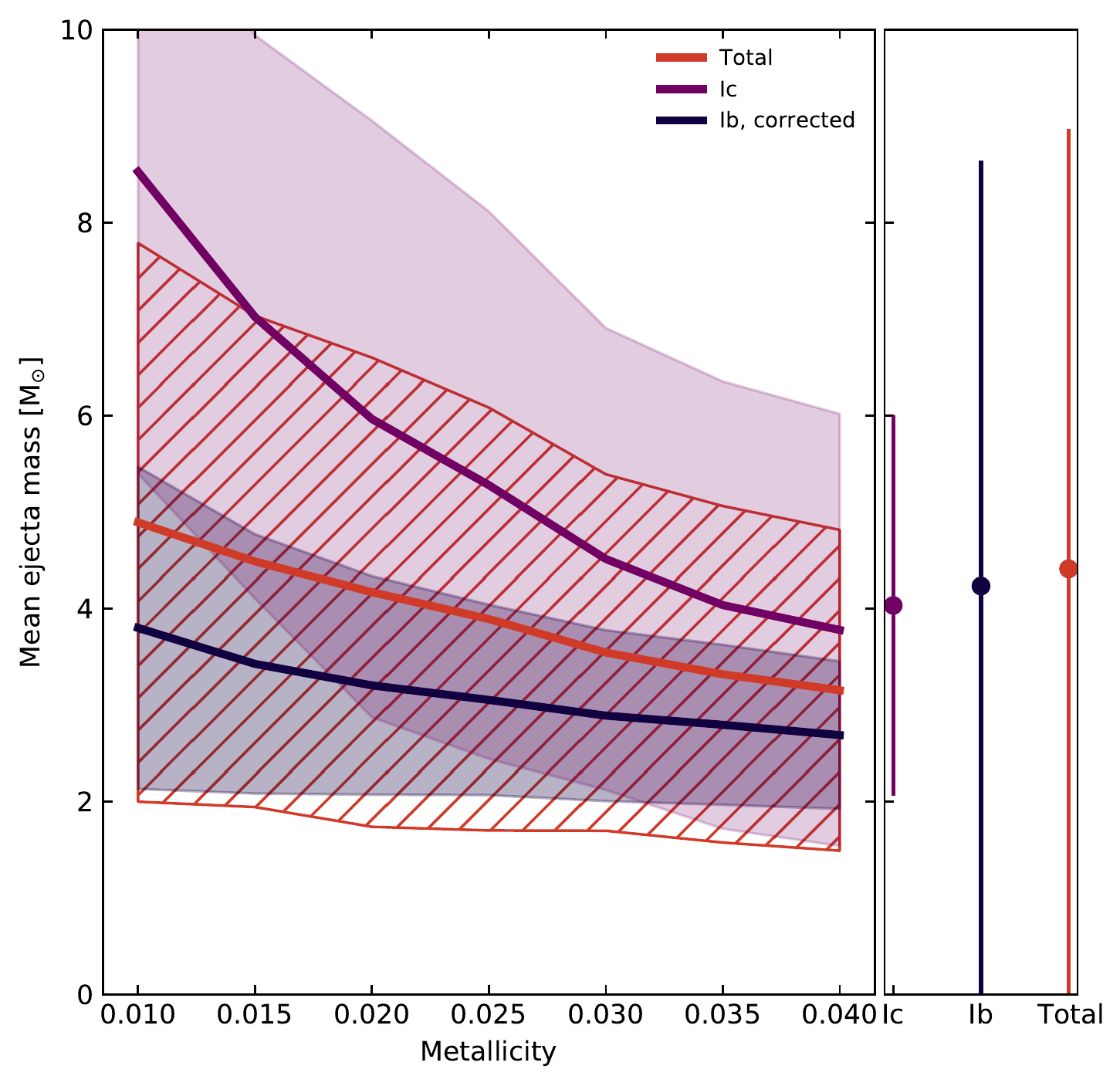}
\includegraphics[width=6cm]{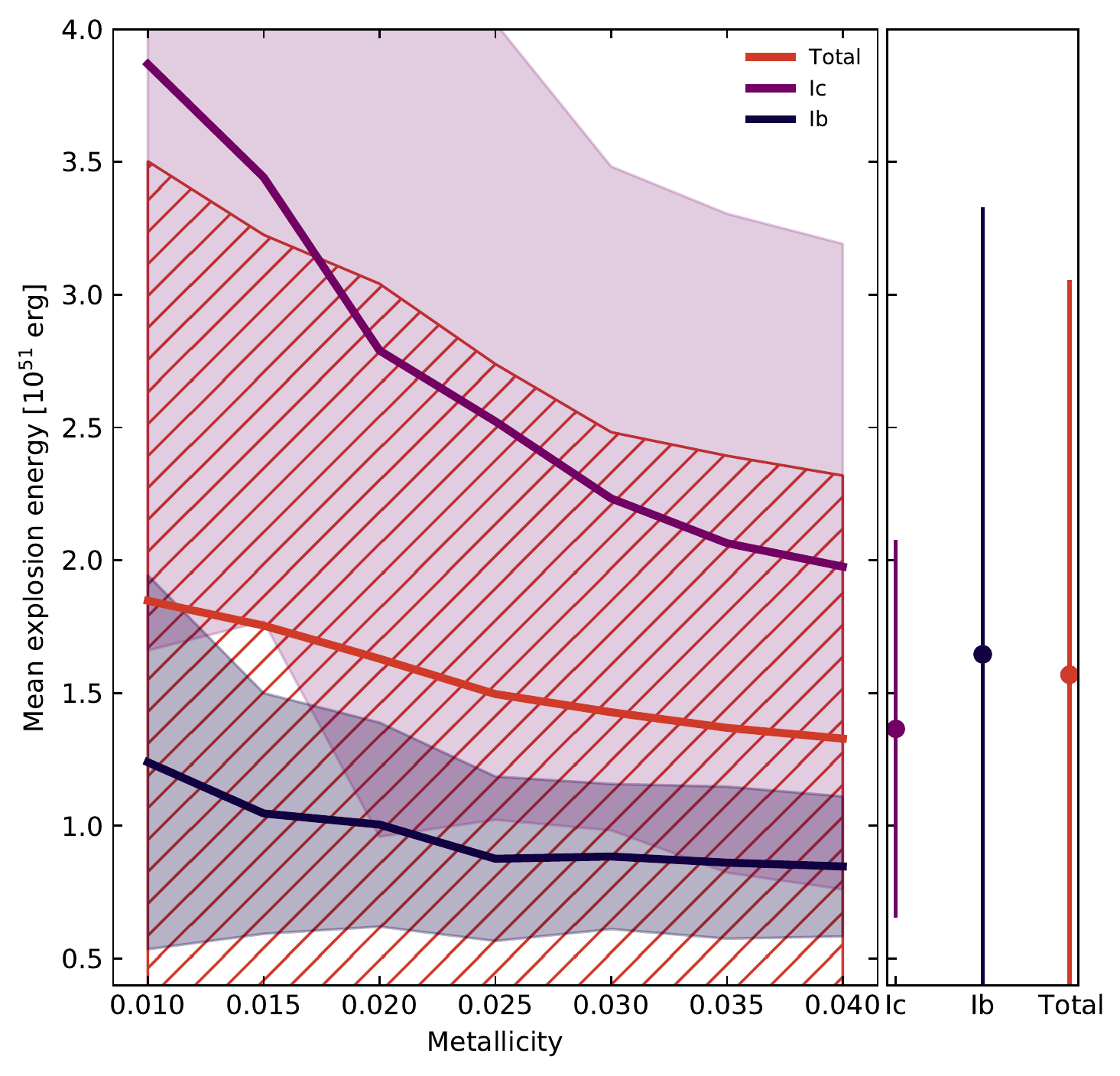}
\includegraphics[width=6cm]{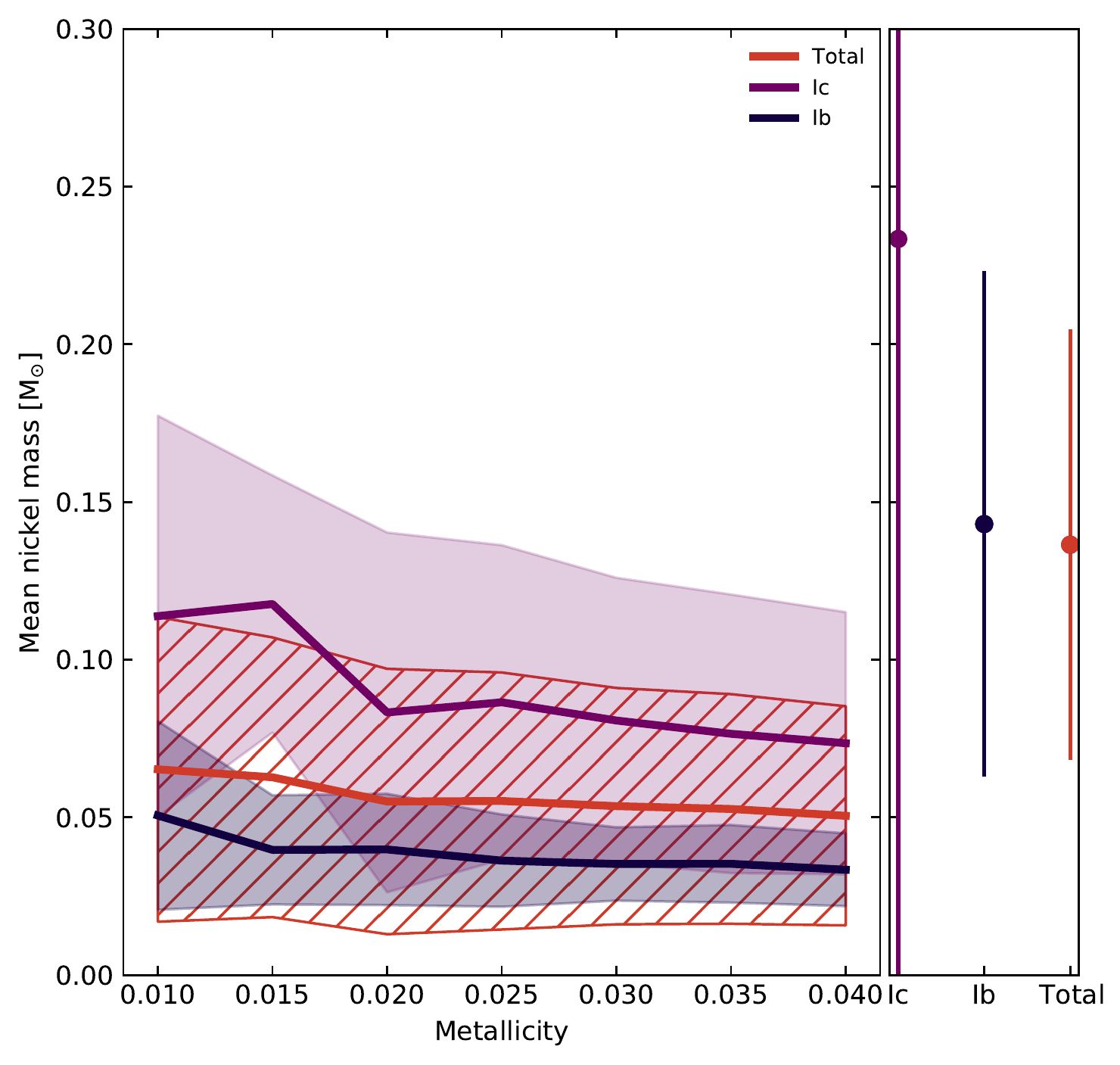}
\includegraphics[width=6cm]{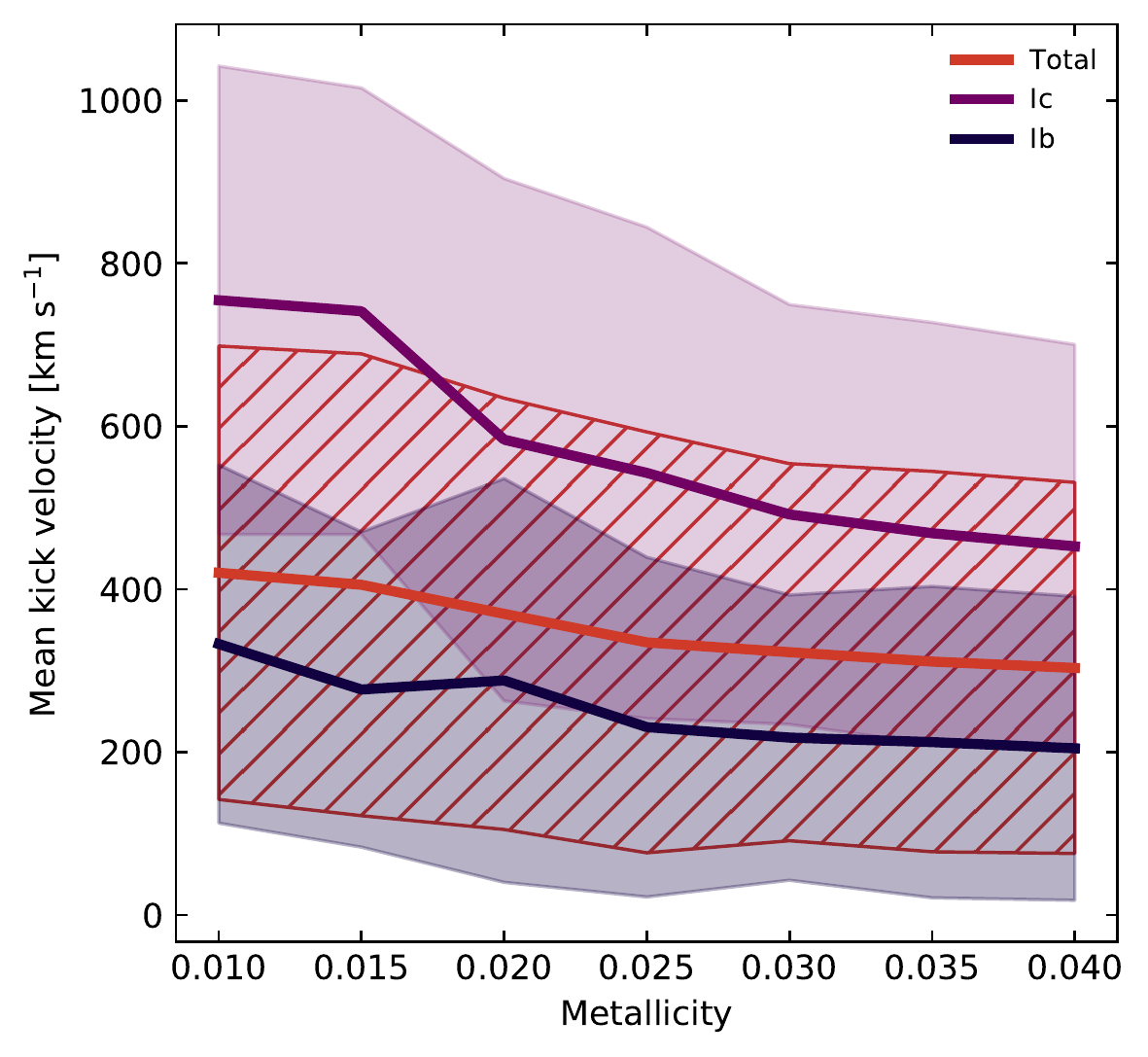}
\caption{Mean values and standard deviation of ejecta masses for both explosion energies, nickel masses and kick velocities, from the full sample as well as for Type Ib and Type Ic SNe individually, at different metallicities, that result from analyzing our core-collapse models with the \cite{2020MNRAS.499.3214M} model, weighted by the IMF from \cite{1955ApJ...121..161S}. 
\label{fig:means}}
\end{figure*}

As expected, we find that the ejecta masses of stripped-envelope SNe, both in Type Ib and Type Ic, decrease as metallicity increases. As shown by the mean values and standard deviations of the distributions in Fig. \ref{fig:means}, ejecta masses for both type of SNe decrease steeply as a function of metallicity, but even applying a correction for the progenitors of Type Ib SNe with very low mass, the range at which these events are found cannot be reproduced from our models.

We find the average explosion energies estimated for our models ranges between about 1.3 and 1.8 $\times 10^{51}$ erg. We find that this average decreases as a function of metallicity. The average values of Type Ib SNe tend to be below the observed average energy at every metallicity, whereas our Type Ic SN models have larger average energies than the observed samples. The explosion energies of Type Ib SNe vary relatively little as a function of metallicity, whereas those of Type Ic SNe tend to decrease quite drastically with increasing metallicity. The energies estimated for Type Ib SNe are considerably lower, but we have not corrected these values for the different wind mass loss rates for low-mass helium stars. The energies predicted for Type Ic SNe at high metallicity are in good agreement with the observed values, although the number of explosions below 10$^{51}$ erg are underestimated.

We find that the amount of $^{56}$Ni synthesized in SNe decreases as metallicity increases. The distributions of $^{56}$Ni mass peak at around 0.05 $\mso$ and few cases are observed with nickel masses above 0.1 $\mso$. For both SN types, we find that the amount of nickel synthesized during the explosion is found to decrease with increasing metallicity. The values we find are systematically lower than the observed values, which also have a much broader distribution, and are found to have closer to 0.15 $\mso$ of nickel on average.

The distribution of kick velocities obtained from our models peaks at around 300--400 km s$^{-1}$, and we find that kick velocities tend to decrease with increasing metallicity. We also find that Type Ic SNe tend to produce SNe with considerably larger kicks, on average, than those coming from Type Ib SNe.

Except for the emergence of a new, low-mass peak in the BH mass distribution, fallback SNe are found to be too rare to have a percievable effect on the distribution of SN observables in large samples. This is due to the fact that these events are predominantly originating in very high mass stars, which are therefore very rare. 

\begin{figure}[h!]
\centering
\resizebox{\hsize}{!}{\includegraphics{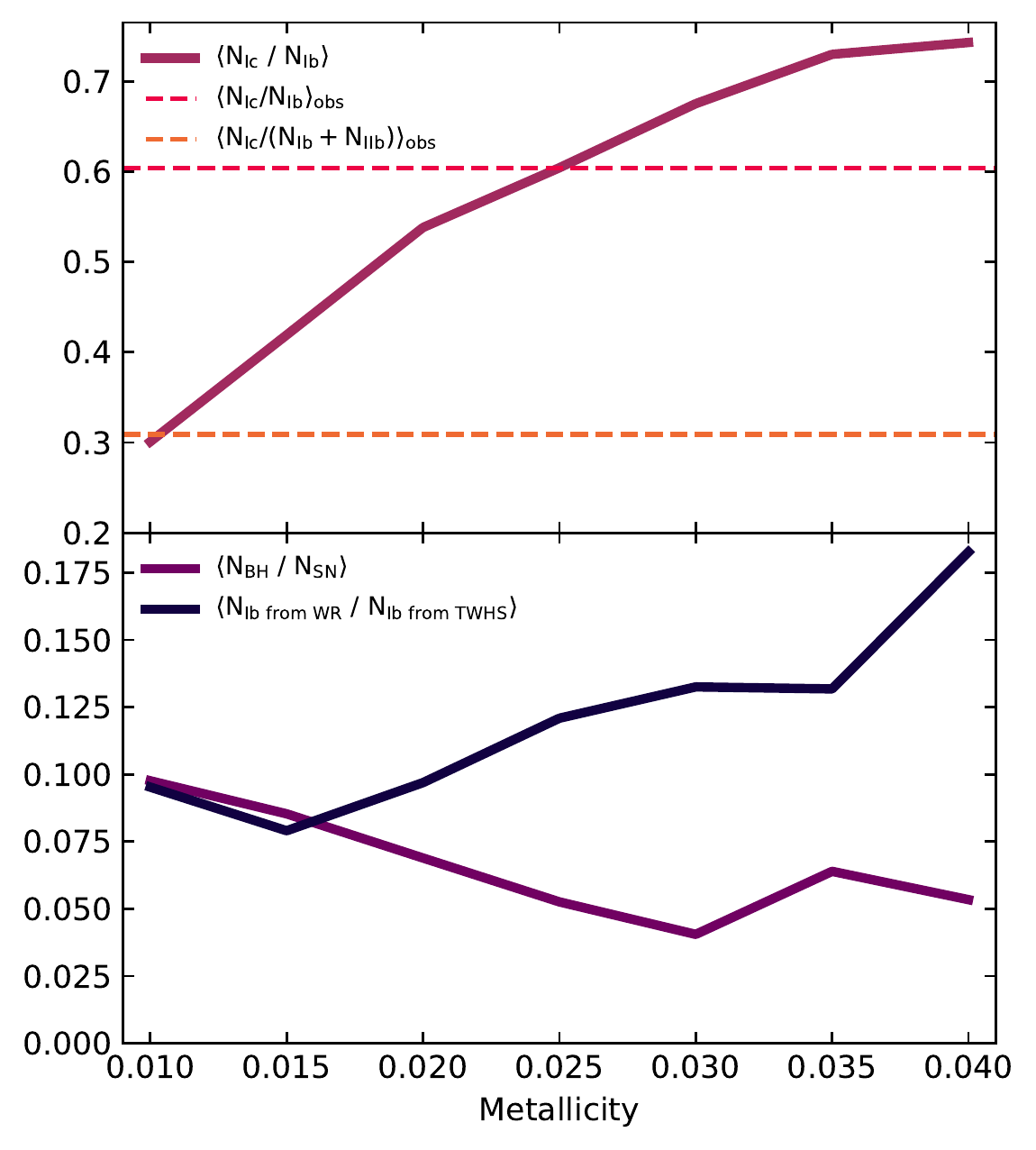}}
\caption{Number ratios of different quantities as a function of metallicity, as obtained from our model population. Top: Number ratio of Type Ib SNe to Type Ic SNe, $\langle \mathrm{N}_\mathrm{Ic}$ / N$_\mathrm{Ib}\rangle$, compared with the observed ratio of Type Ib SNe to Type Ic SNe $\langle\mathrm{N}_{\mathrm{Ic}}/\mathrm{N}_{\mathrm{Ib}}\rangle_{\mathrm{obs}}$ and of Type Ib SNe plus Type IIb to Type Ic SNe $\langle\mathrm{N}_{\mathrm{Ic}}/(\mathrm{N}_{\mathrm{Ib}}+\mathrm{N}_{\mathrm{IIb}})\rangle_{\mathrm{obs}}$ in the local Universe, inferred from the relative rates of \cite{2017PASP..129e4201S}. Bottom: Number ratio of BHs to SN explosions $\langle \mathrm{N}_\mathrm{BH}$ / N$_\mathrm{SN}\rangle$ and number ratio of Type Ib SNe with WR progenitors to Type Ib SNe with transparent wind helium star progenitors $\langle \mathrm{N}_\mathrm{Ib \ from \ WR}$ / N$_\mathrm{Ib \ from \ TWHS}\rangle$. 
} \label{fig:Ic_Ib_ratio}
\end{figure}

Another immediate consequence of the metallicity dependence of stellar winds is an increase of rates of SNe Type Ic with respect to SNe Type Ib for higher metallicity. We estimate the number ratio of Type Ic SNe to Type Ib SNe as a function of metallicity, and find it to be strongly metallicity dependent, as shown in Fig. \ref{fig:Ic_Ib_ratio}.

According to our models, the number ratio of Type Ic to Type Ib SNe should increase as a function of metallicity, starting from approximately 0.2 at a metallicity of 0.01, to about 0.7 at a metallicity of 0.04. This rising trend is mainly due to the fact the initial helium mass threshold for the production of a Type Ic supernova decreases with increasing metallicity. We compare our calculation to the observed ratio of stripped envelope SNe using the observed rates of different SN types in the local Universe from \cite{2017PASP..129e4201S}. We find that the value estimated from our models around solar metallicity is between the observed ratio of Ic to Ib in the local Universe, which represents an upper limit to our model; and the ratio of Type Ic SNe and the rest of stripped-envelope SNe (Types Ib and IIb), which is a more realistic quantity to compare our results to.

Two final quantities that we derive from our population model, and present in the right y-axis of Fig. \ref{fig:Ic_Ib_ratio}, are the relative ratio of BHs to SNe, and the relative ratio of Type Ib SNe that have WR type progenitors to those that have transparent wind helium stars as progenitors (i.e. stars that have luminosities below $\mathrm{L}_\mathrm{min,WN}^\mathrm{tau}$. 

We find that, as metallicity increases, the number of massive stars that produce BHs tends to decrease. At a metallicity of 0.01, we find that around 16\% of massive stars produce BHs, whereas the number decreases to about 8\% in the metallicity range 0.025 -- 0.04. This is due to the reduction in final mass as a function of initial helium star mass (see Fig. \ref{fig:finalmass}), combined with the location of the islands of explodability (e.g. Fig. \ref{fig:compactness}). On the other hand, as metallicity increases, the number of Type Ib SNe that have WR stars as progenitors tends to increase. This result is non trivial because, even though the minimum luminosity at which we expect stripped-envelope stars to be WRs decreases as a function of metallicity, so does the minimum luminosity at which our models produce WC-type stars, meaning that even though there are more WN-type stars at lower luminosities, the threshold at which they explode as Type Ic SNe also decreases. In the metallicity range of our models, this number tends to increase, but the behaviour is likely different in metallicities beyond those in our grids. However, the absolute value of this ratio at all metallicities in our grid is below 18\%, implying that most Type Ib SNe originate from low mass helium stars with transparent winds, and not WR stars. The number is likely to be even lower at metallicities corresponding to environments like the SMC and below, and this trend should be reflected in the methods that are employed in the search of direct imaging of Type Ib SN progenitors.

\section{Discussion}\label{sec:discussion}

The results presented in this work were derived from evolutionary calculations of helium stars with metallicities in the range 0.01 -- 0.04, and their resulting core collapse models. The aim of studying helium stars in this metallicity range is to explore a possible evolutionary channel that can help explain the observed prevalence of Type Ic SNe in high metallicity environments \citep{2011ApJ...731L...4M}, and address the discrepancy between predicted final masses of stripped-envelope SN progenitors \citep[e.g.][]{1993ApJ...411..823W,2005A&A...429..581M,2010ApJ...725..940Y,2012A&A...542A..29G}, and the distribution of ejecta masses observed in SNe, particularly of Type Ic \citep[e.g.][]{2018A&A...609A.136T,2021A&A...651A..81B}. 

We divide our core collapse models into possible progenitors of either Type Ib or Type Ic SNe according to whether their surface has a helium abundance of $\mathrm{Y}= 1-Z_{\mathrm{init}}$ and high nitrogen abundance, or whether it has decreased helium abundance, and is instead enriched with the products of helium burning, respectively. This dichotomy has been found to be at the heart of the differences in spectral features of Type I core collapse SNe by \cite{2020A&A...642A.106D}.

We then analyze our results according to the explodability tests of \cite{2011ApJ...730...70O}, \cite{2016ApJ...818..124E} and \cite{2016MNRAS.460..742M}, the latter including the effect of fallback following \cite{2020MNRAS.499.3214M}. The latter explodability test consists of a parametrized, semi-analytical model of the explosion, which yields predictions of NS gravitational masses, ejecta masses and nickel masses in successfully exploding models. We also estimate the distribution of kick velocities following \cite{2020MNRAS.499.3214M}. Finally, we construct a simple synthetic population of stripped envelope SNe, to contrast our results with observations.

In this Section, we discuss and interpret the results obtained from analyzing our core collapse models and the resulting population, as well as the uncertainties that might have an impact in our results.

\subsection{Uncertainties in core collapse stellar models}

Some uncertainties in our models are inherent to our physical and numerical assumptions, and others stem from poorly constrained physical parameters. Many are discussed in detail in \citetalias{2021arXiv211206948A}, but here we list a few, paying particular attention to those that affect our models at core collapse.

\subsubsection{Uncertainties in the mass at core collapse}

Our evolutionary calculations are modeled using the empirical mass loss rates suggested by \cite{2017MNRAS.470.3970Y}. However, stars with luminosities below $\mathrm{L}_\mathrm{min,WN}^\mathrm{tau}$ are believed to lose mass at a much lower rate. The precise mass loss rates of these stars are unknown, but the simulations of \cite{2017A&A...607L...8V} suggest that the mass loss rates of such stars are orders of magnitude below those of WR stars.

Since we extrapolate WR mass loss rates into this regime, the final masses of models with luminosities below $\mathrm{L}_\mathrm{min,WN}^\mathrm{tau}$ are overestimated. Furthermore, the core evolution of these models does not correspond to that of helium stars of the appropriate final mass. This likely has no effect in whether they explode or not, since models with luminosities below $\mathrm{L}_\mathrm{min,WN}^\mathrm{tau}$ have low initial masses, and are expected to have low values of $\xi_{2.5}$, leading to successful SN explosions. The approach we take to overcome this limitation in our population models is to assume that the helium stars that produce SNe in this regime have final masses that can be approximated by their initial masses, since they have very small mass loss rates. We further justify this assumption \textit{a posteriori} by arguing that this makes our estimates for ejecta masses resemble the observed distribution more closely. For these stars, we assume that the NS gravitational mass is 1.22--1.3 $\mso$, and we assume that the minimum mass at which stars explode as SNe is 3 $\mso$, following \cite{savvas2021}. 

The mass loss rates of WR stars are also uncertain. Recent theoretical estimates of WR winds \citep{2020MNRAS.491.4406S,2020MNRAS.499..873S} suggest that WR winds have a weaker dependence on metallicity, and that WC star mass loss rates are lower than the empirically obtained values of \cite{2016ApJ...833..133T}. This would imply that the final masses of WC stars in our models are underestimated, as well as those of models with high Z.

The masses of our models at core collapse may also be overestimated due to the implementation of MESA's \texttt{mlt++}, which enhances the efficiency of energy transport by convection, and results in more compact stars that show very little or no envelope inflation \citep[e.g.][]{2015A&A...580A..20S}. The mass loss rates of WR stars should not be affected by this assumption, since they only depend on the luminosity and metallicity of helium stars, which are well constrained by our models. However, if they exceed the Eddington limit in their envelopes by a large amount, it may lead to episodic mass loss \citep[e.g.][]{2015ASSL..412..113O}. Furthermore, if inflation causes helium stars to expand significantly beyond their initial radius, it might trigger a second episode of mass transfer with their companion, which will also reduce their final mass.

Another uncertainty in our models comes from ignoring the main sequence and envelope stripping phases, and modelling directly fully stripped helium stars. As discussed in detail in \citetalias{2021arXiv211206948A}, envelope stripping is not always efficient in removing all of the hydrogen from the surface of a star. Therefore, they are likely to spend a fraction of their helium burning lifetime with a hydrogen envelope of varying mass, which can result in higher final and ejecta masses than those predicted in our models.

\subsubsection{Uncertainties in the core structure}

In addition to the uncertainties introduced in our modelling by ignoring the main sequence evolution of the progenitors of stripped-envelope stars, and by their mass loss rates; other issues affect the accuracy of numerical simulations of massive star evolution, which are not unique to helium stars, but affect stellar evolution models in general. Numerical resolution and the choice of nuclear network are known to lead to different core structures at the core collapse, which in turn may affect the explodability of stellar models \citep{2016ApJS..227...22F}. We use a nuclear network with only 21 isotopes, which leads to uncertainties in the final core structure of our models. We have made this choice for practical reasons, since we wanted to cover a large mass and metallicity parameter space. We argue that, while the exact masses at which models transition from convective to radiative nuclear burning is uncertain due to our choice in nuclear network, our calculations cover a broad enough range that this uncertainty will likely not affect the main conclusions we draw from our results.

The efficiency of the $^{12}$C($\alpha$,$\gamma$)$^{16}$O reaction in particular remains uncertain \citep{2017RvMP...89c5007D}, and can have a large effect on the evolution and outcome of stripped-envelope stars, as it may influence the amount of carbon remaining at the end of helium burning, which in turn influences whether carbon burning proceeds radiatively or convectively \citep{2001ApJ...558..903I}. This is known to have a large effect on the explodability of massive stars \citep{2014ApJ...783...10S,2020MNRAS.499.2803P,2021A&A...645A...5S}, as well as on the upper limit for BH formation \citep{2019ApJ...887...53F,2020ApJ...902L..36F}. The effect of these uncertainties is beyond the scope of this work, but they can have an effect on the exact predictions of our analysis. However, we argue that because of the size of our model grids, the effect of metallicity dependent mass loss on the evolution of helium stars explored in this work will have a larger impact than that of the uncertainties inherent to stellar evolution models.

The efficiency of convection and the presence of overshooting in helium stars are also uncertain. Varying these parameters may lead to convective helium-burning cores of different mass. This will result different lifetimes, and lead to significant differences in evolution, particularly in the lower mass regime. Overshooting in low mass helium stars will most significantly affect the locations of the boundaries in initial mass that lead to different final outcomes \cite{savvas2021}. This uncertainty is perhaps the largest in our population models, as it might imply that the minimum mass at which stars become core collapse SNe is dependent on metallicity. We believe that variations in this limit will likely have an effect on real populations of stripped-envelope SNe, but the conclusions derived from our simplistic model will remain broadly valid. However, we highlight the need for a more detailed population synthesis model of stripped-envelope stars, that additionally includes the efficiency of stripping as a function of both mass and metallicity.

\subsubsection{Determination of supernova properties}

Using the \cite{2016MNRAS.460..742M} SN explosion model, we estimate properties of the resulting explosions of individual models. These properties are relatively uncertain on their own, and accurate estimations of the explosion outcome can only be more accurately determined through detailed 3D simulations. However, this model can yield predictions in much shorter timescales, and predicts trends in the behaviour of large sets of models, revealing the consequences of the assumptions made to produce them. These explosion models, however, depend on a series of free parameters (see Sect. \ref{sec:methods}, and Appendix \ref{sec:app_parameters}) and variations in their value produce different results. 

We have performed several tests to characterize the effect of variations of these parameters. We find that, within a reasonable range of values, the outcome of the models is broadly similar. The distributions produced by simulating SNe explosions with a different set of parameters changes the individual values of the predicted outcomes, and the range where explosions are expected to occur, so we chose the set of parameters that more closely resembles the predictions of the \cite{2016ApJ...818..124E} explodability test.

\subsection{Comparison to previous stellar evolution models}

Numerical studies to investigate evolutionary channels for the progenitors of stripped envelope SNe have been carried out extensively in the context of single star evolution \citep[e.g.][]{2003A&A...404..975M,2009A&A...502..611G,2012A&A...542A..29G,2013ApJ...764...21C}, massive stars in binaries \citep[e.g.][]{1992ApJ...391..246P,2017ApJ...840...10Y,2021A&A...645A...5S,2021A&A...656A..58L}, and through the evolution of helium stars \citep[e.g.][]{2020A&A...642A.106D,2020ApJ...890...51E}.

Single star models where wind stripping is the only component to produce hydrogen free stars at core collapse require very large initial masses ($\gtrsim 25 \mso$), which may be in conflict with the rate of stripped-envelope SNe, and predict high final masses, which is in conflict with the observed distribution of ejecta masses \citep[e.g.][]{2015PASA...32...15Y}. Binary stellar models are much less restrictive in the minimum mass that is required to produce stripped-envelope stars, but often grids of binary models have very few suitable progenitors for Type Ic SNe \citep[e.g.][]{2010ApJ...725..940Y,2017ApJ...840...10Y}.

\cite{2017MNRAS.470.3970Y} suggested that modifying the mass loss rates of WR stars might contribute to produce models in the right mass and compositon range, and these models were found to reproduce the spectral properties of both Type Ib and Type Ic SNe by \cite{2020A&A...642A.106D}. The models in this work have very similar properties and evolution as those from \cite{2020A&A...642A.106D}, but extend the applicability of their predictions by covering a larger mass and metallicity range, and analysing their explodability properties.

We find that the core properties of our models at core collapse are consistent with previous models of stripped envelope stars coming from binary calculations, and --like other single helium star models \citep[e.g][]{2019ApJ...878...49W,2020ApJ...890...51E}-- form a distinct family of transient progenitors with properties that differ from those obtained for single stars: stripped-envelope stars tend to have lower compactness, lower central carbon mass fraction at the end of helium burning, and lower carbon-oxygen core masses than stars with similar main sequence masses that did not become stripped \citep{2019ApJ...878...49W,2021A&A...656A..58L,2021A&A...645A...5S}.

Studying several metallicities, however, allows us isolate the effect of mass loss in core and surface properties of helium stars. We find that the effect of distinct mass loss in the WN and WC phases can have a strong effect on the precollapse properties of WR stars. Most notably, we find features that were either absent or not reported in previous models: compactness parameters of helium stars with final masses above $\sim$16-25 $\mso$ bifurcate, creating two families of solutions that have typical compactness parameters of $\sim$ 0.4--0.6 and $\sim$ 0.6--0.8. Explodability tests \citep{2011ApJ...730...70O,2016ApJ...818..124E,2016MNRAS.460..742M} predict these models to form BHs with approximately the final stellar mass, but using the model of \cite{2020MNRAS.499.3214M}, we find that many of the low compactness models are candidates for forming fallback SNe (i.e. a fallback island of explodability). These models may be similar to the successfully exploding models of M>60 $\mso$ of \cite{2016ApJ...818..124E}, but the connection between them and the reason for the bifurcation are unclear.

\subsection{Transients and compact objects}

\subsubsection{Explodability of stripped-envelope stars}

We find that the distribution of $\xi_{2.5}$ is similar for helium stars of different metallicities that have similar final masses. Differences between the distributions of $\xi_{2.5}$ as a function of final mass, however, likely arise due to differences in CO core masses and central carbon abundances at the end of helium burning, and result in variations in the core structure. These discrepancies are reflected in the different values of $M_4$ and $\mu_4$ that we find for stars of similar final mass, as well as in the results obtained from applying the \cite{2016MNRAS.460..742M} and \cite{2020MNRAS.499.3214M} model to our stellar models at core collapse (see Fig. \ref{fig:everything}).

Difference in core structure in models of equal final mass imply that the ejecta mass and remnant mass distribution at different metallicities will be affected by more than just a displacement in the final mass distribution of models (see Figs. \ref{fig:panel_mueller} and \ref{fig:panel_fallback}. The imprint of these differences is summarized in Figs. \ref{fig:cdf} and \ref{fig:BH_hist}, and are further discussed below.

\subsubsection{Ejecta masses of stripped-envelope supernovae}

Many of our predictions are agnostic to the assumptions and uncertainties inherent in the SN model. In particular, the masses of models that explode or implode are in agreement with the predictions obtained from the \cite{2016ApJ...818..124E} test. This suggests that the distributions of BH masses and of SN ejecta masses are roughly independent of the way that they were obtained, and only reflect the metallicity dependence of WR mass loss.

We find that the distribution of ejecta masses for Type Ic SNe is reproduced by our models, preferentially at very high metallicity (0.03 -- 0.04). This suggests several interesting possibilities for the progenitors of these SNe. \cite{2011ApJ...731L...4M} found that most Type Ic are indeed produced in environmnets of supersolar metallicity, but it's unlikely that the typical metallicity of their progenitors is twice solar and above. Instead, this might reflect that the mass loss rates inferred for either WN-type stars, WC-type type stars, or both, are larger than what is reported in the literature, in stark contrast with the predictions of \cite{2020MNRAS.491.4406S} and \cite{2020MNRAS.499..873S}.

Another possibility is that stripped-envelope stars experience short episodes of mass loss at the end of their lives, likely originating from their proximity to the Eddington limit. Another explanation is that, since many of them are likely to be formed in binaries, they may experience episodes of mass transfer after they have been stripped, and therefore reach the end of their lives with lower masses. Because helium stars lose mass at such a high rate, both their radii and their luminosities decrease during their core helium burning lifetime. This implies that, even if they are in a close binary, helium stars will not interact with their companions during core helium burning, and can only interact in later stages if helium shell burning drives a radial expansion.

We also find that the ejecta masses of our models, even at the highest metallicities in our grid, fail to reproduce the lowest observed ejecta masses. This phenomenon might be explained by the reasons stated above, or it might be caused by a larger fraction of their progenitors experiencing late time fallback during explosion, leading to the formation of either massive NSs or low mass BHs, and low ejecta mass explosions (although fallback SNe are not predicted in the relevant mass regime by \citealt{2020ApJ...890...51E}, but see \citealt{2021ApJ...920L..17V} and Appendix\,\ref{sec:app_parameters}).

The observed distribution of Type Ib SNe is closer to the results from our models at low metallicity, but the predicted distributions of ejecta masses for Type Ib SNe obtained by taking the output from the models at face value are overestimated. Helium stars with initial masses below or near the minimum mass at which stripped-envelope stars are observable as WN-type stars lose less mass than their WN counterparts. Through our simplistic population model, we find that most progenitors of Type Ib SNe will in fact not be WR stars, but rather helium stars with transparent winds that lie below this limit. We propose this as an explanation to the stark discrepancy between the ejecta masses of Type Ib SNe predicted from our models, and the observed sample from \cite{2018A&A...609A.136T}, who find Type Ib SNe with significantly larger ejecta masses than those that result from our models, and find that correcting for this produces a distribution of ejecta masses that is more resemblant to the observed distribution. However, Type Ib SNe progenitors that come from stars that are stripped of their hydrogen envelope after a significant fraction of their helium burning lifetimes has taken place, may also contribute to increasing the number of Type Ib progenitors with high ejecta mass, and to increase the scatter in the distribution.

\subsubsection{Explosion energies, nickel yields and kicks}\label{sec:energies}

The values predicted for SN explosion energies, nickel yields and kick velocities from our helium star models rely on the \cite{2016MNRAS.460..742M} model. We have calibrated the input parameters of the model to agree with the predictions from the explodability test of \cite{2016ApJ...818..124E}, and to produce explosion energies and values of $\mathrm{E}/\mathrm{M}$ in a range that we found reasonable.

For many of the models, we find explosion energies that are larger than those predicted by the more detailed study of \cite{2020ApJ...890...51E}, even when only neutrino-driven explosions are considered. However, recent 3D simulations of SN explosions from massive progenitors \citep[e.g.][]{2020MNRAS.494.4665P} have found that producing SNe with energies in excess of about $3\times 10^{51}$ erg is plausible. The highest energies in our models come from fallback SNe. The models where we predict fallback SNe to take place are typically of higher mass, so a larger energy is required to unbind the envelope of the progenitor. As the fallback prescription of \cite{2020MNRAS.499.3214M} has been calibrated based on 3D fallback simulations of relatively weak explosion, the extrapolation to the very energetic fallback supernovae predicted in this study is quantitatively uncertain, and the largest explosion energies, which well exceed any 3D models of neutrino-driven models, may be overestimated. On the other hand, late time fallback accretion provides an additional energy source which allows explosions to occur, even after the central proto-NS has reached its maximum mass. This is not included in our analytic treatment.

Furthermore, the nickel masses in fallback explosions are more uncertain than the energetics. Whereas the transfer of energy from the engine to the ejecta in a fallback explosion can be approximated analytically in a quasi-spherical picture, the amount of Ni that survives fallback will depend sensitively on mixing instabilities, for which we presently do not have an effective treatment in our analytic supernova models.

Regardless of the high energies that some individual models reach, the predicted distribution of energies in our population models does not contain many of these very high energy explosions, as they originate predominantly in high mass models. The energy distribution we obtain for Type Ib SNe at Z=0.01 seems to be in rough agreement with the observed sample of \cite{2018A&A...609A.136T}, although it is skewed to slightly lower energies (see Fig. \ref{fig:cdf}). A possible explanation for this behaviour might be that the progenitor mass of the majority of our models is underestimated, since they correspond to helium stars below the minimum WN luminosity limit. A larger final masses will result in larger carbon oxygen core masses at core collapse, which is, in turn, likely to shift the distribution to higher energies. Furthermore, progenitors of Type Ib SNe that come from stars that were not fully stripped after the main sequence may also result in more energetic explosions since they are expected to have higher final mass, and higher core mass.  This conjecture can be tested when accurate progenitor models of Type Ib SNe become available.

The energy distribution of Type Ic SNe on the other hand has the opposite effect. Since most of the progenitors of Type Ic SNe at low metallicity come from stars of higher mass (and are also therefore more rare), their energy distribution we obtain is skewed to much larger energies. It is only at higher metallicity that we find a broad energy distribution that also includes lower energy explosions, and resembles the observed distribution more closely.

Like ejecta masses, explosion energies of Type Ic SNe are only reproduced by very high metallicity models, and even at our highest metallicity, we find that the SNe with the lowest kinetic energies in the observed sample are missing from our predictions. This supports our claim that Type Ic SN progenitors likely experience additional mass loss, either due to winds or to a different effect that acts on shorter timescales.

It must be noted, however, that ejecta masses and explosion energies inferred from light curves of stripped-envelope SNe are degenerate \citep[i.e., several combinations of these two parameters can produce identical light-curves. See][]{2020A&A...642A.106D}. However, comparing the value of $\mathrm{E}/\mathrm{M}_\mathrm{ej}$ predicted by our models to the observed one results in a distribution that is similar from those obtained in observations. We highlight, however, that there is a large scatter in the value of $\mathrm{E}/\mathrm{M}_\mathrm{ej}$ resulting from our models. The distribution of this quantity ranges from about $0.3\times10^{50} \ \mathrm{erg}/ \mso$ in models with the lowest masses, monotonically increasing with initial mass up to about 2 in the models with the highest initial mass, while some models with initial masses in the range of 20--40 $\mso$ reach slightly larger values, as well as those that are predicted to become fallback SNe.

Nickel masses in our models tend to be considerably lower than those found by \cite{2018A&A...609A.136T} and \cite{2021A&A...651A..81B} at every metallicity. Our population model predicts that the median nickel mass of Type Ib SNe is about a factor of 3 lower than that of the analysed sample. In Type Ic SNe, the difference is less stark, about a factor 2, but is still widely underestimated. We find that nickel mass has a tendency to decrease with metallicity.

The nickel mass in the observed samples is obtained by means of Arnett's rule \citep{1974ApJ...194..373A}. It has been found that this estimate overestimates the nickel mass produced in stripped-envelope SNe of all types \citep{2016MNRAS.458.1618D,2021ApJ...918...89A}, which may help explain the discrepancy between our estimations and the observed sample. As with explosion energies and ejecta masses, in the case of Type Ib SNe, an additional cause for the prediction of low nickel masses might be that their final mass is underestimated, since we find that most Type Ib SNe are originated in stars with luminosities below the minimum WR luminosity, and therefore have overestimated mass loss rates.

Kick velocities are computed following \cite{2018MNRAS.481.4009V}. They depend on the explosion energy and the remnant mass (as well as in the mass enclosed between the acceleration region and the point where the shock exceeds the escape velocity of the material below it), and are therefore correlated (see Fig. \ref{fig:everything}). The distribution of kick velocities in our models resembles that from \cite{2020MNRAS.499.3214M}, and resembles the velocity distribution of pulsars \citep{2005MNRAS.360..974H}. However, we find that the distribution has a tendency to decrease with increasing metallicity, likely as a consequence of the decrease in mass of SN progenitors.

\subsubsection{Neutron star and black hole mass distributions}

We find that Type Ic SNe produce more massive NSs than Type Ib SNe. In agreement with \cite{2020ApJ...890...51E} and \cite{2021A&A...645A...5S}, we find that NS gravitational mass is correlated to $\xi_{2.5}$ (see Fig. \ref{fig:everything}). This implies that the prediction is also independent of the SN explosion model, since most Type Ic progenitors come from stars that are more massive than those of Type Ib SNe at the time of core collapse, and also tend to have a larger value of $\xi_{2.5}$.

Variations in the locations of the ``islands of explodability'' as a function of metallicity are clearly reflected in different BH mass distributions. Our results resemble those of \cite{2020ApJ...896...56W}, but we find that variations in metallicity result in more drastic changes in the locations of the peaks and valleys of the BH mass spectrum.

The emergence of a low mass peak at around $3 \mso$ in some of our grids arises from fallback SNe that result from some high mass models with relatively low values of $\xi_{2.5}$. The intermediate mass peak in our distributions becomes displaced from around $9\mso$ at Z=0.01 to around $6\mso$ at Z=0.04, and has a structure that results from the variations in core properties as a function of mass in models in this regime, consequence of the transition between convective and radiative core carbon burning \citep{2014ApJ...783...10S}.

The peak mass in the distribution that corresponds to high mass BHs, above $10\mso$, also decreases as a function of metallicity. Due to variations in explodability that arise as a function of metallicity, even in models with the same final mass, it is enough to bring this peak from about $15\mso$ in the Z=0.01 grid, down to about $11\mso$ at Z=0.04. The highest mass BH at each metallicity corresponds to the highest mass model, and carries no information about whether the location of the BH mass gap depends on metallicity. 

This behaviour reflects the need to take metallicity dependence of BH mass production into account when generating predictions for the BH mass distributions in the Universe, and when interpreting the findings of the LIGO/VIRGO collaboration \citep[e.g.][]{2021ApJ...912...98F} and future gravitational wave observatories.

\subsubsection{The metallicity evolution of progenitors of stripped-envelope supernovae}

We find that the number ratio of Type Ib SNe to Type Ic SNe increases as a function of metallicity due to the metallicity dependent mass loss rates of WR stars. Our findings are in broad agreement with the relative rates found in the local Universe \citep{2017PASP..129e4201S}, and with the fact that Type Ic SNe are found in environments of higher metallicity than other core collapse SNe \citep{2011ApJ...731L...4M}. According to our analysis in \citetalias{2021arXiv211206948A}, stars in our sample that evolve to become progenitors of Type Ic SNe will be luminous enough to be observable as WC-type stars, meaning that they are unaffected by the uncertainty in the mass loss rates of low luminosity helium stars. 

On the other hand, we find that although the relative number of Type Ib SNe with transparent wind stripped-envelope stars as progenitors relative to those that have WR progenitors both tends to increase as a function of metallicity, it is below 0.2 even at the highest metallicity in our grid, and about 0.06--0.08 at approximately solar metallicity. Even though the value of the minimum luminosity of WR stars as a function of metallicity is approximate, this implies that, in our simplistic model, the overwhelming majority of progenitors of Type Ib SNe are in fact not WR stars, but rather stripped-envelope stars with transparent winds, and luminosities below the WR threshold. This result is further backed by the fact that, making this assumption, the distribution of ejecta masses we obtain for Type Ib SNe resembles the observed distribution more closely.

Because all of our explodability tests roughly coincide in their results, the inferred values of ratios presented in Fig. \ref{fig:Ic_Ib_ratio} are independent of our SN calculations, and their uncertainties stem from uncertainties in our determination of the minimum WR luminosity, the minimum mass at which helium stars produce core-collapse SNe and the stripping efficiency of stars as a function of metallicity.

The number of SNe coming from low luminosity helium stars is not expected to be affected by a reduction in the mass loss rate of their progenitor, since their low initial mass also implies a low value of $\xi_{2.5}$. Therefore, we argue that our calculations on the relative rate of SN types and the spectral class of the progenitors of Type Ib SNe (Fig. \ref{fig:Ic_Ib_ratio}) are not dependent on the effect of decreasing winds for helium stars with luminosities below the minimum WR luminosity, but they are uncertain inasmuch as the minimum WR luminosity and the minimum mass at which stars produce SNe are uncertain.

Although WR stars are not very luminous in optical bands, since they are compact and hot, the problem becomes more dramatic for low mass stripped-envelope stars, as they do not show strong emission lines, and are less luminous and comparably compact. The fact that we predict that most Type Ib SNe to form in transparent wind helium stars might be the reason why many searches for archival images of their progenitors in SN explosion sites have been unfruitful \citep[e.g.][]{2009ARA&A..47...63S}. During the core helium burning phase, low mass helium stars with transparent winds are not very bright in optical wavelengths \citep{2018A&A...615A..78G}, where the searches for their progenitors have taken place. Models predict that, below a certain mass, helium stars expand during the late phases of their evolution. This expansion is caused by a combination of inflation and the presence of a helium burning shell \citep[e.g.][]{2010ApJ...725..940Y,2019ApJ...878...49W}. This implies that the temperatures of helium stars after core helium depletion will decrease, leading to a larger brightness in the optical band \citep{2012A&A...544L..11Y,2015ApJ...809..131K}. The transition mass at which expansion becomes important cannot be directly assessed from our models, since it requires a more detailed treatment of the envelope (i.e. without the use of \texttt{mlt++}). However, our current models indicate that this is metallicity dependent, and may have consequences for the fate of these stars, particularly if they are in binaries \citep{2010ApJ...725..940Y,2021ApJ...920L..17V}.

Direct observations have been so far limited by the fact that most of these explosions have not been detected in the nearby Universe. Regardless, new atmosphere and stellar evolution models of low mass stripped-envelope stars are required to constrain their properties at core collapse. Their mass loss rates and the response of their envelopes to expansion (both due to inflation and helium shell burning) as a function of metallicity, are paramount to understand current and future identifications of progenitors of Type Ib SNe, and currently available upper limits on their optical brightness. Together with direct observations of stripped-envelope stars may reveal phenomena that occur in the final few thousand years of the evolution of stripped-envelope stars, which are so far poorly constrained.

\subsection{Implications for double compact object assembly and gravitational-wave sources}
Massive binaries are believed to be the progenitors of double compact object mergers and gravitational-wave sources.
Moreover, interacting massive binaries are believe to be common in the field \citep[e.g.,][]{2012Sci...337..444S}, and these interactions usually result in mass transfer episodes and subsequent envelope stripping \citep[e.g.][]{1991PhR...203....1B,1993MNRAS.260..675T,2016Natur.534..512B}.
Additionally, wind stripping can also occur in massive binaries, particularly at high metallicities \citep[e.g.][]{2012A&A...542A..29G}.
Therefore, the evolution of stripped helium stars play a fundamental role in the assembly and characterisation of double compact objects.

\subsubsection{Compact object formation and islands of explodability}
The predictions of the properties of a compact object remnant is a difficult problem.
The main question, in the context of gravitational-wave sources, is: what is the mass of a remnant given a stellar progenitor?
While for decades the answer seemed to be that increasing masses lead to increased remnants \citep[e.g.][]{2012ApJ...749...91F}, more recent and more detailed studies suggest that the structure of the star at core collapse plays a crucial and complex role \citep[e.g.][]{2011ApJ...730...70O,2016ApJ...818..124E,2020MNRAS.499.2803P,2020ApJ...890...51E}.
One of the main features that has emerged from the discovery of ``islands of explodability'' can be interpreted as models that are similar in mass at core collapse but lead to significantly different compact object remnants, the most extreme case scenario being a light neutron star and a heavy black hole. 
Additionally, it represents regions on the progenitor mass space that are very discontinuous in the remnant mass space. According to our results, the picture may be further complicated by the presence of fallback SNe resulting from the evolution of very massive stars (see Fig. \ref{fig:all_tests_fallback}), that lead to the formation of remnants with very small masses.
This probably has important implications in the rates of double compact objects.
If these islands of explodability are real, i.e., not a numerical artifact, that would decrease rates of binary black hole mergers and likely increase the rate of neutron star/black hole mergers (modulo disruption via neutron star natal kicks).

Double compact object mergers with highly asymmetric mass ratios, such as the $\approx$23.0-2.6 $\rm{M_{\odot}}$ gravitational wave event GW190814 \citep{2020ApJ...896L..44A}, might be the right candidates to probe the explodability island hypothesis. 
Currently, rapid population synthesis models struggle predicting such asymmetric mergers \citep[e.g.][]{2020ApJ...899L...1Z}, even when considering the possibility of explodability islands \citep{2021MNRAS.500.1380M}. This idea is explored further by \cite{2022A&A...657L...6A}.

\subsubsection{High mass X-ray binaries}
High mass X-ray binaries are observed systems that can potentially become double compact objects. 
The outbursts of Be X-ray binaries \citep[e.g. ][]{1982IAUS...98..327R}, comprised of Be stars \citep[e.g.][]{2013A&ARv..21...69R} and a compact accretor, have been observed in the Milky Way and the Magellanic Clouds \citep{2011Ap&SS.332....1R}.
The compact object progenitor is believed to have been stripped in a mass transfer episode with the stellar companion \citep[e.g.][]{2020MNRAS.498.4705V}, and therefore likely spend a fraction of its lifetime as a stripped-envelope star. Its behaviour will be determined by whether or not its luminosity is above the minimum WR luminosity; i.e. if it is a WR star or not. Furthermore, the presence of inflation will determine whether or not binaries that include a stripped-envelope star will have episodes of accretion, and the question remains whether this accretion will be stable or lead to a merger.

If the star in an X-ray binary is indeed a WR, its winds are expected to widen the orbit. 
Moreover, higher mass loss rates, as those expected at higher metallicities, lead to longer period orbits.
Heavier X-ray binary systems are comprised of O-type stars with black hole companions, which have been the archetype of interacting compact object binaries since the discovery of Cygnus X-1 \citep[e.g.,][]{1965Sci...147..394B,1972Natur.235...37W,1972Natur.235..271B}.
Recently, the inferred black hole mass of Cygnus X-1 has been corrected from $14.8\ \rm{M_{\odot}}$ to $21.2\ \rm{M_{\odot}}$ \citep{2021Sci...371.1046M}, suggesting that winds from massive stars at high metallicities have been overestimated in the last decades.
The canonical evolutionary pathway in the formation of Cygnus X-1 involves a helium-star with a stripped (initial) mass between $22-30\ \rm{M_{\odot}}$, depending on the metallicity \citep{2021ApJ...908..118N}.

Having a better understanding of the evolution of stripped-envelope stars is paramount to understand the observable time of X-ray binary progenitors as WR stars \citep{2017MNRAS.471.4256V}, the evolution of the orbit and assembly of the X-ray binary phase \citep[e.g.][]{2017MNRAS.467..298R}, and the fate of the system as a double compact object \citep[e.g.][]{2019IAUS..346..417K}.

\subsubsection{Gravitational wave sources}
Most double compact objects formed in isolation are expected to engage in at least one mass transfer episode throughout their evolution.
The canonical binary black hole formation scenario from isolated binaries includes stripping via the common envelope phase \citep[e.g.][]{2016Natur.534..512B,2017NatCo...814906S}. However, more recent studies suggest that stable mass transfer might be frequent during binary black hole formation 
\citep{2019MNRAS.490.3740N,2021A&A...650A.107M}. 
The details of stripping are complex and not well understood \citep[e.g.][]{2022MNRAS.511.2326V}, but it is not unreasonable to expect that stripped stars will (eventually) be stripped-envelope stars.
The fate of a binary system is deeply entangled with the evolution of stripped helium stars.
The post-stripping radial evolution dictates whether or not they will engage in an additional mass transfer episode, as well as if they can merge within the age of the Universe.
Stellar winds not only widen the orbit, but directly affect the mass of the stellar core of what will become the compact object remnant.
This combined effect suggests that higher stellar winds of stripped stars, which are likely to widen the orbits and decrease remnant masses, increases the coalescence time of the double compact object.
Less frequent and less massive double compact object mergers are less likely to be detected by ground-based gravitational-wave observatories.

Progenitors of binary neutron stars \citep[e.g.][]{1991PhR...203....1B,2018MNRAS.481.4009V} or neutron star/black hole \citep[e.g.][]{1994ApJ...423L.121L} binaries are also believed to host stripped stars at point in their evolutionary history.
The initially most massive star is usually stripped first and leads to configurations such as the aforementioned X-ray binaries.
The stripping of the companion is very important in the context of gravitational-wave sources:
in principle, light stripped stars will lead to binary neutron stars, and heavy stripped stars to neutron star/black hole binaries.
However, as previously mentioned, the mass and radial evolution of stripped stars is determinant in the fate of double compact objects.
In the canonical formation channel of neutron star binaries \citep[e.g.][]{1991PhR...203....1B,2017ApJ...846..170T}, light stripped stars expand and are likely to engage in a final (stable) mass transfer episode which will further strip the donor star and might spin-up the pulsar companion.
Slightly more massive stripped stars, which also expand, have been predicted to lead to a common envelope which results in short orbits ($\sim$ min) and leads to fast ($\sim$ Myr) double compact object mergers \citep{2018MNRAS.481.4009V,2020MNRAS.496L..64R}.

Recently, \cite{2021ApJ...920L..17V} used the results from this paper to show that heavy stripped stars, which do not expand significantly before core-collapse, will not have a final mass transfer episode and will form radio quiet double compact objects.
\cite{2021ApJ...920L..17V} also showed that the fate of the remnant, to become either a neutron star or a black hole, doesn't depends only on the mass at core collapse but also on the explosion energy.
It also showed that this formation avenue is a plausible formation channel to form neutron stars with heavy neutron star or light black hole companions.

In the context of neutron star binaries,
there is a particular scenario of interest for low-mass stripped stars: the double-core common-envelope scenario \citep{1995ApJ...440..270B}.
In this scenario, a nearly equal mass binary experiences an early common envelope phase in which both stars are giants.
The common envelope is now comprised of both envelopes and the ejection leads to a close binary with two stripped stars. 
Rapid population synthesis studies predict that $\approx$20\% of binary neutron stars formed at solar metallicity form in this way \citep{2018MNRAS.481.4009V}.
Moreover, this formation channel might even be more abundant at lower metallicities.

A more exotic avenue predicts the late, double core, common envelope episode of stripped stars \citep{2001ApJ...550L.183B}.
This formation channel results in very short period ($\lesssim$ few hr) binaries where both stars will explode as Type Ic SNe.
The remaining lifetime of stars that have been stripped late is short, and they conform less than 1\% of binary neutron stars at high metallicity environments \citep{2018MNRAS.481.4009V}.
However, they remain interesting candidates to the models presented in this paper.

\section{Conclusions}\label{sec:conclusions}

We have estimated the final outcome after core collapse of the models of helium stars presented in \citetalias{2021arXiv211206948A}. These models approximate the late evolution of stripped-envelope stars with the simplifying assumption that they are fully stripped at the onset of helium burning. We analysed the resulting core collapse models using several explodability tests. The outcome of these tests allowed us to constrain the observables of Type I core collapse SNe that result from the evolution of stripped-envelope stars, and to construct a synthetic population model of the obtained properties to compare with observations.

These stars have a helium and nitrogen rich envelope and a convective helium burning core enriched in carbon and oxygen rich. As they lose mass, some of them lose their nitrogen-rich envelope due to mass loss. As we employed the mass loss rate of \cite{2017MNRAS.470.3970Y}, this change in surface abundance is also accompanied by an increase in mass loss rate, as they transition from WN to WC-type WR stars, which creates a bifurcation in the resulting final masses of our models. This dichotomy is also associated with the different spectral properties of Type Ib and Type Ic SNe \citep{2020A&A...642A.106D}. Using the method derived in \citetalias{2021arXiv211206948A}, we also recognized which stars will be observable as WRs, and which ones will have transparent winds and low mass loss rates.

We find that the core properties of our models are different from those of stars that do not experience stripping, but are rather consistent with recent works of stripped-envelope stars \citep[e.g.][]{2019ApJ...878...49W,2020ApJ...890...51E,2020MNRAS.491.4406S,2020arXiv201106630L}. However, we find that core properties of stripped-envelope stars such as core compactness, carbon-oxygen core mass and central carbon fraction after helium burning are metallicity dependent, and are also altered by the dichotomy in mass loss rate.

We find that core properties that are related to explodability of massive stars, such as $\xi_{2.5}$, $M_4$ and $\mu_4$, tend to have similar values for stars of similar final mass. However, subtle differences in relative core mass and core composition alter the explodability landscape enough to have an effect in the resulting distribution of SN properties and remnant masses that result from our models.

After assessing the explodability of our models, we construct a synthetic model for the distribution of SN properties and BH masses as a function of metallicity. We find that NSs produced by helium stars tend to have a distribution that resembles the low mass peak of the distribution, as inferred by \cite{2016arXiv160501665A}, and the total distribution of NS masses decreases only weakly  with increasing metallicity. However, we find that Type Ic SNe preferentially form more massive NSs than those that result from Type Ib SNe.

We find that although the overall structure of $\xi_{2.5}$, $M_4$ and $\mu_4$ is similar as a function of final mass, the resulting BH mass distributions differ. The location and width of the intermediate mass and high mass peaks, and therefore also the mass gaps where we do not expect to find any BHs, become displaced to lower masses as a function of metallicity. We also find a low-mass peak in the distribution of BH masses, at around $3\mso$, that emerges from rare fallback SNe that come from very high mass stars (with initial helium star masses $\gtrsim$ $50\mso$). Convolving this behaviour with the metallicity evolution of the Universe is therefore important to be able to draw any conclusions from the observed distributions that will become available as the sample of known stellar mass BHs with measured masses increases.

Using the value of the minimum luminosity of WC-type stars in Eq. \ref{eq:fit_wc}, we find that the progenitors of Type Ic SNe are observable as WC-type stars in the metallicities covered in our simulations. Comparing the distribution we predict for ejecta masses in Type I SNe to observations, we find that the observed distribution of ejecta masses of Type Ic SNe resembles our results most closely at very high metallicities, while they are underestimated at $\sim \zso$ and below. This may imply that either WC mass loss rates, WN mass loss rates, or both, are higher than what we have accounted for. Alternatively, it might suggest that late binary interactions or eruptive mass loss episodes are common in the later phases of evolution of these stars. We find that the explosion energy distribution of Type Ic SNe is also better reproduced by high-metallicity models, which we interpret as confirmation that their progenitors experience subsequent mass loss beyond what is predicted by WR winds.

In our simplified population model, we find that the vast majority of progenitors of Type Ib SNe are below the minumum luminosity of WN-type stars; i.e., they are not WR stars but low mass helium stars with transpratent winds. Assuming that Type Ib SN progenitors have very small mass loss rates does a better job at reproducing the distribution of observed Type Ib ejecta masses. The distributions of explosion energy and nickel mass of Type Ib SNe are underestimated in our model compared to the observed sample. We conclude that this is further evidence that the progenitors of most Type Ib SNe must have low mass loss rates. This highlights the need to characterize, both theoretically and observationally, the properties of low mass helium stars with optically thin winds. Such systems can be studied during their core helium burning phase, where they are most likely to be found, but the final years of their evolution deserve special attention in order to understand why campaigns to directly detect Type Ib SNe progenitors have been generally unsuccessful.

We argue that detailed modeling of stripped-envelope stars, both as individual systems, as components in binary systems, and in rapid population synthesis codes must include the considerations utilized in this paper to have a better understanding of several important phenomena. SN properties and populations depend not only on the efficiency of binarity and wind mass loss to create stripped-envelope stars, but also in the subsequent wind mass loss of stripped-envelope stars. The metallicity dependence of WR winds and the existence of transparent wind helium stars with very low mass loss rates will shape the resulting islands of explodability, and therefore the populations and properties of Type I core collapse SNe and the mass distributions of their remnants. Differences in their evolution and outcome will affect systems such as X-ray binaries and gravitational wave sources.

The physics of stripped-envelope stars is still uncertain. Their mass loss rates are unconstrained in some regimes, it is unknown if they commonly experience outbursts or second episodes of mass transfer, and it is unknown how common they are, and what is their mass distribution. A holistic approach, combining theory and observations of stripped-envelope stars and stripped-envelope SNe, can be a very powerful tool to understand many far-reaching open questions in massive star physics.

\begin{acknowledgements}
D. R. A.-D. and J. A. are supported by the Stavros Niarchos Foundation (SNF) and the Hellenic Foundation for Research and Innovation (H.F.R.I.) under the 2nd Call of ``Science and Society’' Action Always strive for excellence – ``Theodoros Papazoglou’' (Project Number: 01431). B. M. acknowledges support by ARC Future Fellowship FT160100035. S.-C. Y. has been supported by the National Research Foundation of Korea (NRF) grant (NRF-2019R1A2C2010885). A.V.-G. received support through Villum Fonden grant no. 29466.
D. R. A.-D. would like to acknowledge valuable discussions with Eva Laplace, Jakub Klencki and Mathieu Renzo.

This research made extensive use of NASA's ADS and \textsc{mesa}\footnote{\url{http://mesastar.org}} \citep{MESAI,MESAII,MESAIII,2018ApJS..234...34P}.
\end{acknowledgements}

\bibliographystyle{aa}
\bibliography{references}

\begin{appendix} 
\section{Summary of explodability tests}\label{sec:app_extra_figs}

\begin{figure*}[ht!]
\centering
\includegraphics[width=6cm]{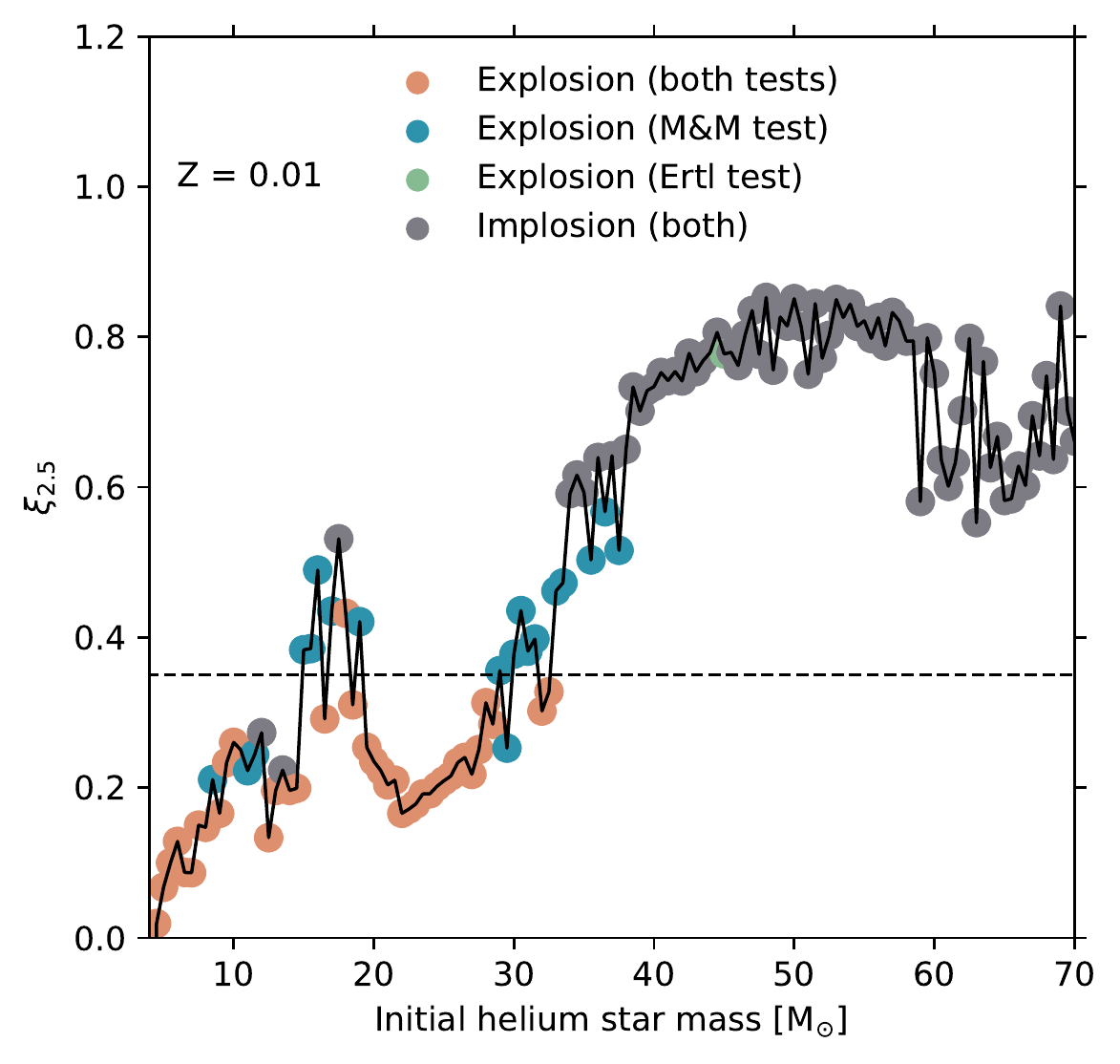}
\includegraphics[width=6cm]{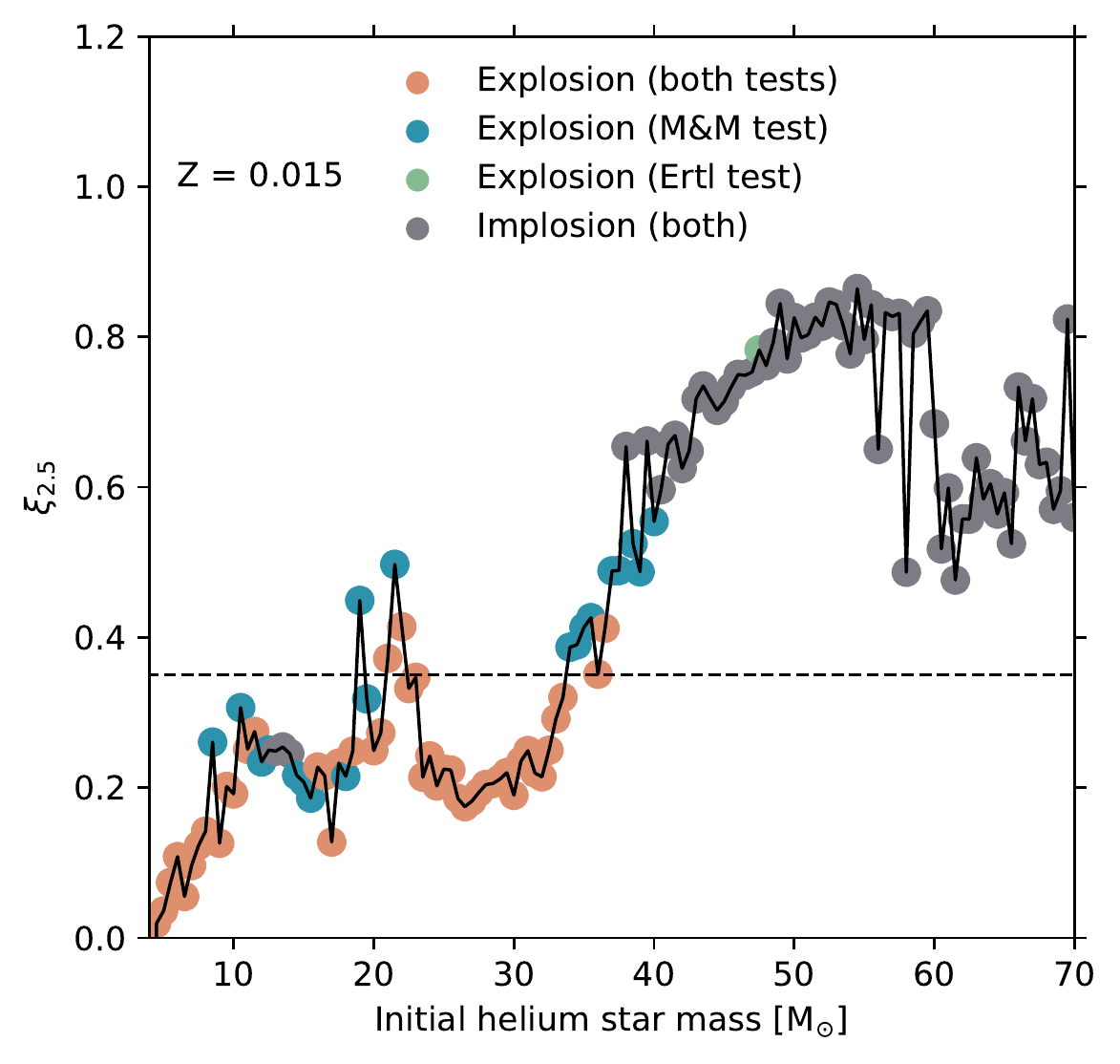}
\includegraphics[width=6cm]{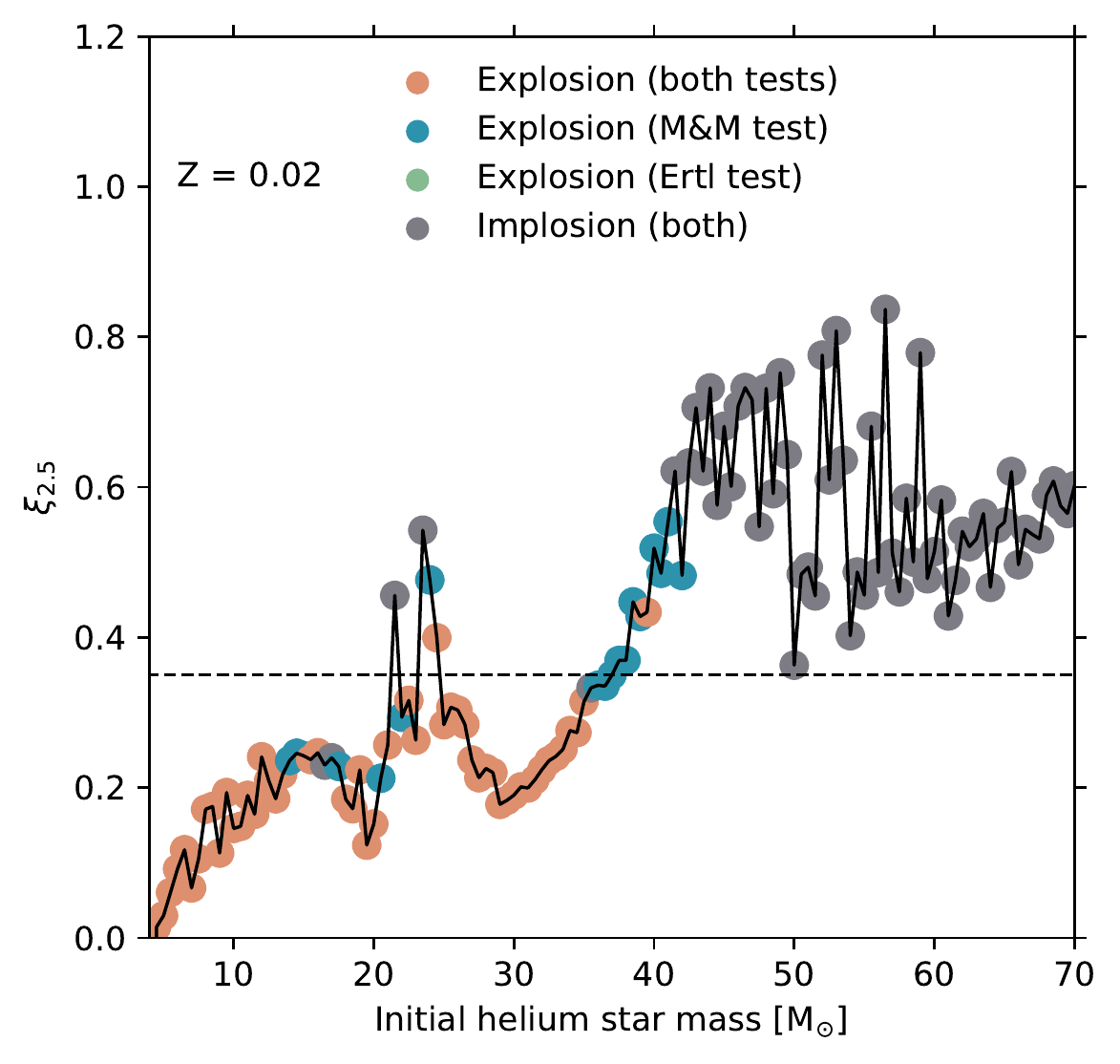}
\includegraphics[width=6cm]{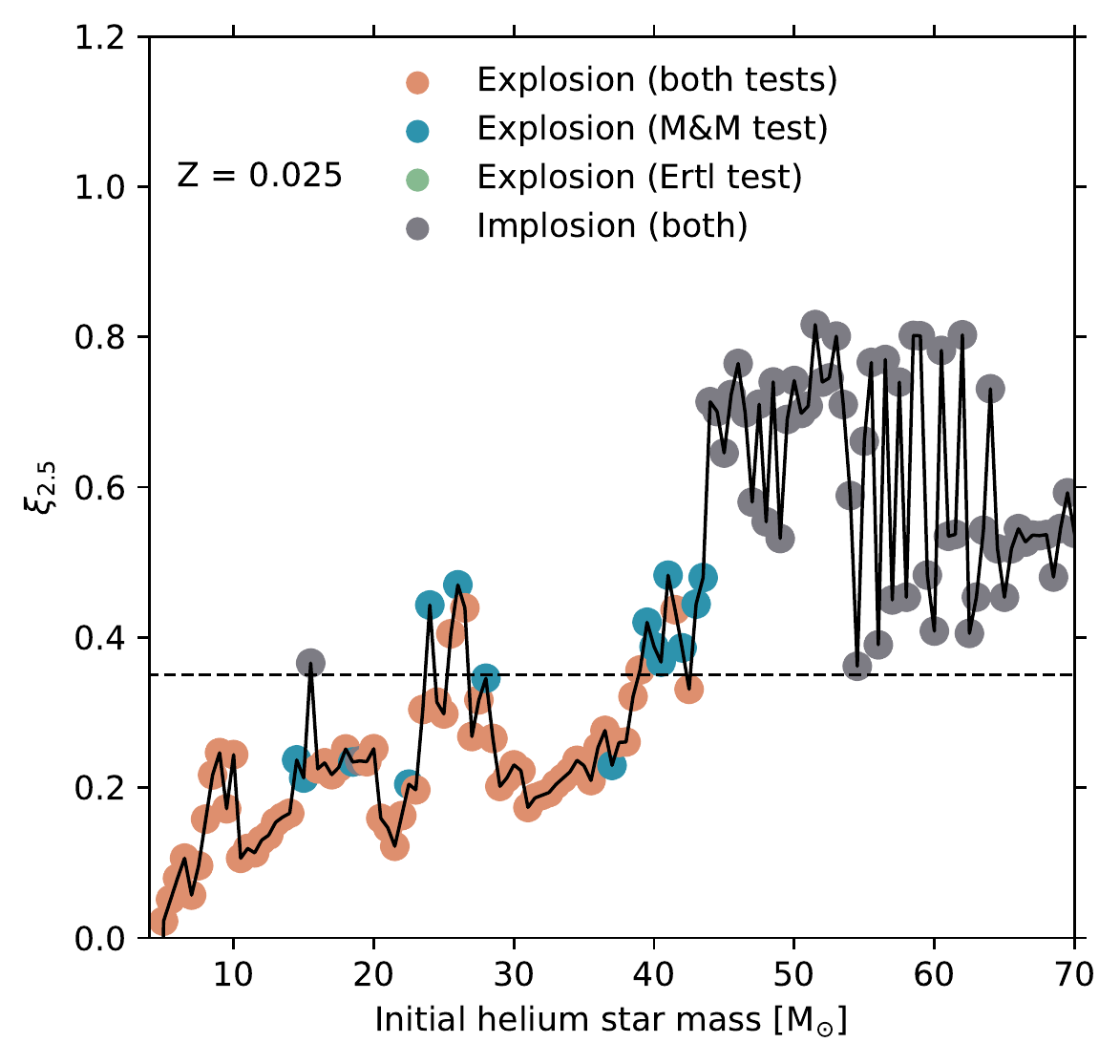}
\includegraphics[width=6cm]{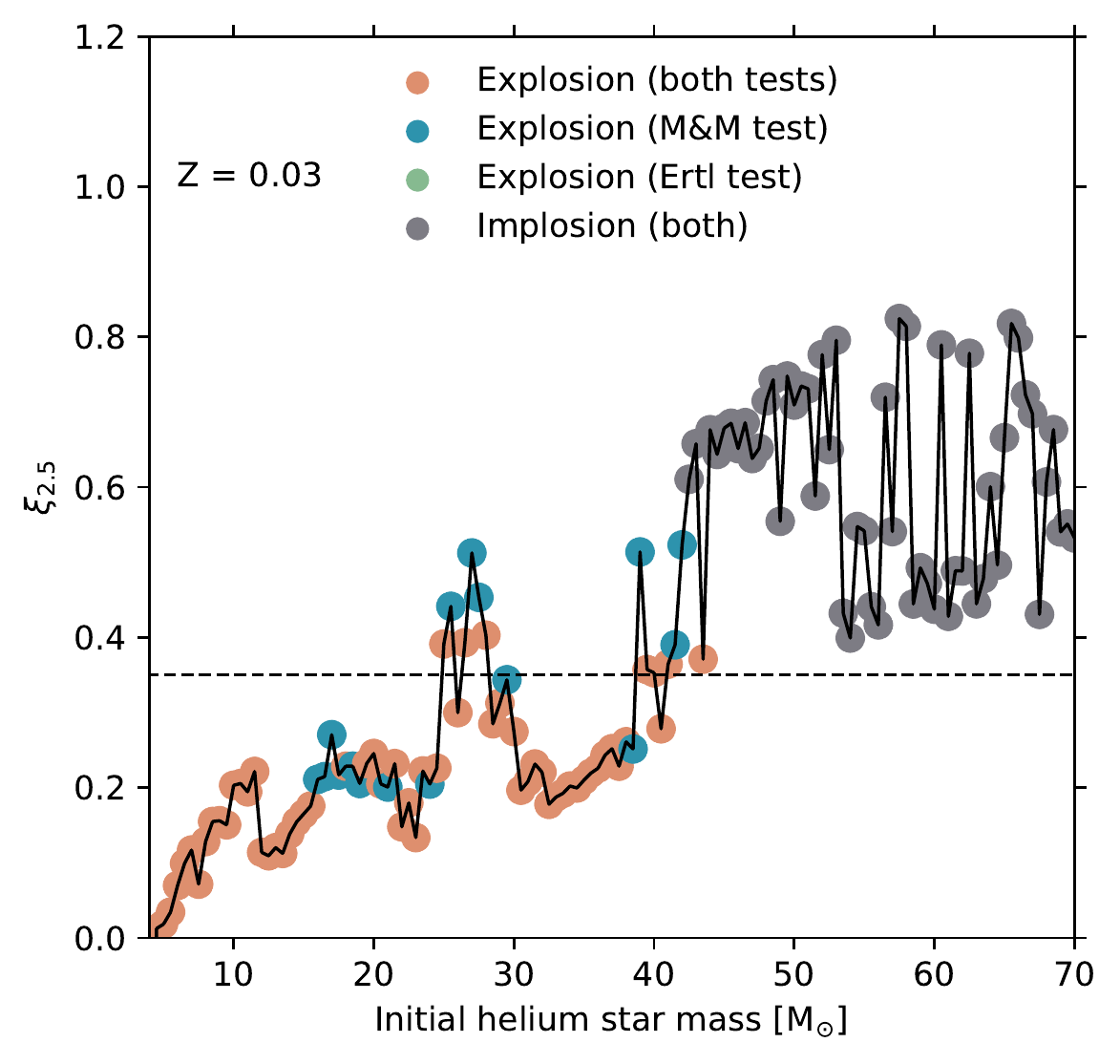}
\includegraphics[width=6cm]{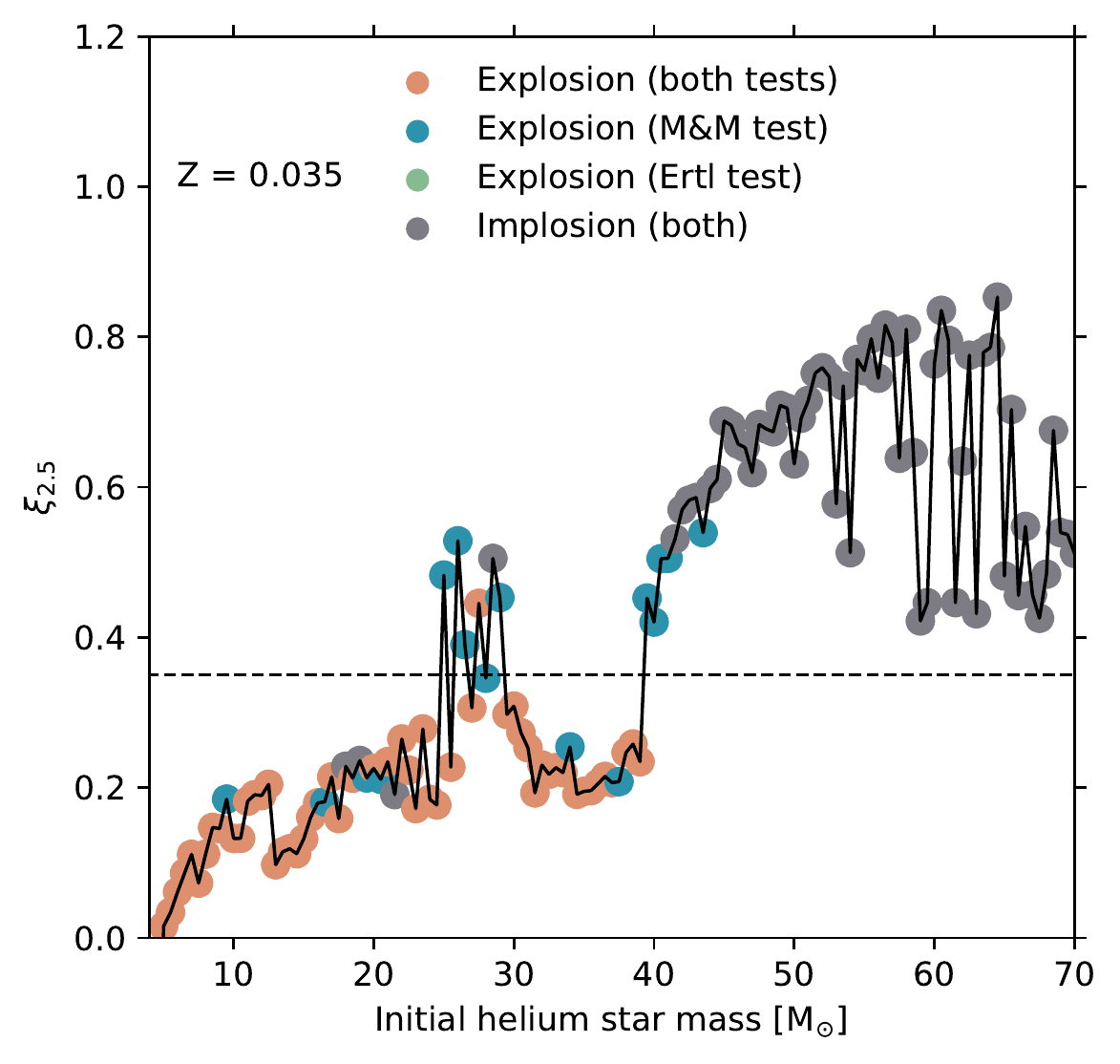}
\includegraphics[width=6cm]{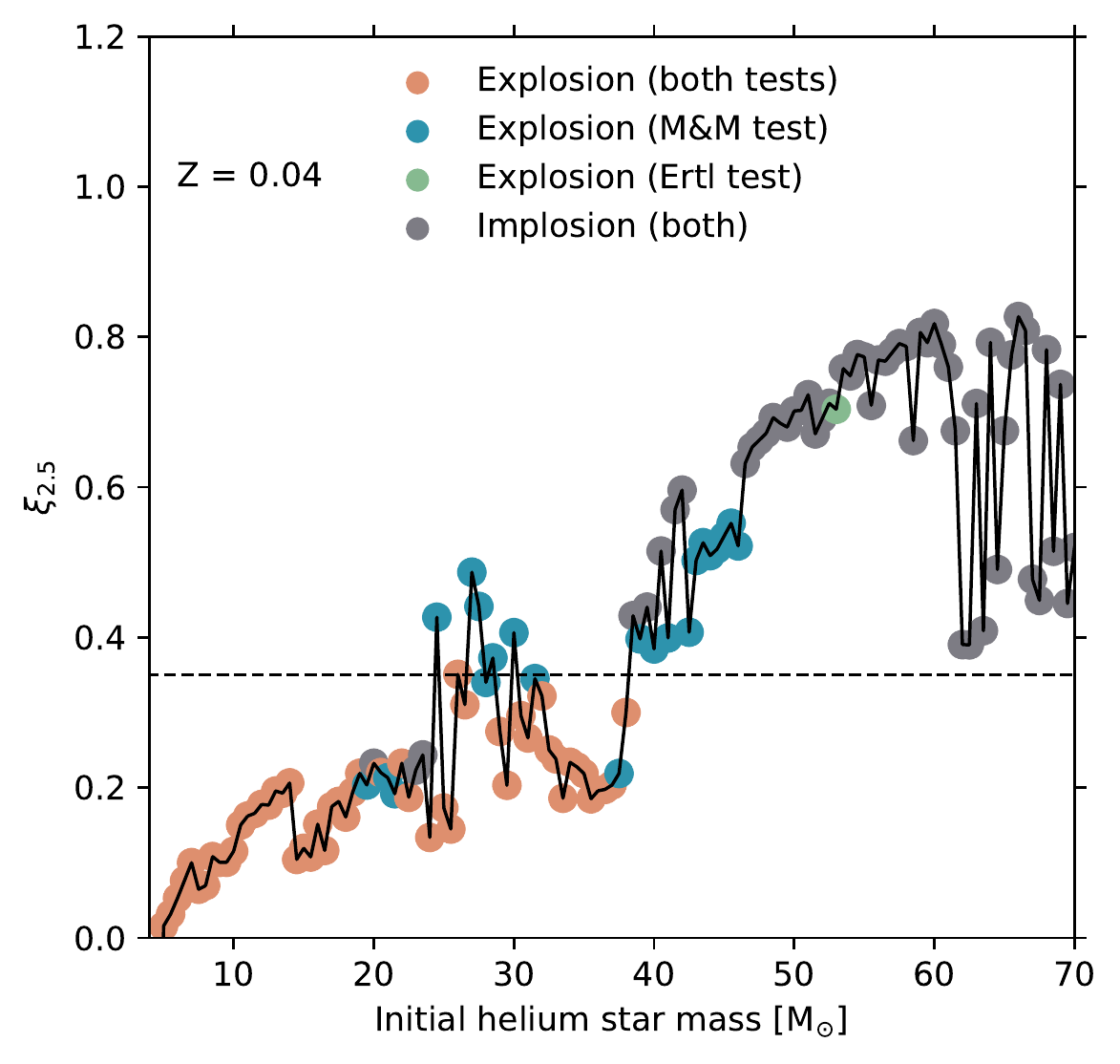}
\caption{Compactness parameter $\xi_{2.5} $of the core-collapse models in this study as a function of their initial helium star mass. The dotted line at $\xi_{2.5} = 0.35$ separates most models that might explode (below the line) or implode (above the line). Orange points indicate models that are predicted to explode according to the explodability of the \cite{2016ApJ...818..124E} and \cite{2016MNRAS.460..742M} tests (using their 
``standard'' parameters), blue points are models that are predicted to explode according the \cite{2016MNRAS.460..742M} test but not the \cite{2016ApJ...818..124E} test, green points are models that is predicted to explode according to the \cite {2016ApJ...818..124E} test but not the \cite{2016MNRAS.460..742M} test, and gray points are models that would not successfully explode according to both tests. \label{fig:all_tests}}
\end{figure*}

\begin{figure*}[ht!]
\centering
\includegraphics[width=6cm]{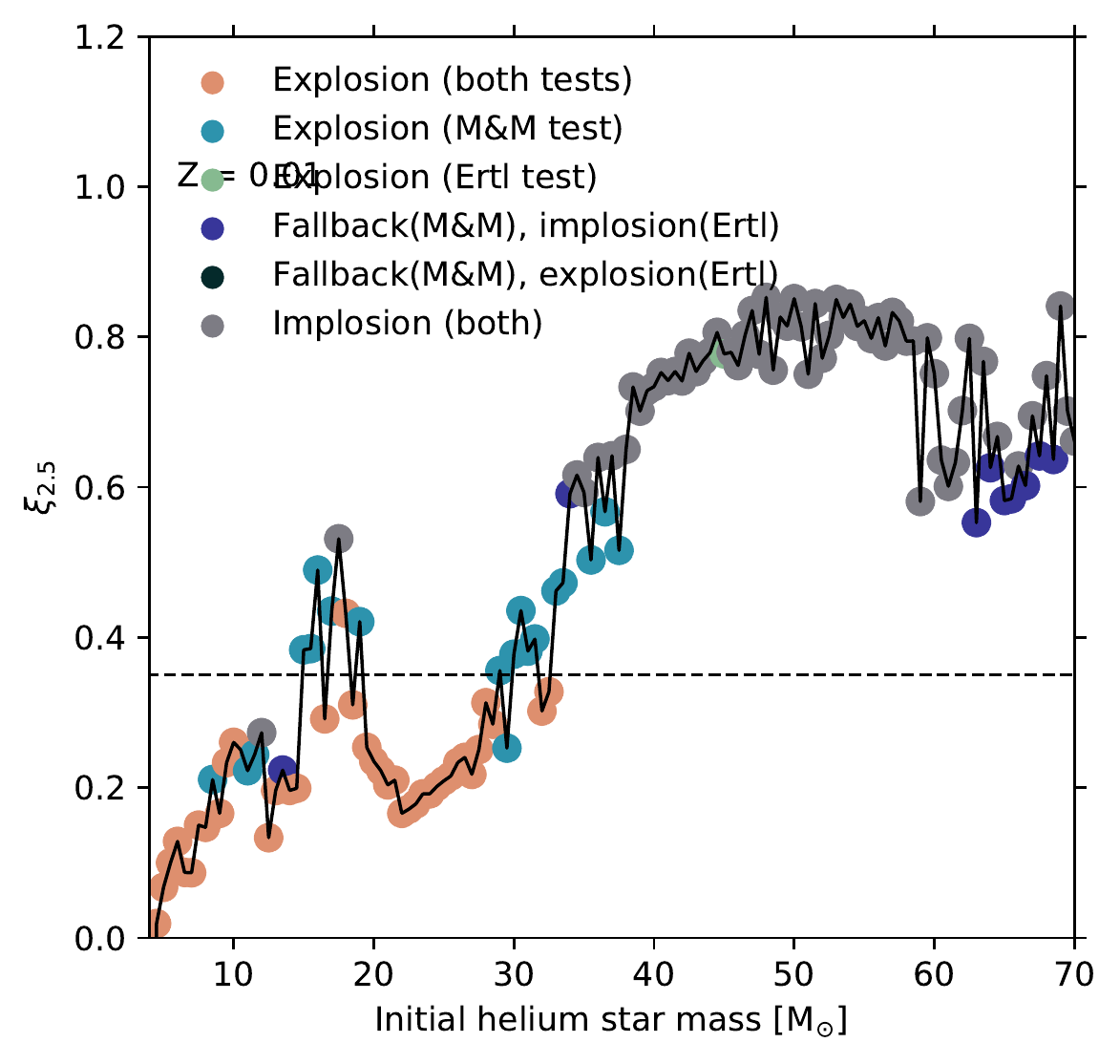}
\includegraphics[width=6cm]{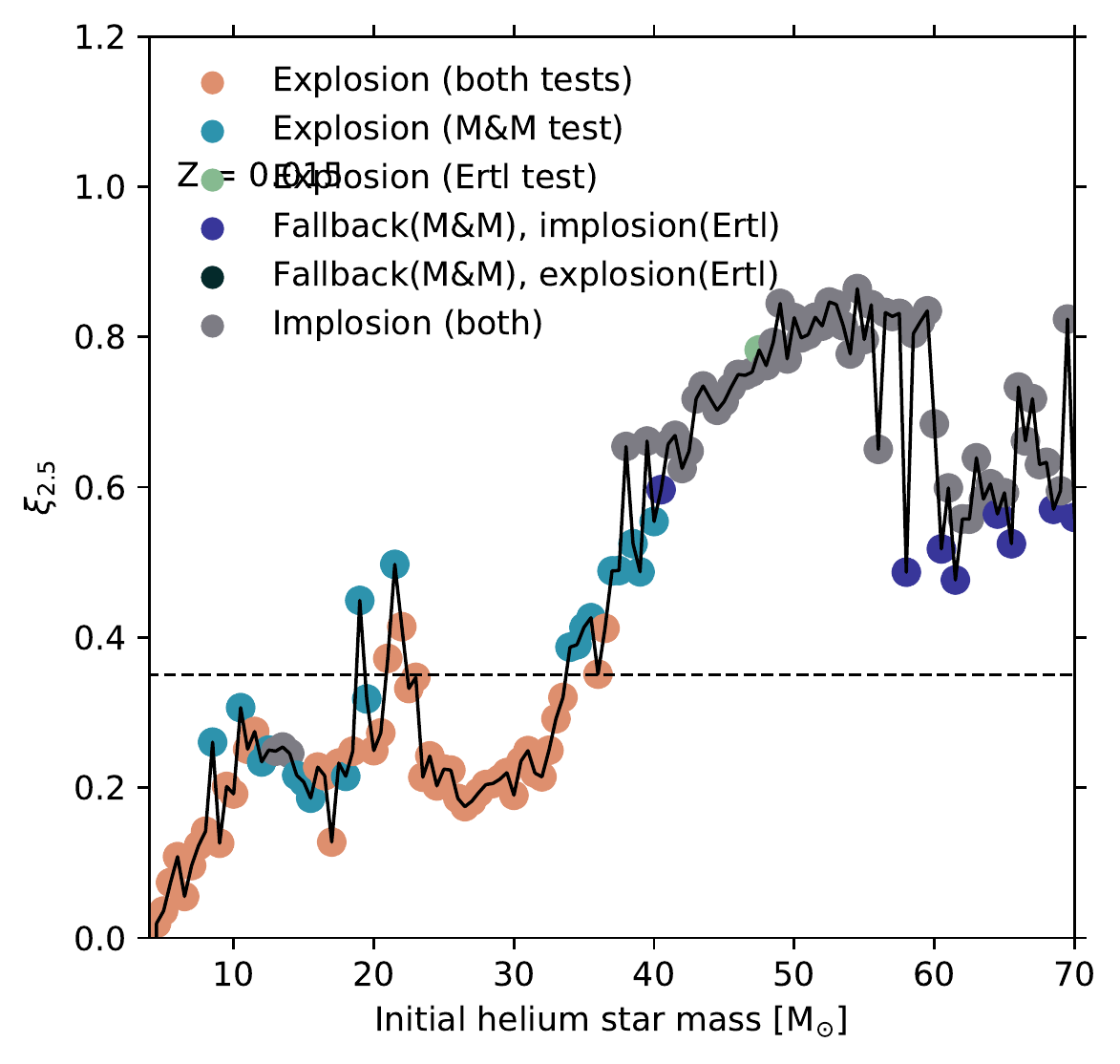}
\includegraphics[width=6cm]{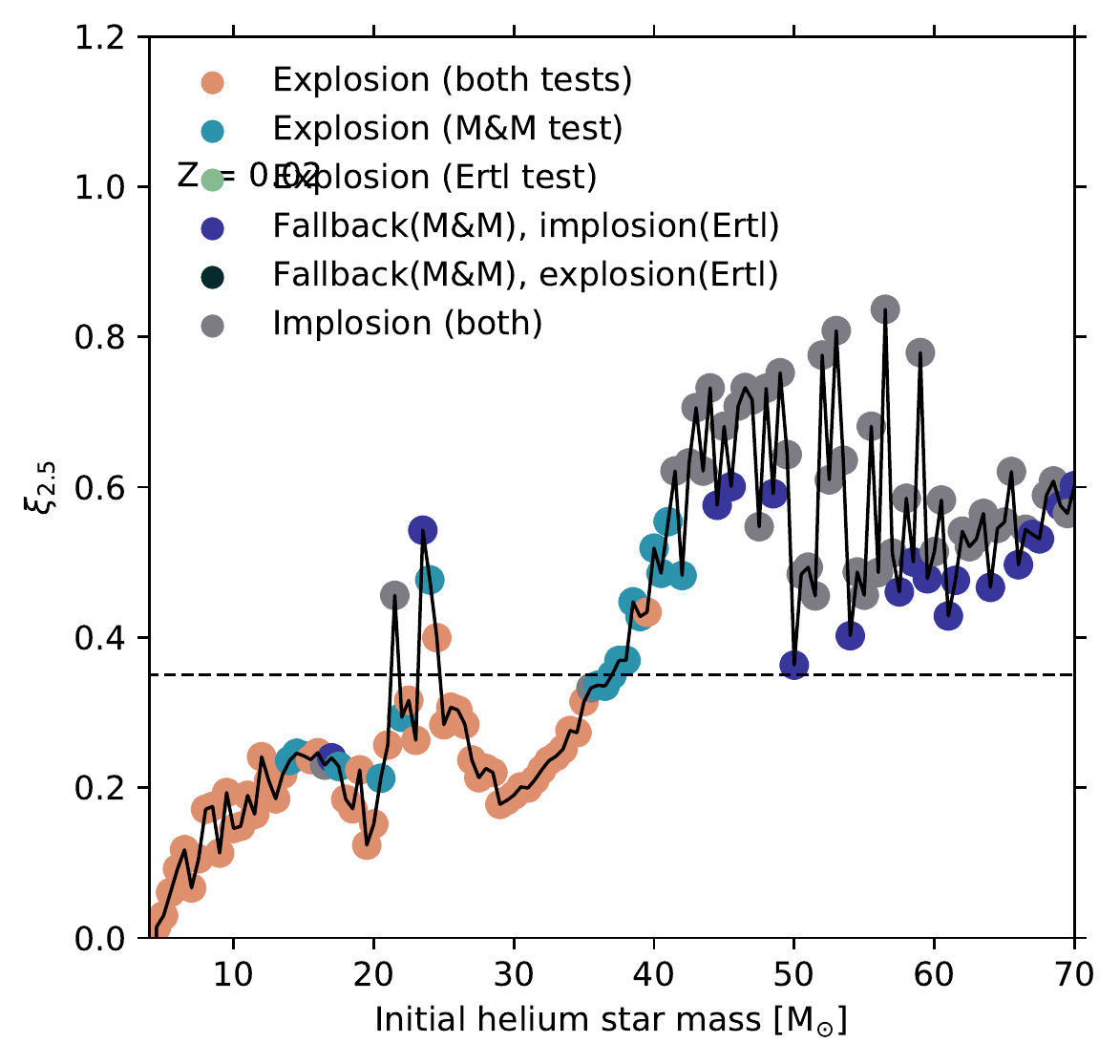}
\includegraphics[width=6cm]{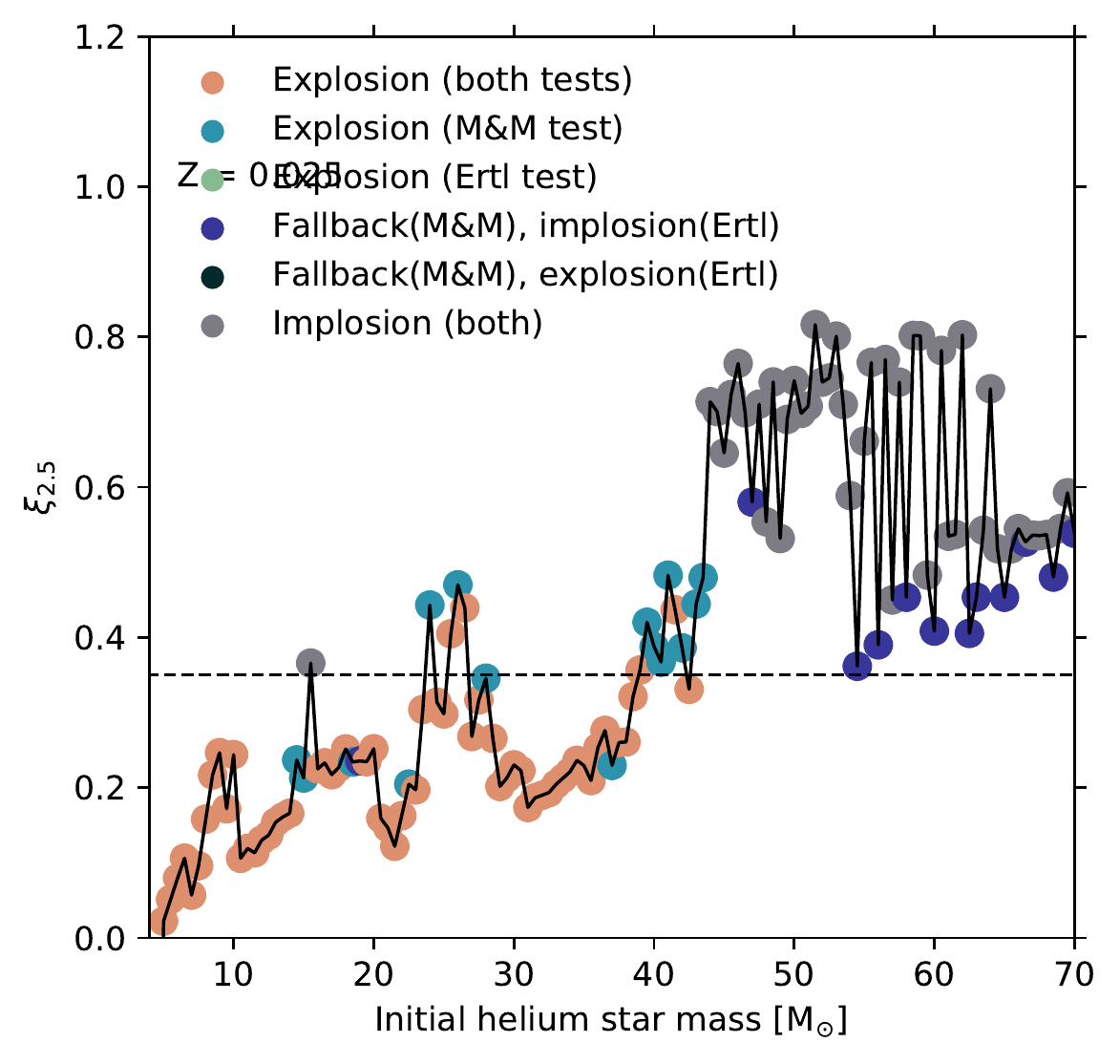}
\includegraphics[width=6cm]{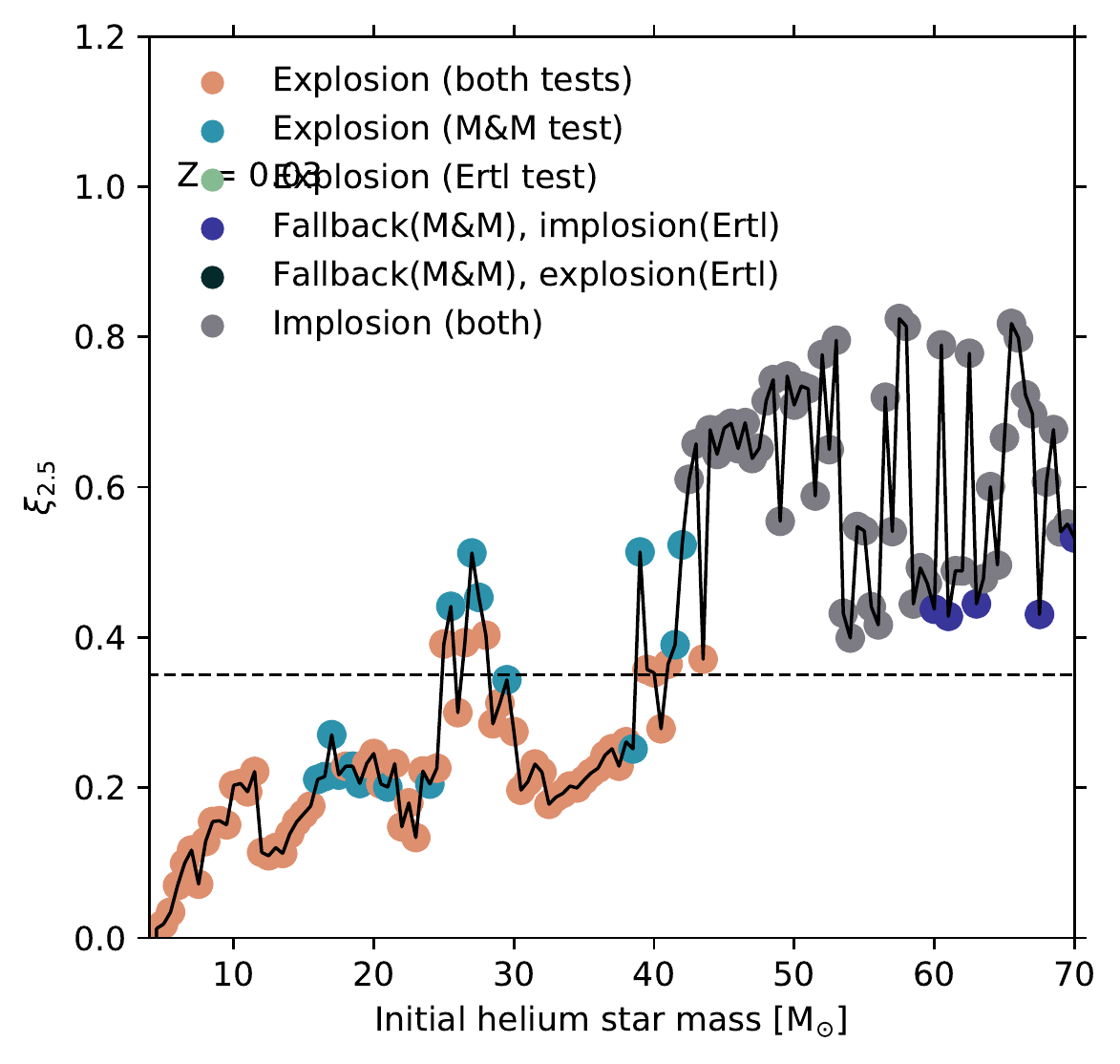}
\includegraphics[width=6cm]{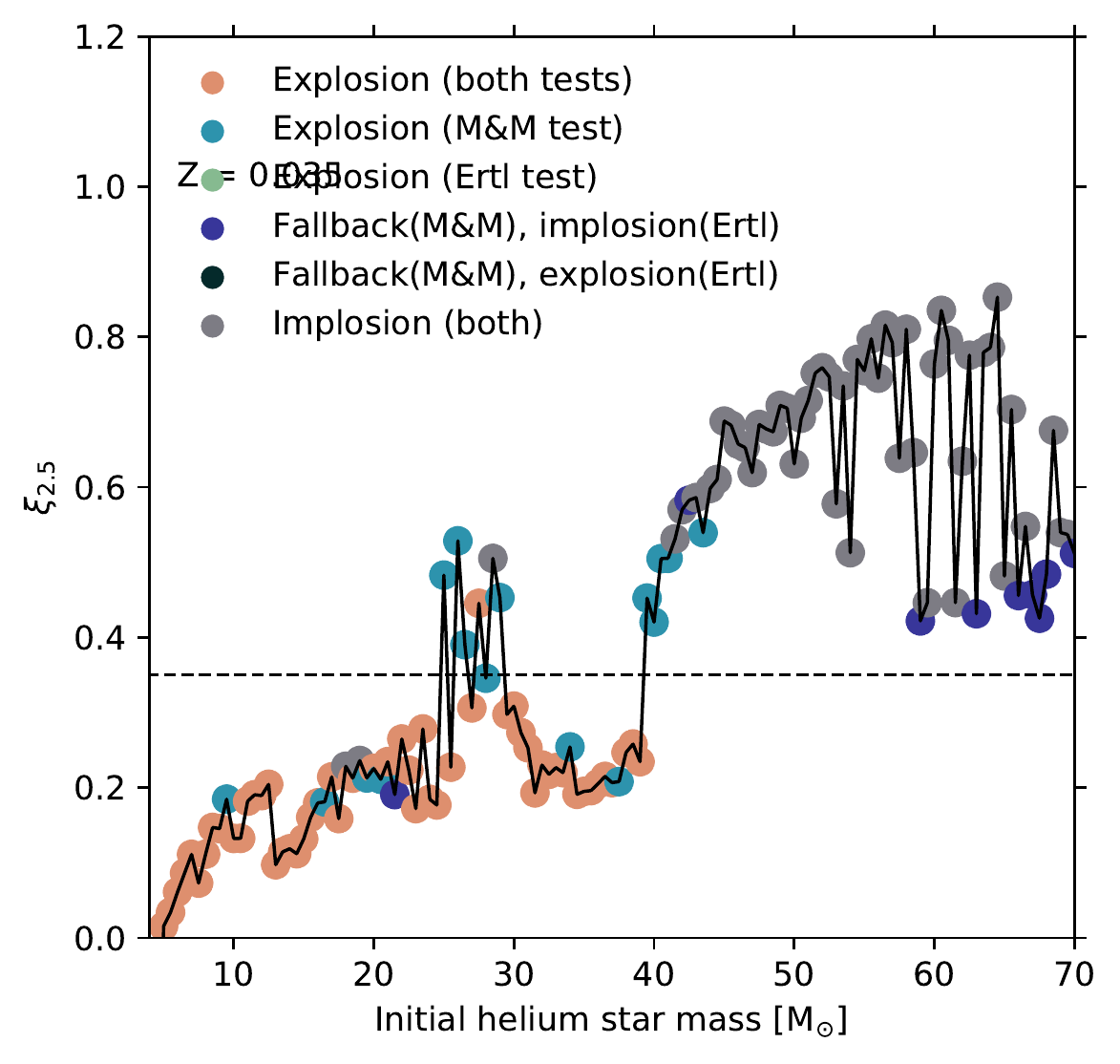}
\includegraphics[width=6cm]{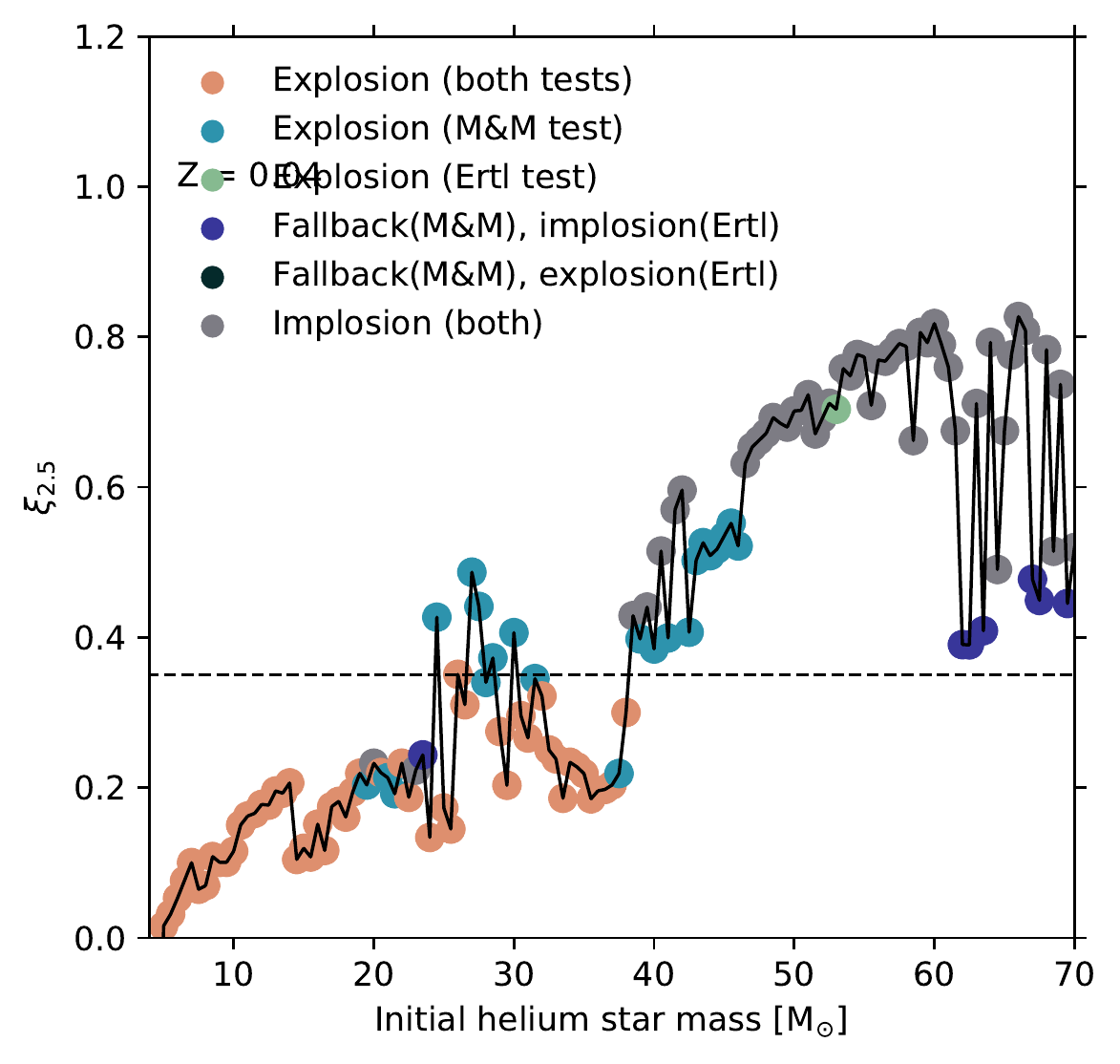}
\caption{Compactness parameter $\xi_{2.5} $of the core-collapse models in this study as a function of their initial helium star mass. The dotted line at $\xi_{2.5} = 0.35$ separates most models that might explode (below the line) or implode (above the line). Orange points indicate models that are predicted to explode according to the explodability of the \cite{2016ApJ...818..124E} and \cite{2016MNRAS.460..742M} tests (using their 
``standard'' parameters), blue points are models that are predicted to explode according the \cite{2016MNRAS.460..742M} test but not the \cite{2016ApJ...818..124E} test, green points are models that is predicted to explode according to the \cite {2016ApJ...818..124E} test but not the \cite{2016MNRAS.460..742M} test, and gray points are models that would not successfully explode according to both tests. \label{fig:all_tests_fallback}}
\end{figure*}

\begin{figure*}[h!]
\centering
\resizebox{\hsize}{!}{\includegraphics{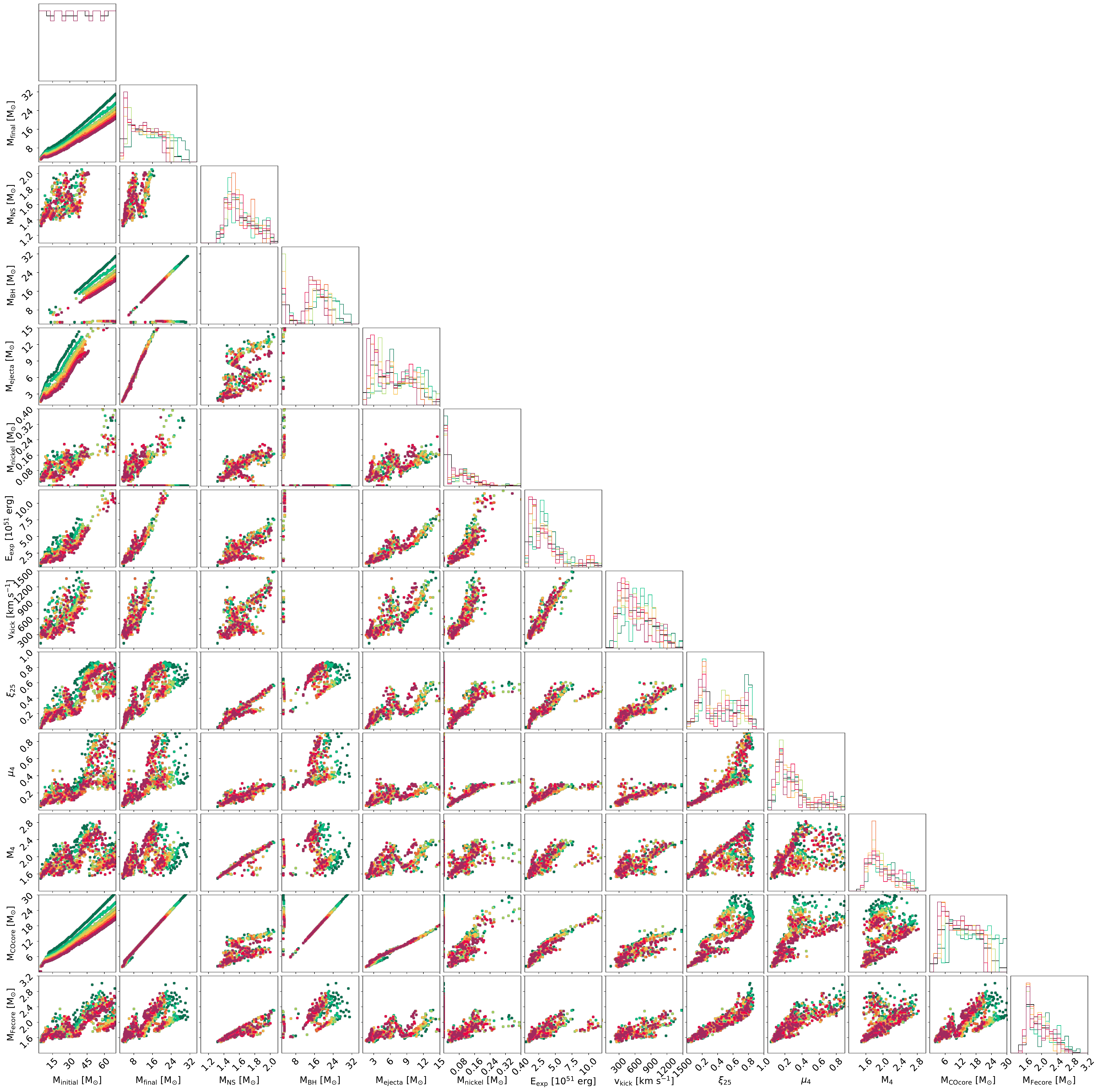}}
\caption{Summary of the distribution and relationships between a selection of stellar parameters from our simulations, and parameters resulting from the analysis of core collapse models with the \cite{2020MNRAS.499.3214M} model.
} \label{fig:everything}
\end{figure*}

Additional information is provided about the analysis performed on our models using the \cite{2011ApJ...730...70O}, \cite{2016ApJ...818..124E}, \cite{2016MNRAS.460..742M} and \cite{2020MNRAS.499.3214M} explodability tests in Figs. \ref{fig:all_tests}, \ref{fig:all_tests_fallback} and \ref{fig:everything}.

\section{Effect of input parameter variation in the explosion model}\label{sec:app_parameters}

In this Section, we describe the effect of varying the binding energy and the parameters that go into the \cite{2016MNRAS.460..742M} and \cite{2020MNRAS.499.3214M} explosion models.

In the main text, we calculate the outcome of the explosion models using a set of parameters given by $\alpha_\mathrm{out}=0.4$, $\alpha_\mathrm{turb}=1.18$, $\beta_\mathrm{expl}=4$, $\zeta=0.75$ and $\tau_{1.5}=1.2$. The binding energy that is input into the model is calculated as the sum of the (negative) gravitational energy, and the (positive) thermal energy at core collapse. The effect of kinetic energy is neglected (but kinetic energy is small compared with the other two components), and rotation is not included.

In Sect.\,\ref{sec:app_ener}, we show the effect of artificially increasing the binding energy of the core collapse models, and on Sect.\,\ref{sec:app_params}, we briefly describe the effect of varying the model parameters.

\subsection{Effects of binding energy variation on supernova explosions}\label{sec:app_ener}

To measure the effect of the binding energy of a stellar model at core collapse in the outcome of the \cite{2016MNRAS.460..742M} and \cite{2020MNRAS.499.3214M} explosion models, we compare the outcome of our fiducial set of parameters (Fiducial Set) with the outcome that results from calculating the explosion properties using the same set of parameters, but using the approximation of $E_\mathrm{bind} = 3GM^2/5R$ as a proxy for the binding energy (Set B). This approximation effectively decreases the absolute value of the binding energy outside of the iron core.

\begin{figure}
\centering
\resizebox{\hsize}{!}{\includegraphics{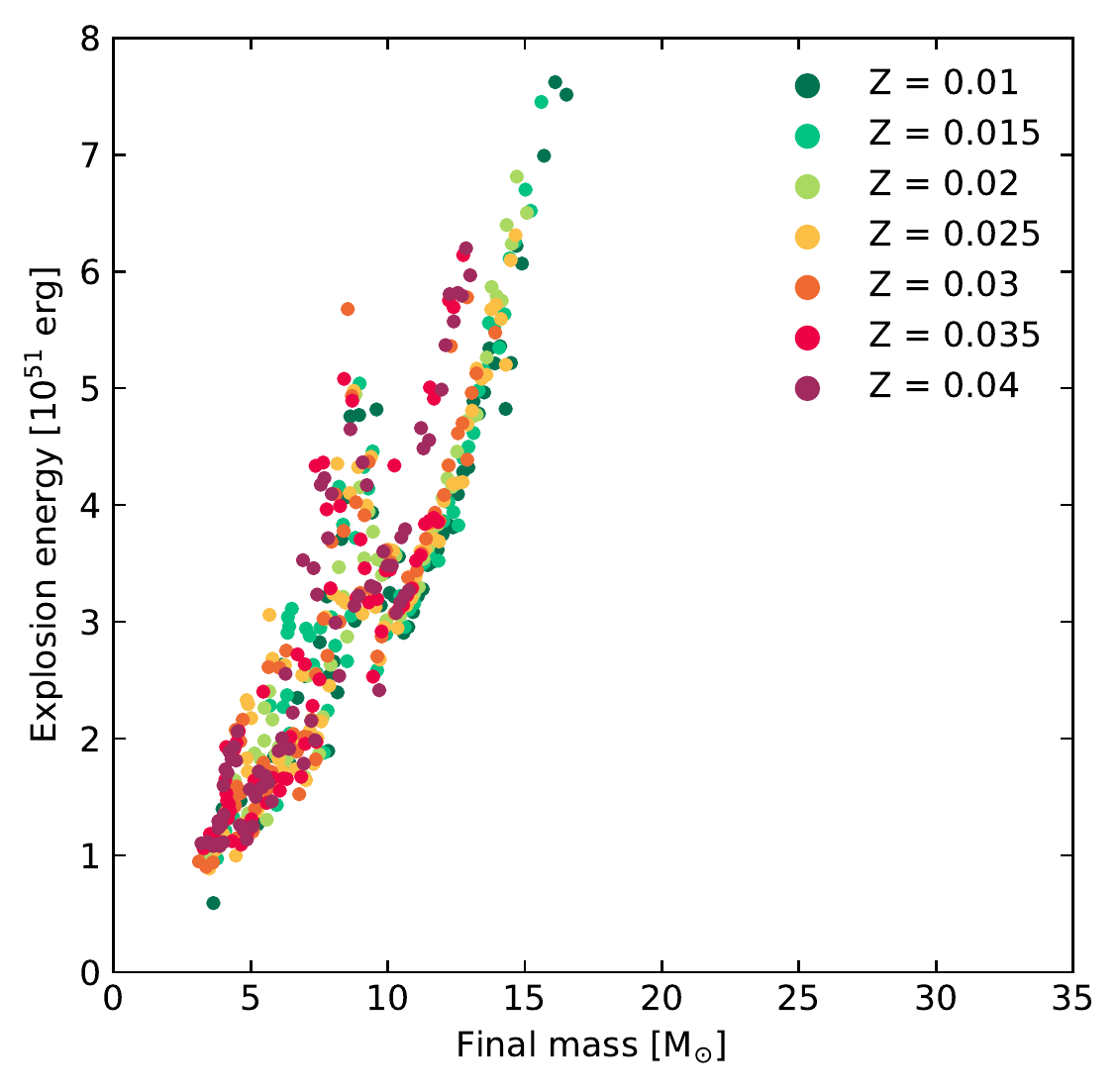}}
\resizebox{\hsize}{!}{\includegraphics{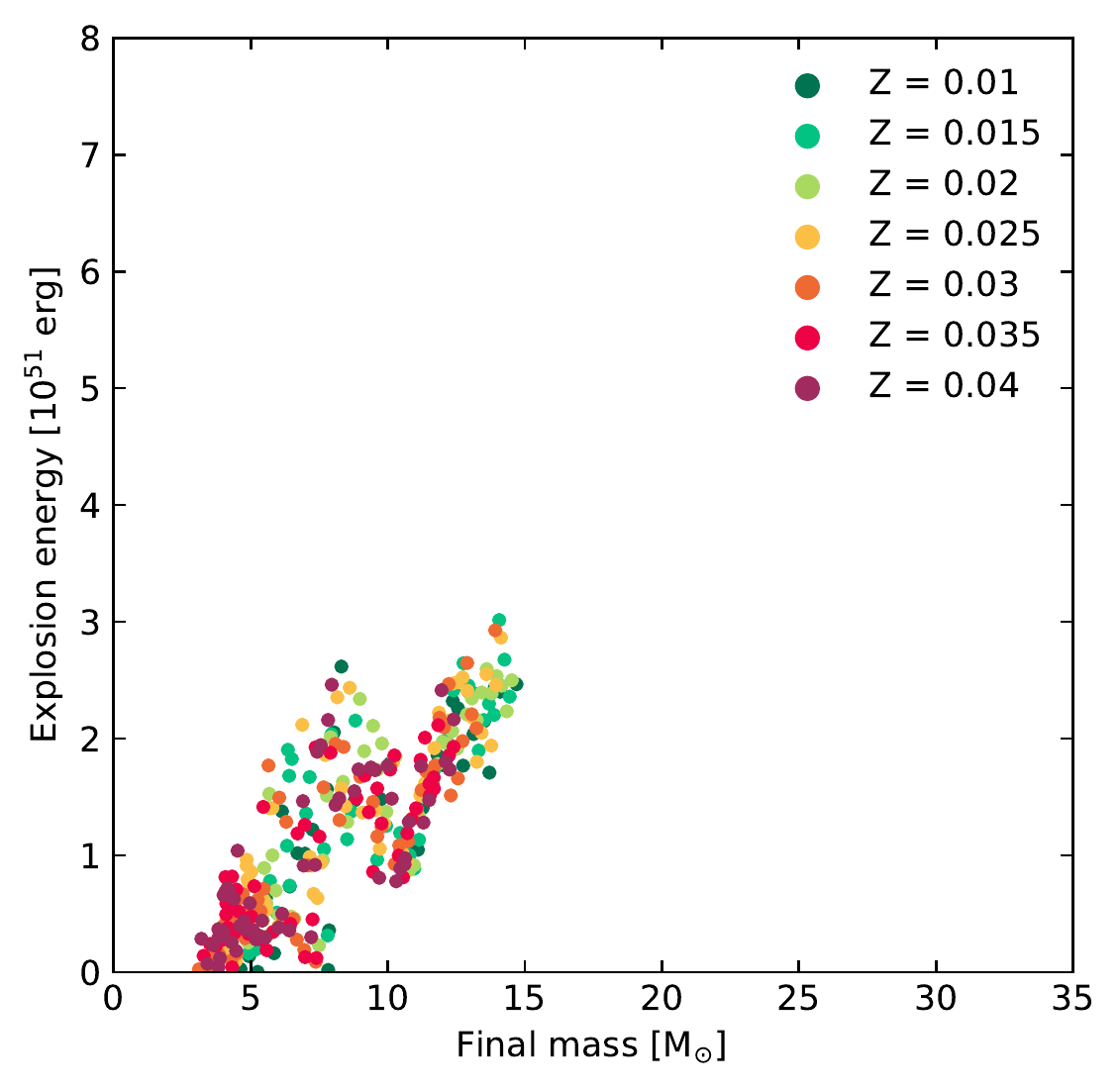}}
\caption{Explosion energy as a function of final mass from our Fiducial Set (top), and from Set B, calculated with the \cite{2016MNRAS.460..742M} model.
}\label{fig:energy_comp}
\end{figure}

As shown in Fig.\,\ref{fig:energy_comp}, the explosion energies are reduced to lower values in Set B, with values that are more in line with the results of \cite{2016MNRAS.460..742M}. These models have a lower $\mathrm{E}/\mathrm{M}$ ratio, ranging from about 0.2 and up to $4\times 10^{50} \ \mathrm{erg}/ \mso$, with most models having a value of around 2.

The distribution of NS masses is not heavily impacted, although a few models with initial masses in the range 4--10 $\mso$ are predicted to produce NSs of more than 1.5 $\mso$, whereas they produce NSs with masses lower than 1.4 $\mso$ in the Fiducial Set. However, fewer models result in successful explosions. In most cases, the models that result in successful explosions in the Fiducial Set, but do not produce explosions in Set B, are expected to produce explosions according to the \cite{2016ApJ...818..124E} criterion. This implies that the resulting distribution of BH masses, and the number of BHs per SN is impacted.

\begin{figure}
\centering
\resizebox{\hsize}{!}{\includegraphics{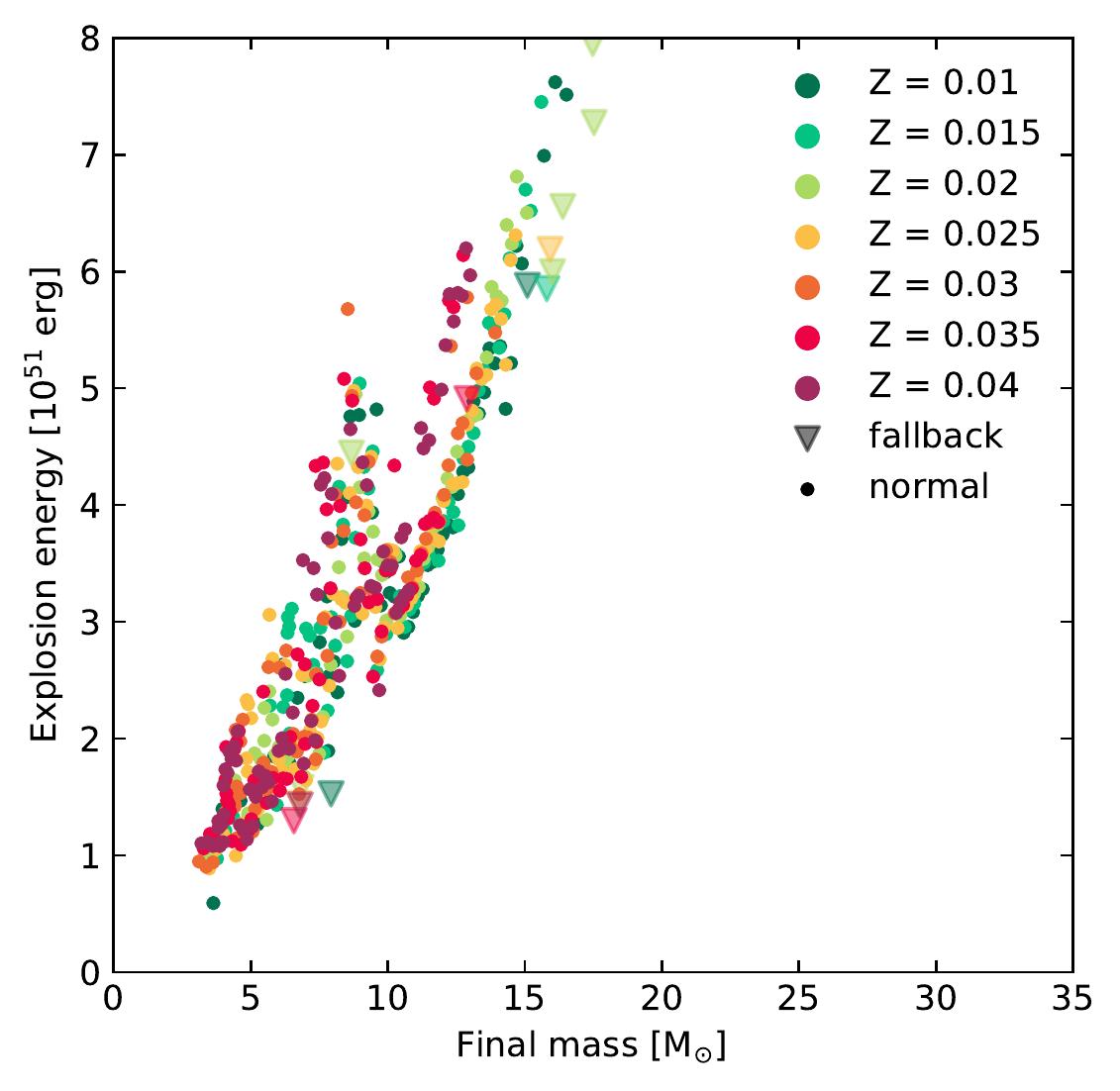}}
\resizebox{\hsize}{!}{\includegraphics{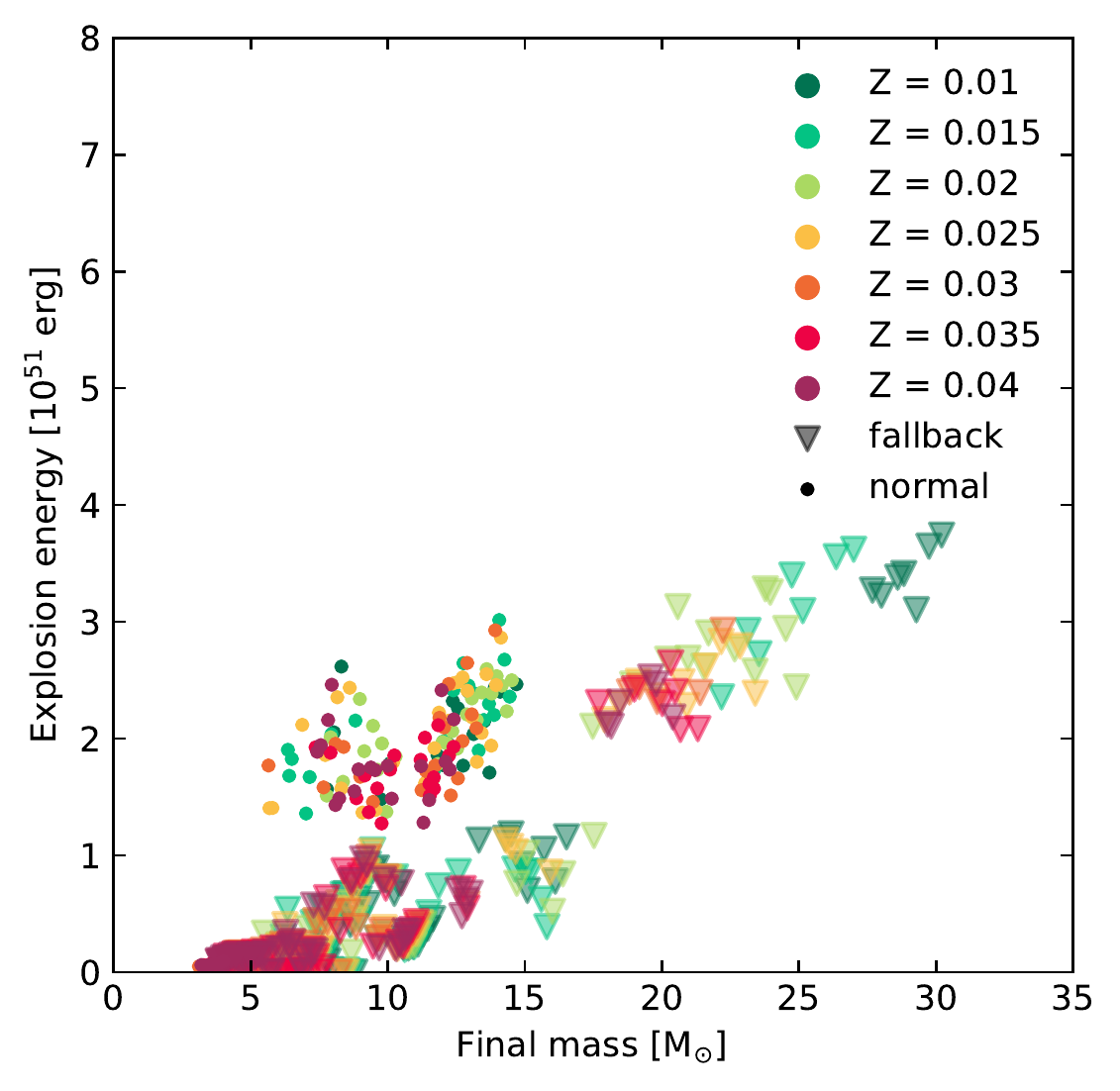}}
\caption{Explosion energy as a function of final mass from our Fiducial Set (top), and from Set B (bottom), calculated with the \cite{2020MNRAS.499.3214M} model. Circles in the figures correspond to normal SNe, whereas triangles correspond to fallback SNe.
}\label{fig:energy_comp_fb}
\end{figure}

When the effect of fallback is included in the calculations, as shown in Fig.\,\ref{fig:energy_comp_fb} many models that produce NSs in the Fiducial Set produce low energy explosions that leave behind mass gap BHs instead. In general, explosion energies are lower in Set B than in the Fiducial Set. In the latter case, fallback SNe from models with initial masses above about 15 $\mso$ produce explosions with energies that range from $6 \times 10^{51}$ erg, but that can reach energies of up to a few times $10^{52}$ erg (not shown in Fig\,\ref{fig:energy_comp_fb}). However, as shown in Sect. \ref{sec:sn_pop}, they are statistically rare and could be representative of rare, very luminous transients. On the other hand, fallback SNe in this regime in Set B produce explosions that are only a few times more energetic than typical SNe, and in many cases in the lower mass regime, such explosions are considerably less energetic than typical SNe.

The initial masses of the majority of progenitors of fallback SNe in the Fiducial Set range from about 35 and up to 70 $\mso$, whereas in Set B, models of all masses result in fallback SNe. This has a very large effect on the expected BH and NS mass distributions. In contrast to the Fiducual Set (see Figs. \ref{fig:NS_hist} and \ref{fig:BH_hist}) in Set B we expect most of the NSs resulting from a population of stripped-envelope stars to be around the more massive peak of the NS mass distribution, and most BHs produced are expected to be mass gap objects with masses of about 3 $\mso$. Notably, many fallback SNe from models with initial mass lower than 15 $\mso$ produce low energy explosions, have low nickel masses, and no kicks.

Nickel masses in Set B are lower than in the Fiducial Set, regardless of whether we include the effect of fallback or not. Kick velocities in Set B tend to be lower in Set B, and in particular some models with low explosion energy are also predicted to have 0 kick velocity.

\begin{figure}
\centering
\resizebox{\hsize}{!}{\includegraphics{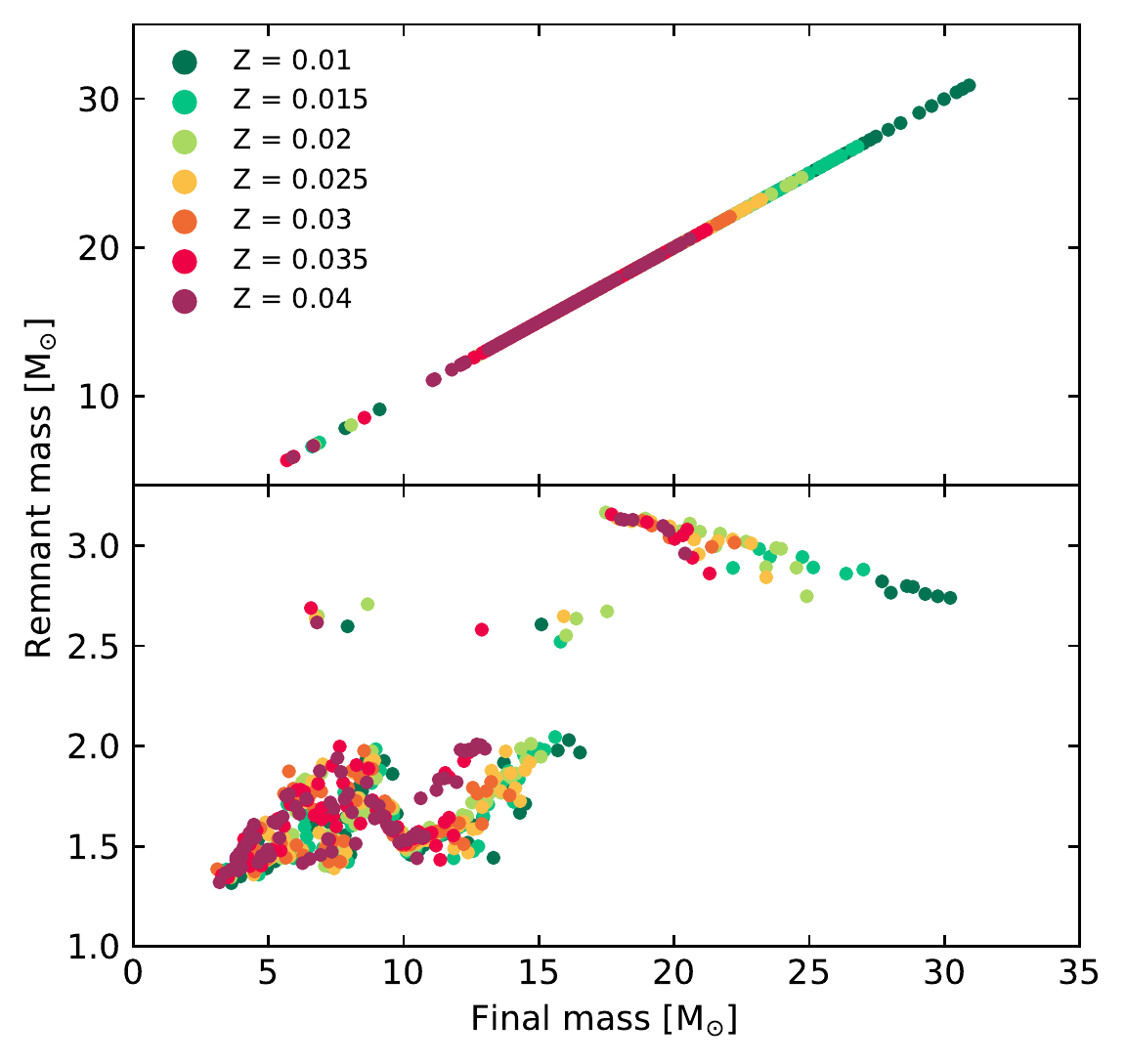}}
\resizebox{\hsize}{!}{\includegraphics{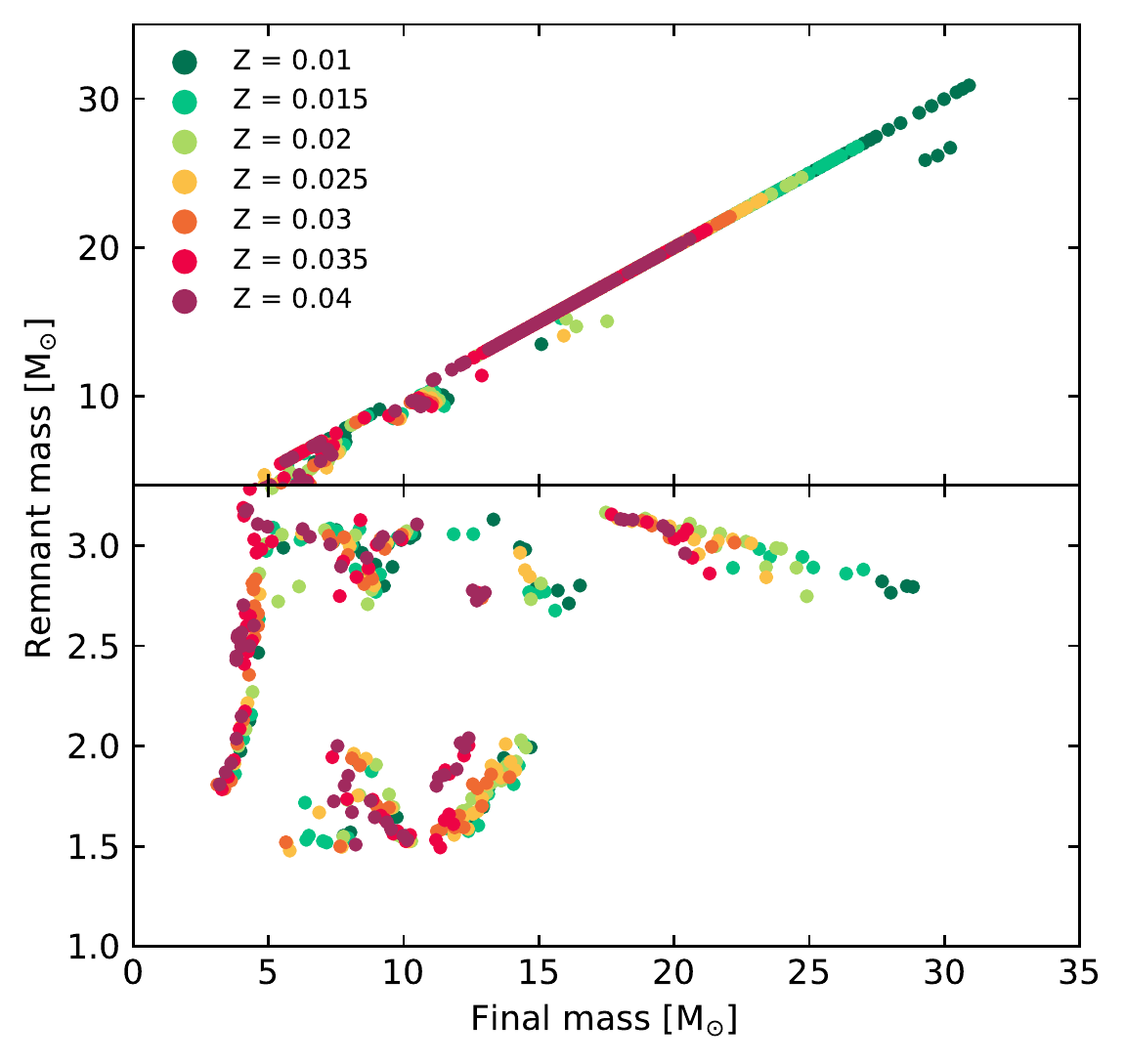}}
\caption{Remnant masses as a function of final mass from our Fiducial Set (top), and from Set B, calculated with the \cite{2016MNRAS.460..742M} model.
}\label{fig:rem_masses}
\end{figure}

Regardless of the choice of model, however, the resulting remnant mass distributions, shown in Fig.\,\ref{fig:rem_masses}, are similar to each other, and very different from typically assumed relations, such as the often employed relationships of \cite{2012ApJ...749...91F}. This demonstrates the importance of including detailed, physically motivated explodability prescriptions in population synthesis studies.

\subsection{Effects of parameter variation on supernova explosions}\label{sec:app_params}

The outcomes of the \cite{2016MNRAS.460..742M} and \cite{2020MNRAS.499.3214M} explosion models depends somewhat sensitively on the choice of input parameters. However, many trends and correlations persist in the explosion outcome regardless of the parameter choices.

In this Section, we describe general trends observed when varying the explosion parameters in both models. Besides output from the Fiducial Set and Set B, we calculated several sets of calculations varying each parameter independently, and keeping all others fixed, with the same base values as those employed by \cite{2016MNRAS.460..742M}.

Variations in $\alpha_\mathrm{out}$ result in a significant variation in the number of models that produce successful explosions. We changed the value of $\alpha_\mathrm{out}$ from 0.3 to 0.7 in steps of 0.1, and found that lower values result in fewer explosions, removing explosions that come from models with low $\xi_{2.5}$ (below the 0.35 threshold shown in Fig.\,\ref{fig:compactness}), ant that have low energy in comparison to the rest of the population. Similarly, explosions with low nickel mass and low kicks are similarly removed when increasing the value of $\alpha_\mathrm{out}$. The resulting explosion energies have lower energies than models of similar mass in the Fiducial Set, and a few NSs are formed that are more massive than expected from the correlation between $\xi_{2.5}$ and NS mass.

We varied the value of $\zeta$ from 0.5 to 0.9 in steps of 0.1. We found that as $\zeta$ is increased, the initial mass threshold between models that produce successful explosions and those that produce BHs is increased. These explosions have a similar distribution of $\mathrm{E}/\mathrm{M}$ than the rest of the explosions with the same given set of parameters, and therefore are increasingly more energetic. Similarly, such explosions have a higher nickel mass and kick velocity.

We varied the value of $\alpha_\mathrm{turb}$ from 1.14 to 1.22 in steps of 0.02, and of $\tau_{1.5}$ from 1.2 to 1.5 in steps of 0.1, and found no significant difference in the outcome of our calculations.

The set of parameters selected as our Fiducial Set was therefore calibrated varying $\alpha_\mathrm{out}$ and $\zeta$ to yield a better agreement with the explodability landscape predicted by the \cite{2016ApJ...818..124E} criterion, resulting in the landscape shown in Fig.\,\ref{fig:all_tests}.

\end{appendix}
\end{document}